\documentclass[aps,prb,a4paper,twocolumn,amssymb,superscriptaddress,amsfonts,amsmath,floatfix]{revtex4-1}
\usepackage{docs}%
\usepackage{bm}%
\expandafter\ifx\csname package@font\endcsname\relax\else
 \expandafter\expandafter
 \expandafter\usepackage
 \expandafter\expandafter
 \expandafter{\csname package@font\endcsname}%
\fi
\usepackage{graphicx,epsfig,color}
\usepackage[utf8]{inputenc}
\usepackage{amssymb,amsfonts,amsmath}
\usepackage{verbatim}
\usepackage{tikz}
\usepackage{bbold}
\usepackage{hyperref}
\usepackage{fancyhdr}
\usepackage{hyperref}
\usepackage{makecell}
\usepackage{lastpage}
\usepackage[all]{xy}

\usepackage{bbm}
\usepackage{chngcntr}
\counterwithin*{equation}{section}
\usepackage{float}

  \newcommand{\cL}{{\cal L}}

\newcommand{\be}{\begin{equation}} \newcommand{\ee}{\end{equation}}
\newcommand{\bea}{\begin{eqnarray}} \newcommand{\eea}{\end{eqnarray}}
\newcommand{\beann}{\begin{eqnarray*}}  \newcommand{\eeann}{\end{eqnarray*}}
\newcommand{\bfig}{\begin{figure}} \newcommand{\efig}{\end{figure}}
\newcommand{\ba}{\begin{array}} \newcommand{\ea}{\end{array}}
\newcommand{\bcen}{\begin{center}} \newcommand{\ecen}{\end{center}}
\newcommand{\btab}{\begin{tabular}} \newcommand{\etab}{\end{tabular}}

\newcommand{\Eq}[1]{(\ref{#1})}
     \def\diag{\operatorname{diag}}

\newcommand{\vev}[1]{\left\langle{#1}\right\rangle}

\newcommand{\e}{{\rm e}}

\newtheorem{Proposition}{Proposition}[section]

\newtheorem{Theorem}{Theorem}[section]
\newtheorem{Lemma}{Lemma}[section]
\newtheorem{Corrolary}{Corrolary}[section]

\newcommand{\bp}{\begin{Proposition}}   \newcommand{\ep}{\end{Proposition}}
\newcommand{\bt}{\begin{Theorem}}   \newcommand{\et}{\end{Theorem}}
\newcommand{\bl}{\begin{Lemma}}     \newcommand{\el}{\end{Lemma}}
\newcommand{\bc}{\begin{Corrolary}} \newcommand{\ec}{\end{Corrolary}}

\def\e{\eta}
\def\ep{\epsilon}

\def\a{\alpha}

\def\s{\sigma}
\def\m{\mu}
\def\n{\nu}
\def\r{\rho}

\newcommand{\refeq}[1]{(\ref{#1})}

\relax

\def\be{\begin{equation}}
\def\ee{\end{equation}}
\def\bea{\begin{eqnarray}}
\def\eea{\end{eqnarray}}
\newcommand\fverb{\setbox\pippobox=\hbox\bgroup\verb}
\newcommand\fverbdo{\egroup\medskip\noindent%
                        \fbox{\unhbox\pippobox}\ }
\newcommand\fverbit{\egroup\item[\fbox{\unhbox\pippobox}]}

\newcommand{\bwt}{\begin{widetext}}
\newcommand{\ewt}{\end{widetext}}
\newcommand{\bear}{\begin{eqnarray}}
\newcommand{\eear}{\end{eqnarray}}

\newcommand{\bsea}{\begin{subeqnarray}}
\newcommand{\esea}{\end{subeqnarray}}
\newbox\pippobox

\def\6{\partial}

\def\a{\alpha}

\def\pa{\partial}

\def\e{\epsilon}
\def\m{\mu}
\def\n{\nu}
\def\r{\rho}
\def\s{\sigma}

\def\spa{\;,\;}

\def\sq
\def\a{\alpha}

\def\h{\eta}
\def\e{\epsilon}

\def\cL{{\cal L}}

\begin{document}

\begin{abstract}

    In the context of describing electrons in solids as a fluid in the
    hydrodynamic regime, we consider a flow of electrons in a channel of
    finite width, i.e.~a Poiseuille flow.  The
    electrons are accelerated by a constant electric field. We develop
    the appropriate relativistic hydrodynamic formalism in 2+1 dimensions and show
    that the fluid has a finite dc conductivity due to
    boundary-induced momentum relaxation,  even in the absence of
    impurities. We use methods involving the AdS/CFT correspondence to
    examine the system in the strong-coupling regime. We calculate and study
    velocity profiles across the channel, from which we obtain the
    differential resistance $dV/dI$. We find that $dV/dI$ decreases
    with increasing current $I$ as expected for a Poiseuille flow, also 
    at strong coupling and in the relativistic velocity regime. Moreover, we vary the coupling strength by varying
    $\eta/s$, the ratio of shear viscosity over entropy density. We
    find that $dV/dI$ decreases when the coupling is
    increased. We also find that strongly coupled fluids are more
    likely to become ultra-relativistic and turbulent. These conclusions are insensitive to the presence of impurities.
    In particular, we predict that in channels which are clearly in the hydrodynamic regime already at small currents, the DC channel resistance strongly depends on $\eta/s$.
   

\end{abstract}

	\author{Johanna Erdmenger, Ioannis Matthaiakakis, Ren\'e Meyer, David Rodr\'iguez Fern\'andez}
	\affiliation{Institute for Theoretical Physics and Astrophysics, \\
	Julius-Maximilians-Universit\"at W\"urzburg, 97074 W\"urzburg, Germany}
	\title{Strongly coupled electron fluids in the Poiseuille regime}
	\date{\today}
	\maketitle
\newpage



\section{Introduction}\label{sec:introconc}

In recent years, hydrodynamic behavior in electron systems has
received considerable interest from both theoretical and experimental
condensed matter physics.~\cite{Narozhny:2017vc,Lucas:2017wa}. Experimentally, the
hydrodynamic regime was first  reached in (Al,Ga)As high mobility
wires more than two decades ago when the Gurzhi effect was
observed \cite{Molenkamp:1994kb,Molenkamp:1994ii,deJong:1995bn}, and
more recently in several other materials including graphene
\cite{Crossno:2015iy,Moll:2016ju,bandurin2016negative,nam2017electron}. The
Gurzhi effect \cite{Gurzhi:1968ji} in such wires is a crossover
between boundary-dominated scattering (the Knudsen regime) at low
densities and hydrodynamic behavior at higher densities (the
Poiseuille regime). It manifests itself as a crossover in the
channel resistance as a function of the applied current: The channel resistance rises
in the Knudsen regime, while it falls in the Poiseuille regime.  
In spite of the experimental successes mentioned, 
observing clear signs of viscous hydrodynamic transport or exactly
measuring the value of the viscosity turns out to be challenging in 
electronic systems in general \footnote{In order to meaningfully
  extract the hydrodynamic shear viscosity, the conditions for
  hydrodynamics need to apply as a low-energy effective theory. These
  conditions are e.g. spelled out in sec.~\ref{sec21}. Otherwise the
  result for the viscosity from e.g. the Kubo formula for the
  viscosity in quantum field theory or from effective approaches such as kinetic theory will not have a meaningful hydrodynamic interpretation as the viscosity of a fluid in local thermal equilibrium.}. One key reason is that the
interpretation of averaged observables such as e.g. voltages in
nonlocal measurements (proposed e.g. as a signature of hydrodynamic
whirlpools \cite{levitov2016electron}) depend crucially on the boundary
conditions \cite{Pellegrino:2016kp} of the measurement setup. 
Nevertheless, theoretical proposals exist to measure the 
viscosity using non-local transport in Hall bars \cite{Torre:2015eja},
rheometers \cite{Tomadin:2014by}, and AC transport
\cite{PhysRevB.97.161112}. Also, in the presence of impurity
scattering, the interesting hydrodynamic regime may actually be
restricted to a narrow range of parameters, as can e.g. be seen from the  
weak coupling phase diagram of the relativistic Gurzhi effect derived
from kinetic theory\cite{2018arXiv180502987K}. 

In order to make progress towards resolving some of the issues
mentioned, in the present paper we investigate hydrodynamic behavior as realized
by a Poiseuille flow in a relativistic strongly coupled system. This is
conveniently done by using methods of the AdS/CFT 
correspondence \cite{Maldacena:1997re,Gubser:1998bc,Witten:1998qj}. In
particular, we  present a model where we are able to investigate
the dependence of the flow on the coupling strength starting from very
large couplings by varying only one parameter, which is the ratio of
shear viscosity over entropy density.

The AdS/CFT conjecture provides universal predictions for observables of
strongly coupled and correlated systems~\cite{CasalderreySolana:2011us,Ammon:2015,zaanen2015holographic,Hartnoll:2016apf}. 
 One such prediction is that the low-energy excitations of strongly coupled gapless systems described via AdS/CFT are governed by the laws of hydrodynamics. The hydrodynamic equations of motion are conservation laws for the only long-lived excitations expected in a strongly interacting system. These are the long-wavelength, low-energy excitations of conserved quantities such as e.g. energy, momentum, or charge. 
 This is not surprising since a very strongly coupled gapless system at low energies is expected to have a very short equilibration time and mean free path beyond which the system behaves hydrodynamically. In the absence of additional conserved quantities or other parametric suppressions by ratios of scales,\footnote{There are examples where hidden conserved quantities together with a long-range disorder potential invalidate the hydrodynamic approximation~\cite{Lucas:2017vlc}.} the equilibration time will be of order \cite{zaanen2004superconductivity}
\be\label{taueq}
\tau_{eq} \sim  \frac{\hbar}{k_B T}\,.
\ee
In particular, \eqref{taueq} holds in all rotationally invariant holographic models of electronic transport with a stable ground state.  In these models, the approach to local equilibrium is governed by the  most long-lived non-hydrodynamical quasinormal mode of the dual black brane, which has a lifetime of order \eqref{taueq}\footnote{For an explanation of this fact, c.f.~e.g.~sec.~3.5~of~\cite{Hartnoll:2016apf}.}. 
However, an equilibration time of order \eqref{taueq} is not a necessary condition for hydrodynamic behaviour. 
The conditions of applicability of hydrodynamics as a low-energy effective theory will be discussed in sec.~\ref{sec21}.

The AdS/CFT correspondence  allows not only to derive hydrodynamics from perturbations of black branes in AdS space-time \cite{Bhattacharyya:2008jc,Bhattacharyya:2008xc,Bhattacharyya:2008ji,Blake:2015epa}, but also the values of the transport coefficients (viscosities, conductivities, etc.). Calculating the shear viscosity from AdS/CFT  for rotationally invariant systems \cite{Policastro:2001yc}, the ratio of shear viscosity to entropy density takes a universal value 
\be\label{etasAdSCFT}
\frac{\eta}{s} = \frac{1}{4\pi}\frac{\hbar}{k_B}\,,
\ee
independent of the other parameters of the system such as
e.g. temperature or chemical potential. It is conjectured that\cite{Kovtun:2004de}   \eqref{etasAdSCFT} constitutes a universal
minimal bound for all interacting quantum systems,\footnote{For
  rotationally non-invariant systems, there can be violations of the
  bound,~\cite{Mateos:2011ix,Rebhan:2011vd}. However, all
  such violations that have been found so far to leading order in the
  inverse string tension  $\alpha'$ (dual to the coupling constant) or
  the gauge group rank $N$ (controlling the large $N$ limit) occur in  unstable ground state,~\cite{Mateos:2011ix,Mateos:2011tv}. On the other hand, in systems with rotationally non-invariant but obviously stable ground states, the KSS bound \protect\eqref{etasAdSCFT2} seems to be observed for all components of the viscosity tensor \cite{Erdmenger:2012zu}. To subleading order in $\alpha'$ (corresponding to finite coupling corrections), violations of \protect\eqref{etasAdSCFT2} induce causality violations in the dual field theory \cite{Brigante:2008gz,Camanho:2014apa}. Presumably only  string theories including all order $\alpha'$ corrections is causal \cite{Camanho:2014apa}, but $\eta/s$ has never bee calculated in such a setup. Similar remarks apply to finite $N$ corrections \cite{Kats:2007mq} which also correspond to higher derivative terms in the dual gravitational theory. In summary, there is good evidence\cite{Cremonini:2011iq} that \eqref{etasAdSCFT} holds in the stable ground states of all consistent models.} 
\be\label{etasAdSCFT2}
\frac{\eta}{s} \geq \frac{1}{4\pi}\frac{\hbar}{k_B}\,.
\ee
This so-called Kovtun-Son-Starinets (KSS) bound can be motivated  by e.g. scaling arguments in quantum critical phases \cite{zaanen2004superconductivity,Delacretaz:2018cfk,Hartman:2017hhp}, and ultimately should be related to fundamental properties of transport in interacting quantum systems. 

It is of great interest to experimentally test the predictions
\eqref{taueq}-\eqref{etasAdSCFT2}. In heavy-ion collision
experiments, $\eta/s$ was
experimentally found to be of the same order as \eqref{etasAdSCFT} \cite{song2011200}, albeit with large error bars.  The main issue in extracting $\eta/s$ in heavy-ion collisions is the very indirect dependence of the final state particle distributions measured in the detector on the value of the hydrodynamic transport coefficients. The transport properties of strongly coupled and correlated electron systems in the hydrodynamic regime could hence provide a more direct window into the strong coupling regime of interacting quantum systems.\footnote{For electrons in solids we have direct access to the hydrodynamic regime. In QCD, we need to reconstruct the hydrodynamic regime from the data collected by particle detectors back through the hadronization crossover.} In particular, two-dimensional Dirac materials such as graphene \cite{novoselov2005two} or the surface states of topological insulators such as e.g. HgTe \cite{konig2007quantum}, as well as Weyl- and Dirac semimetals in three spatial dimensions \cite{Klinkhamer:2004hg,Volovik:2003fe,PhysRevB.83.205101} are of interest here due to their relativistic band structures. If electron-electron interactions are the fastest way for electrons to redistribute their energy and momentum and equilibrate, the electrons should form a relativistic fluid governed by the equations of relativistic hydrodynamics.

\begin{center}
 	\begin{figure}
 		\includegraphics[scale=0.5]{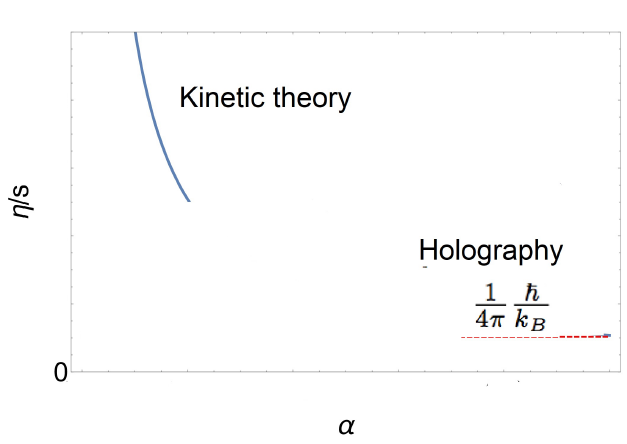}
 		\caption{Sketch of $\eta/s$ as a function of the coupling strength. At low $\alpha$, perturbation theory predicts a $1/\alpha^2$ behavior, which asymptotes to a constant value in the large $\alpha$ limit.\cite{} Holography predicts the value of this constant to be $\hbar/4\pi k_B$ .\label{fig:etas_alpha} }
 	\end{figure}
 \end{center}

The ratio $\eta/s$ depends explicitly on the coupling constant, and hence serves as a measure of the interaction strength itself. In the weak coupling regime, $\eta/s$ can be calculated in perturbation theory. Increasing the interactions, one expects the system to enter the strongly coupled regime and saturate the bound \eqref{etasAdSCFT2}. Kinetic theory calculations of $\eta/s$ for Coulomb interacting two-dimensional Dirac materials support this picture: At charge neutrality and to first order in the effective temperature dependent Coulomb coupling $\alpha(T)$, $\eta/s$ was found to be\cite{Fritz:2008go,Muller:2009cy} 
\be\label{etasFritzetal}
\frac{\eta}{s} =  \frac{C_\eta \pi}{9 \zeta(3)} \frac{1}{\alpha^2(T)} \frac{\hbar}{k_B} \approx \frac{1.64}{4\pi} \frac{1}{\alpha^2(T)}\frac{\hbar}{k_B} \,,
\ee
where $C_\eta \approx 0.449$ is a numerical constant. As $\alpha$ enters the strong coupling regime, $\alpha \simeq 1$, $\eta/s$ is expected to approach the holographic value \eqref{etasAdSCFT}.  The behavior of $\eta/s$ as a function of $\alpha$ is shown in fig.~\ref{fig:etas_alpha}. Hence, if the hydrodynamic regime can be reached experimentally in materials with large Coulomb coupling, and if observables sensitive to $\eta/s$ can be found, it will become possible to experimentally test the AdS/CFT prediction \eqref{etasAdSCFT} in hydrodynamic electron systems.

\begin{figure}[tbp]
		\includegraphics[scale=0.30]{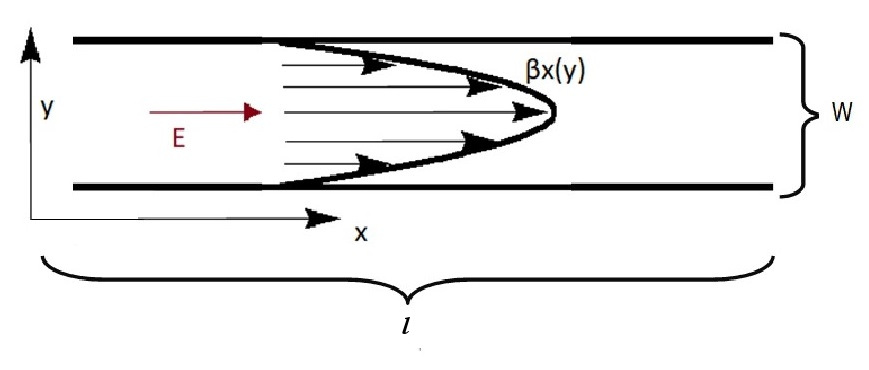}
		\caption{Schematic flow of charged particles along a $xy$ plane in the Poiseuille flow. The fluid in this regime is incompressible and has zero velocity at the boundaries, such that the maximum speed is attained in the middle of the channel. In the $\beta_x\ll v_F$ approximation, the profile acquires a characteristic parabolic shape.\label{fig:pois}}
\end{figure}

In view of further progress in detecting hydrodynamic behaviour,
the channel setup
\cite{Molenkamp:1994kb,Molenkamp:1994ii,deJong:1995bn,2018arXiv180502987K} 
showing the two-dimensional  Gurzhi effect \cite{Gurzhi:1968ji}
described above is probably the simplest setup possible. In
hydrodynamical terms, this corresponds to the simplest 
flow geometry in which a controlled hydrodynamic flow can be
achieved, such that it may be easiest to find observables
sensitive to changes of the value of the ratio $\eta/s$ alone. 
The channel setup consists of two parallel boundaries with prescribed boundary conditions, and a laminar flow in between in the presence of an electric field along the channel direction. Two different choices of boundary conditions are common:\cite{Lucas:2017wa} The boundary condition of vanishing velocity at the channel wall gives rise to the Poiseuille flow, which is depicted in~Fig.~\ref{fig:pois}. The choice\footnote{In experiment, the channels are typically prepared by etching them out of a thin layer of material. During that process, the channel edges are sufficiently disordered, such that zero velocity boundary conditions are applicable.} of vanishing stress at the boundaries leads to an Ohmic flow. 

In our work, we take the channel setup as a starting point and provide a detailed study of parametric effects for a relativistic charged Poiseuille flow in the absence of parity or time reversal breaking. We in particular study the dependence on the value of $\eta/s$ of the Poiseuille flow and observables derived from it such as the differential wire resistance. We vary $\eta/s$ independently of  the impurity relaxation time, channel width, chemical potential, temperature and external electrical field. Varying the value of $\eta/s$ from \eqref{etasAdSCFT} to larger values allows us in particular to interpolate from the strong to the intermediate coupling regime, and in this way to search for physical signatures of strongly coupled fluids.\footnote{In this work, we assume that we can vary $\eta/s$ independently of $s$. This amounts to varying the coupling strength independent of the temperature, in which case, the entropy density $s$ at fixed temperature is a constant in thermal and chemical equilibrium. We leave the investigation on the possible experimental realization for the future.} 

In the original form of the AdS/CFT correspondence, the strong coupling regime associated with \eqref{etasAdSCFT} involves taking the limit in which the coupling strength is infinite. Moving to lower coupling strengths through changing the value of $\eta/s$ requires us to specify the coupling dependence of this ratio. We make the assumption that $\eta/s$ is a monotonically decreasing function of the coupling constant $\alpha$, i.e. we interpolate from the extremely strongly coupled regime towards intermediate weaker coupling strengths. Our guiding principle for this interpolation is the known behavior at weak and very strong coupling, as shown in Fig.~\ref{fig:etas_alpha}. We further assume that hydrodynamics is applicable throughout this interpolation.\footnote{\label{foot:newhydromodes}The appearance of new non-hydrodynamic modes at finite coupling\cite{grozdanov2016strong} may invalidate the hydrodynamic approximation. However, we note that in the present work our analysis is performed at finite density in 2+1 dimensions, while these results\cite{grozdanov2016strong} apply to 3+1 dimensions at vanishing density. Since the structure of higher derivative terms in 3+1 dimensional gravity will be very different from the ones in 4+1 dimensional gravity, these results\cite{grozdanov2016strong} are not directly applicable to our analysis. We note that in the context of applying AdS/CFT to heavy ion physics, the physical implications of these terms are still under debate.\cite{Strickland:2018exs} At present we therefore consider our monotonic coupling interpolation to be in the hydrodynamic regime. We point out that a detailed analysis of this issue will be necessary in 2+1 dimensions and at finite density as well.} Further support for the interpolation is given by the following argument based on the perturbative result \eqref{etasFritzetal}: In the strong coupling regime, higher orders in $\alpha$ will become as important as the lowest order term in \eqref{etasFritzetal}. Nevertheless, since $\eta/s$ has the interpretation of the rate of diffusive momentum transfer between adjacent fluid layers normalized to the effective numbers of degrees of freedom, we expect higher orders only to enhance momentum transfer and hence decrease $\eta/s$. Our analysis does not depend on the exact form of the interpolating function $\eta/s(\alpha)$, only on its monotonicity.

We find that the channel resistance depends on the ratio $\eta/s$
rather than on $\eta$ alone: Solving the fully relativistic equations,
which do not admit a simple analytic solution as in the
non-relativistic limit,\cite{Lucas:2017wa} we confirm that the
differential resistance indeed does decrease with decreasing $\eta/s$
for all velocity regimes. Moreover, $\eta/s$ is a good
measure for the momentum transfer between fluid layers normalized to
the central charge and temperature, i.e. to the number of effective
degrees of freedoms in the momentum transfer. We find that the resistance increases when reducing the coupling. We confirm this by changing $\eta/s$ away from its infinite strong coupling value $\hbar/(4\pi k_B)$.

The main focus of our work is the analysis of the fully relativistic Poiseuille flow. In addition we also studied the effect of momentum relaxation, this however only in the non-relativistic regime. In both cases, we calculate the velocity profile, the differential resistance of the channel and the wall relaxation time defined below.

 In the absence of momentum-relaxing impurities, we find that $\eta/s$ has a strong effect on the fluid flows in both the non-relativistic and relativistic velocity regimes. In particular strongly coupled holographic fluids satisfying \eqref{etasAdSCFT} flow faster compared to their weakly coupled counterparts. The reason is that in the absence of impurities, the only way for the fluid to lose momentum is through the boundaries, which absorb the momentum density. The rate of momentum transfer between adjacent fluid layers is controlled by $\eta/s$, and hence the momentum transfer to the boundaries becomes as inefficient as it can be for fluids satisfying \eqref{etasAdSCFT}. The rate of momentum loss through the boundaries is characterized by an associated wall relaxation time scale $\tau_w$. We calculate the wall relaxation time scale and find that it is longer for strongly coupled holographic fluids than for weakly coupled ones. We furthermore calculate the differential channel resistance $dV/dI$, and find as expected that it decreases with decreasing $\eta/s$. Holographic fluids restricted to channels hence show lower  channel resistance than weakly coupled ones.

We also consider the effect of momentum relaxing impurities on the Poiseuille flow. In microscopic descriptions of real-world condensed matter systems, the physics of impurities in general depends on the nature of the scattering potential (extended, hard, soft, long or short range  etc.). A similar model dependence is also inherent in holographic models, which depend on the precise mechanism used to break translation invariance and hence momentum conservation. Nevertheless, in both cases the relaxation time approximation, i.e. the assumption of a constant relaxation time, often proves to be a good and universal approximation to the weak impurity limit. In our setup, we dial the impurity relaxation time independently of the other parameters of the system, in order to distinguish impurity effects from hydrodynamic ones. Hydrodynamics is only a good description if the impurity relaxation scale is a slow scale in the system, slower than the equilibration time \eqref{taueq} and the time-scale of hydrodynamic fluctuations. The latter is given by the gradients of hydrodynamic fields velocity, temperature, and chemical potential. If the momentum relaxation time scale is short compared to the momentum transport time set by $\eta/s$, we find a crossover to an impurity-dominated Ohmic regime even for the  zero velocity boundary conditions characterizing the Poiseuille flow. The crossover happens if the ratio
\be 
\Pi = \frac{1}{T\tau_{\rm imp}} \frac{\eta}{s}
\ee
is of ${\cal O}(1)$. The Poiseuille regime occurs for $\Pi\ll 1$. In the Ohmic regime $\Pi \gg 1$, hydrodynamics is no longer valid as a low-energy effective theory. 

 To conclude the introduction, we summarize  the new results of our work:
\begin{itemize}
\item We performed a numerical as well as analytical analysis of {\it relativistic} hydrodynamics for a fluid confined in a two-dimensional wire geometry, in addition to considerations for the non-relativistic case. We find Poiseuille behavior for all velocities, and in particular also in the relativistic regime. The importance of considering the relativistic regime can be seen from the fact that the differential resistance drops to zero in the ultrarelativistic limit, in which the fluid velocity approaches the Fermi velocity $|\vec{v}| \rightarrow v_F$, purely as a consequence of the kinematics of special relativity.

\item We find that the channel resistance strongly scales with the
  ratio of shear viscosity over entropy density $\eta/s$, which is the
  properly normalized measure of momentum transfer between fluid
  layers per effective degree of freedom. 
 Holographic strongly coupled 
  fluids, for which  $\eta/s \simeq 1/4\pi$, exhibit smaller
  differential resistance than conventional weakly coupled fluids. The channel resistance  scales with $\eta/s$ also in the fully relativistic regime. 

\item We calculate the  time scale associated to momentum relaxation
  through the walls, i.e. the time scale associated with the rate of
  momentum loss through the channel walls for zero velocity boundary
  conditions, both in the nonrelativistic and relativistic regimes.
  The wall relaxation time is inversely proportional to $\eta/s$, even
  in the relativistic regime. In the non-relativistic regime, this behavior may be seen analytically 

\item We examine the onset of turbulence in the channel by calculating the Reynolds number of our flows. We find that for the input parameters typical in current experimental realizations, the flow is laminar. The Reynolds number may also be significantly increased by increasing the channel width by an order of magnitude. Transitioning to a pre-turbulent regime is, however, not ruled out for setups with channel constrictions or obstacles placed in the flow.  Moreover, we find that impurities affect the onset of turbulence by bounding the values the Reynolds number can reach. We present explicit formulae for these bounds.

\end{itemize}

In addition, our analysis is phenomenological in nature and relies on the applicability of hydrodynamics alone. It is not restricted to any particular Dirac metal such as graphene, but also applies, for example, to strange metals and other strongly correlated materials. In particular the strange metallic phases in high temperature superconductors are expected to behave hydrodynamically,\cite{davison2014holographic} and our analysis is hence applicable.

This paper is organized as follows: In sec.~\ref{sec2} we review the
equations of relativistic hydrodynamics (sec.~\ref{sec21}), as well as
the pieces of the AdS/CFT correspondence necessary to understand our
analysis (sec.~\ref{sec:holo}). In sec.~\ref{sec:FSE}, we then discuss
the effect of the finite channel width leading to the the Poiseuille
flow (sec.~\ref{sec:poisprel}), the effect of momentum relaxation
through the walls (sec.~\ref{sec:momentumrelaxation}). Sec.~\ref{sec:flowdVdI} is devoted
to the analysis of the flow velocity profiles and the differential
channel resistance in the absence of
impurities. Sec.~\ref{sec:impurities} then includes momentum relaxing
impurities into the analysis. In section \ref{sec:turbulence}, we  present preliminary results on the onset of turbulence in Gurzhi-type channel setups.
Finally, in sec.~\ref{sec:disc_outlook} we discuss our results and give an outlook to possible future research. We estimate in appendix ~\ref{app:reynolds} the Reynolds number both in the nonrelativistic and ultrarelativistic limits.

\section{Relativistic hydrodynamics and the AdS/CFT correspondence}\label{sec2}

\subsection{Relativistic hydrodynamics}\label{sec21}

In this section we review elements of hydrodynamics relevant to our analysis.~\cite{Landau:1987,Romatschke:2010jf,Rezzolla:2013} Relativistic hydrodynamics is the effective field theory of long-wavelength low-energy fluctuations of matter in local thermal equilibrium. Its dynamical equations are the conservation laws of energy, momentum and charge. We neglect the imbalance current in relativistic systems, which is approximately conserved at weak coupling and close to charge neutrality. This additional conserved current does not couple to the external electromagnetic field, and hence does not contribute to electric transport. It will however contribute to thermal and thermoelectric transport.\cite{Lucas:2017wa} 

Consider a two-dimensional electron system in a wire geometry of width
$W$ and length $l\gg W$. Local thermalization is dictated by the
average time between electron-electron collisions $\tau_{\rm ee}$. It
is possible to describe the electron gas by means of hydrodynamics if
the electron-electron scattering time is the shortest timescale
present, and if all other external time-scales are much longer than
the time-scale of hydrodynamic fluctuations. Within our setup, we have
two additional relevant timescales: the time between electron-impurity
collisions, i.e. the momentum relaxation time-scale $\tau_{\rm imp}$,
and $W/v_F$. The corresponding length scales are obtained by
multiplying with the (constant) Fermi velocity $v_F$. Therefore, the
two-dimensional electron gas will behave as a fluid if, in natural units, 
\be \label{hydroregime}
\ell_{\rm ee}  \ll ||\partial_\nu  (T,\mu,u^\mu)|| \ll  \ell_{\rm
  imp}, \, W \, .
\ee 
This is the regime we consider in this work (see again the caveat of footnote 48) . 

If the electrons have a relativistic dispersion, the local thermal equilibrium will be described on scales exceeding $\ell_{\rm ee} = v_F \tau_{\rm ee}$ by a local temperature $T(x^\mu)$, a local chemical potential $\mu(x^\mu)$, and a local relativistic velocity $u^\mu(x^\nu)$, where e.g. $x^\mu = (t,x,y)$ for a planar 2+1-dimensional system. The velocity field must be time-like and normalized to the speed of light relevant for the relativistic fluid, which e.g. in case of  graphene is the Fermi velocity $v_F$ appearing in the dispersion relation of the Dirac particles. We focus on 2+1-dimensional hydrodynamics, as it is relevant to two-dimensional electron systems. In terms of the velocities $(\beta_x(x^\mu),\beta_y(x^\mu))$, 
\begin{align}\label{eq:velocity}
u^\mu &= \gamma \left(v_F,\beta_x,\beta_y\right),\qquad \gamma = \frac{1}{\sqrt{1-\frac{\beta_x^2+\beta_y^2}{v_F^2} }} 
\,,\\
u^\mu u_\mu &= u^\mu u^\nu \eta_{\mu\nu} = - v_F^2\,, \quad \eta_{\mu\nu} = \diag(-1,1,1)\,.
\end{align}
The equations of relativistic hydrodynamics are the conservation equations for the Lorentz covariant energy-momentum tensor $T_{\mu\nu}$ and the relativistic charge current $j^\mu$. In absence of impurities, the covariant energy-momentum and current conservation equations for a fluid in the presence of an external electromagnetic field  are 
\bea\label{relhydroeqs}
\partial_\mu T^{\mu\nu} &=& \frac{h}{e^3 c} \, j_\alpha D^{\nu\alpha}\,,\\\label{relhydroeqs2}
\partial_\mu j^\mu &=& 0\,.
\eea
We assume the background metric to be the metric of flat Minkowski space-time, $g_{\mu\nu}=\eta_{\mu\nu}$. \footnote{Introducing a nontrivial metric is useful when e.g. considering perturbations that correspond to a temperature gradient acting on the fluid \cite{1963AnPhy..24..419K}, and can be achieved by replacing partial derivatives in \eqref{relhydroeqs}-\eqref{relhydroeqs2} by the covariant derivative with an appropriately chosen connection, which most of the time will be the metric compatible Christoffel connection.} $D^{\mu\nu}$ is the  electromagnetic displacement tensor, which in 2+1 dimensions takes the form\footnote{We normalized the electric field by means of the fundamental constants $h,e$ and $c$ and in such a way that $\left[ E^\mu \right]=\text{V/m}, \,  \left[j^\mu \right]=\text{A/m}$ and $\left[ D^{\mu\nu}\right] = {\rm A^2 s/m}$.}
\be \label{eq:eqDcomp}
D^{\mu\nu} = e\left( 
\begin{array}{c c c}
	0 & E_x \epsilon_0 \epsilon_r c & E_y \epsilon_0 \epsilon_r c\\
	-E_x \epsilon_0 \epsilon_r c& 0 & -B/\mu_0 \mu_r \\
	E_y \epsilon_0 \epsilon_r c & B/\mu_0 \mu_r & 0 \\
\end{array}
\right), 
\ee 
with  $D_{\nu\mu} = \eta_{\nu\alpha} \eta_{\mu\delta} D^{\alpha\delta}$. The electric field, given by $E^\mu = h/\left(e^3 c\right) u_\alpha D^{\mu\alpha}$, fulfills $u_\alpha E^\alpha =0$. The charge density to which the electric field couples is then $v_F^2 e \rho = - j^\alpha u_\alpha$. Note that while the speed of light relevant for the Lorentz symmetry of the fluid is $v_F$, the speed of light relevant for the coupling to the external electromagnetic field is the usual one, $c$. $\epsilon_0$, $\mu_0$, $\epsilon_r$, and $\mu_r$ are the dielectric constant of the vacuum, the vacuum permeability, the relative permittivity and the relative permeability in medium. 

In order for \eqref{relhydroeqs}-\eqref{relhydroeqs2} to form a set of equations for the hydrodynamic variables $(T,\mu,u^\mu)$, we need to specify the stress-energy tensor $T_{\mu\nu}$ and the charge current $j^\mu$ in terms of these variables. This is formulated as a derivative expansion around global equilibrium, characterized by constant $T$ and $\mu$, and vanishing velocity in the rest frame, $u^\mu=(v_F,0,0)$. We decompose the stress-energy tensor and charge current in terms of tensor structures parallel and perpendicular to the flow field $u^\mu(x)$,
\begin{subequations}
\label{E:decomposition}
\begin{eqnarray}
\label{eq:T1}
  &&   T_{\mu\nu} = {\cal E} u_\mu u_\nu + {\cal P} \Delta_{\mu\nu}
       + (q_\mu u_\nu {+} q_\nu u_\mu) +t_{\mu\nu},  \\
\label{eq:J1}
  &&  j_\mu = {\cal N} u_\mu + \nu_\mu \,,
\end{eqnarray}
\end{subequations}
where $\Delta^{\mu\nu}=u^\mu u^\nu/v_F^2 + \eta^{\mu\nu}$ is the projector onto the space orthogonal to the flow field. 
${\cal E}$, ${\cal P}$, and ${\cal N}$ are Lorentz scalars,
$q_\mu$, $t_{\mu\nu}$, and $j_\mu$ are transverse,
$u_\mu q^\mu=0$, $u_\mu t^{\mu\nu}=0$, $u_\mu \nu^\mu=0$,
and $t^{\mu\nu}$ is symmetric and traceless. As for any effective field theory, redefinitions of the fields $(T,\mu,u^\mu)$ can mix the different terms in \eqref{E:decomposition}. It is hence important to specify conditions that fix the freedom to redefine the hydrodynamical fields, i.e. to chose a hydrodynamical frame. The most common frame is the Landau frame.\footnote{Another common frame is Eckart frame, in which the fluid is comoving w.r.t. the current flow, amounting to $j^\mu=0$ in \eqref{eq:J1}. Another example can be found in \cite{Jensen:2011xb}, where the $M B$ in the presence of a constant magnetic field can either be considered zeroth order in derivatives due to the constancy of $B$ and incorporated into the pressure, or, if $B$ is small, to be of first order due to the fact that the magnetic field is the derivative of the vector potential. It is possible to switch between both frames through field transformations,~\cite{Kovtun:2012rj}.} In Landau frame, the local energy density in equilibrium $\epsilon$ is identified with ${\cal E}$ via the condition $T^{\mu\nu}u_\mu = -\varepsilon u^\nu$, and the density of conserved charge $e\rho$ is identified with ${\cal N}$ via $ j^\alpha u_\alpha = -v_F^2 e \rho $.  The fluid is comoving to the energy flow, amounting to $q^\mu=0$ in \eqref{eq:T1}. We exclusively work in Landau frame in this paper. 

Having fixed the frame, the expansion up to first order in derivatives of $u^\m$ for $T^{\mu\nu}$ and $j^\mu$ for a rotationally invariant charged fluid in $2+1$ dimensions is
\begin{align}
T^{\mu\nu} &= T^{\mu\nu}_{\rm (ideal)} + T^{\mu\nu}_{\rm (viscous)},\\ \label{eq:T2}
T^{\mu\nu}_{\rm (ideal)} &= \varepsilon\, u^\mu u^\nu/v_F^2 + p \Delta^{\mu\nu}\,, \\ \label{eq:T3}
T^{\mu\nu}_{\rm (viscous)} &= -\eta \Delta^{\mu\alpha}\Delta^{\nu\beta}\left(2\partial_{(\alpha} u_{\beta)}-\Delta^{\alpha\beta} \partial_\sigma u^\sigma \right) \nonumber\\
&\quad -\xi \Delta^{\mu\nu} \partial_\gamma u^\gamma ,\\
\label{eq:j}
j^\mu &= e\rho u^\mu + \sigma_Q \left(E^\mu - T \Delta_{\mu\nu} \partial^\nu \frac{(\mu/e)}{T}\right)\,,
\end{align}
where $\varepsilon,p,\rho$ are energy density, pressure and number density of the fluid in local equilibrium. $\sigma_Q$ is the intrinsic quantum critical conductivity due to the counterflow of electrons and holes \cite{Fritz:2008go}. $\eta$ and $\xi$ are the shear and bulk viscosities. For application to Dirac materials such as e.g. graphene, which are approximately scale invariant, we neglect the bulk viscosity, $\xi=0$.\footnote{Scale invariance implies vanishing stress-energy tensor trace, $T_\mu{}^\mu$=0, and hence $\xi=0$. Furthermore, due to renormalization effects the Fermi velocity runs logarithmically at low temperatures $v_F=v_F(T)$ \cite{Lucas:2017wa}. Another reason for neglecting the bulk viscosity term is that we are working with incompressible fluid flows, i.e. $\partial_\mu u^\mu = 0$}

In the presence of momentum relaxing impurities, the classical Drude model alters the right hand side of the Newton equation $F= m\dot{v}$ by adding a friction term,
\be 
m\dot{v} \to m\dot{v} + \frac{m}{\tau_{\rm imp}}v.
\ee 
We first project \eqref{relhydroeqs} into the direction of time given by the velocity field $u^\mu$ and perpendicular to it by multiplying with $u^\mu$ and $\Delta^{\mu\nu}$, respectively,
\bea \label{eq:nseqs1}
u_\nu \partial_\mu T^{\mu\nu} &=& \frac{h}{e^3 c} u_\nu j_\alpha D^{\nu\alpha}, \label{eq:eq1}\\
\Delta_{\rho\nu}\partial_\mu T^{\mu\nu} &=& \frac{h}{e^3 c}\Delta_{\rho\nu} j_\alpha D^{\nu\alpha}. \label{eq:eq2}
\eea
The first equation enforces the conservation of energy, whereas the second equation is the relativistic version of the Navier-Stokes equation.\footnote{The conservation equation for the current \eqref{relhydroeqs2} is trivially satisfied for the flow we will introduce in sec.~\ref{sec:FSE}, and hence we do not discuss it hereafter.} We add a loss term to the right-hand side of the Navier-Stokes equations \Eq{eq:eq2}, 
\bea 
u_\nu \partial_\mu T^{\mu\nu} &=& \frac{h}{e^3 c} u_\nu j_\alpha D^{\nu\alpha}, \label{eq:hydro1} \\
\Delta_{\rho\nu}\partial_\mu T^{\mu\nu} &=& \frac{h}{e^3 c} \Delta_{\rho\nu} j_\alpha D^{\nu\alpha} - \Delta_{\rho\nu} \frac{T^{t\nu}}{v_F \tau_{\rm imp}}. \label{eq:eqrelax}
\eea 
The momentum relaxation time $\tau_{\rm imp}$ can be thought of arising from electron-impurity scattering in the relaxation time approximation. We henceforth assume that this is the only relaxation time entering the relativistic hydrodynamic equations. We in particular assume that energy relaxation is sufficiently fast such that in the non-equilibrium steady states considered, the non-equilibrium energy density is close to the energy density in local thermal equilibrium at the electron temperature. This is supported by quantum criticality arguments \cite{zaanen2004superconductivity}, as well as measurements of scattering times in materials with strong correlations \cite{bruin2013similarity,gooth2017electrical}.
 
\subsection{Strongly coupled fluids in AdS/CFT}
\label{sec:holo}

In order to solve the hydrodynamic equations \Eq{eq:hydro1} and
\Eq{eq:eqrelax}, we need to specify the thermodynamic variables that
enter the constitutive relations
\eqref{eq:T2}, \eqref{eq:T3}, \eqref{eq:j}. For
strongly coupled fluids, it is natural to take these values from the
AdS/CFT correspondence. In appendix \ref{app:holoren} we detail how AdS/CFT determines them for a strongly coupled field theory at finite temperature and density.

In this approach, the thermodynamic variables read
\bea \label{eq:TDvariables}
\varepsilon &=& \frac{L^2}{8 \pi G_4} \hbar \, v_F \frac{r_H^3}{L^6}, \quad p = \frac{L^2}{16\pi G_4}\hbar \, v_F \frac{r_H^3}{L^6},\\\label{eq:thermoads}
\rho &=& \frac{L^2}{64\pi G_4}\left(\frac{r_H}{L^2}\right)\frac{\mu}{\hbar v_F}, \quad s= \frac{k_B}{4 G_4}\frac{r_H^2}{L^2},
\eea 
where $L$ is the $AdS_4$ radius and $G_4$ is the Newton constant, as explained in detail in appendix \ref{app:holoren}. In particular,
\be \label{eq:rH}
\frac{r_H}{L^2} = \frac{1}{6} \frac{k_B T }{\hbar \, v_F} \left(4 \pi+ \sqrt{16 \pi ^2 + \frac{3\mu ^2}{k_B^2 T^2} }  \right),
\ee 
with $r_H$ the radius of the black hole horizon.

The quantities given by \Eq{eq:TDvariables} and \Eq{eq:thermoads} satisfy the Gibbs-Duhem relation
\be\label{eq:GibbsDuhem}
\varepsilon + p = \mu \rho + T s.
\ee 

In the AdS/CFT approach, the charged fluid is strongly coupled, such that $\eta/s$ is small. We take a phenomenological approach of varying the coupling towards weaker and intermediate values as follows: Starting from the holographic value of $\eta/s = \hbar/4\pi k_B$, we then slowly increase $\eta/s$ to investigate the behavior of the fluid when moving from the holographic strongly coupled regime towards weaker intermediate coupling. In the calculation of physical variables presented below, we fix the overall coefficients of physical observables using the AdS/CFT approach given in appendix \ref{app:holoren}.\\

\section{Finite size effects}
\label{sec:FSE}

\subsection{Poiseuille flow} 
\label{sec:poisprel}
We now investigate the physical effects that arise from the motion of
a fluid in a channel of finite size. Consider an incompressible fluid
moving along a wire of width $W$ and length $l$. We may think of the
fluid as a superposition of fluid layers along the width of the
channel. The fluid's flow is laminar if these layers never mix. A particular kind of laminar flow is the Poiseuille flow characterized by the no-slip boundary conditions,

\be \label{eq:poiscond}
u^{i}(y)\Big\vert_{y=W} =  0, \quad u^i(y)\Big\vert_{y=0} = 0.
\ee  

\noindent Simply stated, the velocity vanishes at the boundaries. A parametrization of the velocity $u^\mu$ is given by
\be \label{eq:param}
u^\mu = \gamma(y) \left(v_F,\beta_x(y),0\right), \qquad \gamma = \frac{1}{\sqrt{1-\frac{\beta_x(y)^2}{v_F^2} }}.
\ee
This Ansatz is incompressible, 
\be  \label{eq:inco}
\partial_\sigma u^\sigma = \partial_0 u^0 + \partial_i u^i = \frac{1}{v_F}\partial_t u^t +\partial_i u^i  =0.
\ee
Moreover, the boundary condition \Eq{eq:poiscond} is equivalent to 
\be \label{eq:bound}
\beta_x(y)\Big\vert_{y=W} =  0, \quad \beta_x(y)\Big\vert_{y=0} = 0.
\ee

Now we derive the hydrodynamic equations in terms of $\beta_x$ from \eqref{eq:eqrelax}.  Focusing on the case of a homogeneous external electric field pointing along the $x$-direction, the current density \eqref{eq:j} reads 
\be \label{eq:jcomps}
j^\mu =  e v_F \gamma\left( \rho+ \frac{h \epsilon_0}{e^3}\frac{\beta_x}{v_F} \epsilon_r \sigma_Q E^x , \rho \frac{\beta_x}{v_F} + \frac{h \epsilon_0}{e^3}\epsilon_r \sigma_Q E^x ,0 \right).
\ee
Here $\epsilon_0 \epsilon_r$ is the electric permittivity of the fluid, $\rho$ its number density, $v_F$ its Fermi velocity, and $\sigma_Q$ its quantum critical conductivity. Decomposing \Eq{eq:eqrelax} into components, we obtain two independent equations \footnote{Note that we assume vanishing temperature gradients. If we were to relax this assumption, the spacetime dependence of $\eta$ should also be taken into account in the equations of motion.}
\begin{widetext}
\begin{align}
0  &= \frac{\eta}{s}\left[ \beta_x'' + \frac{2}{v_F^2} \gamma^2 \beta_x \beta_x'^2\right] + \frac{1}{\gamma } \left( v_F \epsilon_r \frac{\epsilon_0 \,h}{e} \frac{ E_x}{\gamma }\frac{ \rho}{s} - \frac{p + \varepsilon}{ s \tau_{\rm imp} } \frac{\beta_x}{v_F^2}\right) ,\label{eq:eqbeta} \\
0 &= p' - \frac{\eta }{\tau_{\rm im} v_F^2}\gamma^3 \beta_x \beta_x' .\label{eq:eqp}
\end{align} 
\end{widetext}

We find a dynamical equation for the velocity profile \Eq{eq:param}, as well as a relativistic form of the Hagen-Poiseuille equation
for the pressure. 
For strongly coupled theories, the ratio $\eta/s$ is fixed to a
constant value, which is why we divided \Eq{eq:eqbeta} by the entropy density $s$. The solution to the pressure equation \Eq{eq:eqp} is given by
\be \label{eq:pre}
p(y) = p_0 + C + \frac{\eta}{\tau_{\rm imp}} \gamma(y).
\ee 
Here $C$ is an integration constant of \eqref{eq:pre} which will be fixed shortly. Note that the ratio ${\eta}/{\tau_{\rm imp}}$ controlling the pressure shift will be small in the parameter regime of applicability of hydrodynamics \eqref{hydroregime}. The non-equilibrium fluid pressure \Eq{eq:pre} will hence be close to the thermodynamic one as e.g. derived from the AdS solution \Eq{eq:thermoads}, $p \approx p_{0}(\mu,T)$. Therefore, we may use the homogeneous thermodynamic variables \Eq{eq:thermoads} even in the presence of impurities, as long as we restrict ourselves to small electric fields and non-relativistic flows. In that limit, the pressure shift becomes constant in the direction $y$ perpendicular to the channel,
\be \label{eq:nr-pre}
p = p_0 + C + \frac{\eta}{\tau_{\rm imp}},
\ee 
since $\gamma \sim 1 $.  A similar constant pressure shift due to impurities has also been observed in holographic models of momentum relaxation\cite{Vegh:2013sk,Andrade2014}. We emphasize that in our work as well as in the cited paper, the equation of state of the fluid is given by the Gibbs-Duhem equation. 
	
 The pressure as defined in (\ref{eq:pre})  must smoothly reduce to the equilibrium pressure in absence of flow as the velocity profile tends to zero, $\beta_x(y) = 0$. We therefore adjust the integration constant in \eqref{eq:pre}, and identify the resulting pressure $p$ in the $\beta_x(y) = 0$ limit as the thermodynamical pressure that enters into the Gibbs-Duhem relation \ref{eq:GibbsDuhem}. The thermodynamic pressure $p$ in presence of a flow gradient is thus
\be \label{eq:finalpre}
p = p_0 + \frac{\eta}{\tau_{\rm imp}}\left[\gamma(y) -1\right].
\ee

We note that the pressure shift in \eqref{eq:pre} can become large with respect to $p_0$ and strongly $y$-dependent. This can lead to observable consequences such as a voltage drop between the middle of the channel and the channel walls. Notice that the pressure shift \eqref{eq:finalpre}, is a genuine relativistic effect.

\subsection{Momentum relaxation through the walls}\label{sec:momentumrelaxation}

In this subsection we will discuss how the Gurzhi channel setup of Fig.~\ref{fig:pois} allows for momentum relaxation through the boundaries.  We will in particular introduce and calculate the relaxation time scale $\tau_w$ associated to the outflow of momentum through the channel walls.  To the best of our knowledge, such an explicit computation has not been performed in the literature before.

For a stationary flow, if we suddenly turn
off the electric field, momentum will diffuse through the walls located at $y=0$ and $y=W$. Therefore, the $x$
component of the fluid momentum along the velocity direction
changes due to momentum loss through the
boundaries and through collisions with impurities, i.e.
\be \label{eq:time1}
\frac{1}{v_F}\partial_t T^{t x}  = \partial_y T^{yx} - \frac{T^{t
    x}}{v_F \tau_{\rm imp}} \, .
\ee 
We parametrize the rate of momentum loss through the boundaries by
$\tau_w$. Analogously to $\tau_{\rm imp}$, the impurity scattering
rate  $\tau_w$ is also related to momentum loss. However, this particular form of momentum diffusion appears only from finite size effects, unlike $\tau_{\rm imp}$ which can also be present in bulk samples.

If we approximate the time derivative that appears in \Eq{eq:time1} by 

\be \label{eq:tauwapprox}
\partial_t T^{t x}\simeq \frac{T^{t x}}{\tau_w},
\ee

and integrate along the $y$ direction, we obtain

\be \label{eq:momentum1}
\frac{1}{v_F} \int^W_0 dy\,  T^{tx}\left( \frac{1}{\tau_w} + \frac{1}{\tau_{\rm imp}} \right)  = T^{xy}\Big\vert_{0}^W .
\ee 
From the no-slip conditions \Eq{eq:bound}, we have that 
\be\label{eq:momentumsink}
T^{xy}\Big\vert_0^W = -\eta \beta'_x(y)\Big\vert_0^W .
\ee
The channel geometry with no-slip boundary conditions allows for momentum relaxation through the walls, via the boundary term on the right hand side of \eqref{eq:momentum1} and \eqref{eq:momentumsink}. Due to the no-slip boundary conditions, $\beta_x'(y)>0$ and hence momentum is not conserved any more. Therefore, the channel walls act as a sink of momentum for the fluid. 
Substituting $T^{tx}$ from \eqref{eq:T2} into \eqref{eq:momentum1}, we obtain
\be \label{eq:momentum2}
\int_0^W dy \,\,  \frac{\beta_x}{v_F}\left(p + \varepsilon\right) \gamma^2 \left( \frac{1}{\tau_w} + \frac{1}{\tau_{\rm imp}} \right) = -\eta \beta'_x(y)\Big\vert_0^W.
\ee

Equation \Eq{eq:momentum2} can be solved for $\tau_w$ once the velocity profile and thermodynamic variables are specified. Mechanisms that lead to momentum dissipation have been widely discussed in the literature\cite{Andrade2014,Davison:2013jba}. However, we stress that the momentum relaxation effects discussed there are not related to wall momentum relaxation, which is different in nature. This can be seen from our hydrodynamic simulations, which lead to a finite and well-defined channel resistance even in the absence of momentum relaxing impurities. This is to be contrasted with the case of the infinitely extended system, where finite charge density without local momentum relaxation leads to an infinite DC conductivity, i.e. vanishing DC resistivity.

\section{Flow and differential resistance in the absence of impurities}\label{sec:flowdVdI}

\subsection{Velocity profiles}\label{sec:flow}

We now proceed to the calculation of the velocity profile $\beta_x$ of the fluid as a function of  the coordinate $y$ perpendicular to the fluid motion. To do this, we numerically integrate the Navier-Stokes equation \Eq{eq:eqbeta} with the boundary conditions \Eq{eq:bound}. This will allow us to determine how the fluid profile depends on the applied electric field, the temperature, the chemical potential and the width of the wire. Moreover, from the velocity profile we will calculate the differential resistance $dV/dI$. In this section, we will consider the case where impurities are absent ($\tau_{\rm imp}\to \infty$), leaving the inclusion of impurity effects to sec.~\ref{sec:impurities}. We shall not impose any restriction on the maximal fluid velocity, which can be arbitrarily close to the Fermi velocity. Therefore, it is necessary to solve the fully relativistic equation \Eq{eq:eqbeta} numerically. First, it is convenient to define the reduced variables
\be \label{eq:changes}
u = \frac{k_B T}{h v_F} y, \,\,\, \mathcal{E}^x = \frac{E_x}{T^2}\frac{v_F^2 h^2\epsilon_0\epsilon_r}{e k_B^2}\,,\,\,\, w = \frac{k_B T}{h v_F}W\,.
\ee
The Fermi velocity $v_F$ and relative permittivity $\epsilon_r$, which are intrinsic properties of the material, need to be fixed. For concreteness, we take values typical for the surface states of the topological insulator, such as HgTe \cite{PhysRevX.4.041045}, 
\be  \label{eq:input1}
v_F = 10^5\, {\rm m/s}, \quad \epsilon_r \sim 3-5.
\ee

The choice of parameters \eqref{eq:input1} coincides  with having a rather strong electron-electron Coulomb interaction. Indeed,  the effective fine-structure constant in a medium with the input parameters \eqref{eq:input1} is rather large, 
\be 
\alpha_{\rm eff} = \frac{\alpha_0 c}{v_F \epsilon_r} \simeq 4.4 \gg \alpha_0 , \qquad \alpha_0 \simeq \frac{1}{137}\,.
\ee
$\epsilon_r$ is similar in graphene \cite{Lucas:2017wa}, but $v_F = 10^6\, {\rm m/s}$. 
For the Fermi velocity and relative permittivity \Eq{eq:input1}, $\mathcal{E}_x=1$ is equivalent to $E_x \approx 1.5\times 10^{-3} $ V/$\mu$m whereas $w=1$ to $W\approx 5$ $\mu$m. 
Then, \Eq{eq:eqbeta} becomes
\be \label{eq:betaeq}
 \frac{\eta}{s}\left( \ddot{\beta_x} + \frac{2 \gamma^2}{v_F^2} \beta_x \dot{\beta_x}^2  \right) +  h \frac{ v_F \mathcal{E}^x}{\gamma^2}\frac{\rho}{s} = 0, \qquad \dot{\beta_x} = \frac{\partial \beta_x}{\partial u}.
\ee
At this point, we remark the high sensitivity of the profile $\beta_x$ on the set of external values $\lbrace \eta/s,\mathcal{E}^x,\mu/k_B T\rbrace$. In particular, from the first term in \eqref{eq:betaeq} it is clear that reducing $\eta/s$ while keeping the electric field fixed, one can reach higher maximal velocities in the middle of the channel. This is due to the fact that $\eta/s$ controls the momentum transfer between adjacent fluid layers. In terms of the reduced variable $u$, the boundary conditions \Eq{eq:bound} are rewritten as 
\be \label{eq:noslip}
\beta_x(u)\Big\vert_{u=0} = \beta_x (u)\Big\vert_{u=w} =0 \,.  
\ee
The set of variables
\be 
\lbrace w, \mathcal{E}_x,\mu,T,\eta/s\rbrace
\ee 
defines the parameter space of possible fluid velocities. Hereafter, we will refer to this set as \emph{input parameters}. Throughout the present work, we will often draw our attention to the ratio $\eta/s$, which controls important physical properties of the fluid. It has been discussed both in high energy physics \cite{Majumder:2007zh,Buchel:2004di} as well as in the context of  condensed matter physics \cite{Muller:2009cy} that this ratio depends strongly on the coupling and, hence, is a measure of the coupling strength itself. In the extreme strong coupling limit, it is conjectured to attain the universal AdS/CFT value \eqref{etasAdSCFT}. 
For a phenomenological investigation of the dependence of the velocity profile and the differential resistance on the coupling strength, we will vary the value of $\eta/s$, increasing it from the AdS/CFT value. This will correspond to moving to smaller couplings beginning from the extreme strong coupling limit. To avoid large corrections, we will remain close to the holographic bound \Eq{etasAdSCFT}, i.e. increase $\eta/s$ by at most a factor of $20$. 

We now list the results for the velocity profiles and conclusions obtained from this analysis. Each plot is obtained by varying one parameter while keeping the others fixed.  

\begin{figure}
	\begin{center} 
		\includegraphics[scale=0.20]{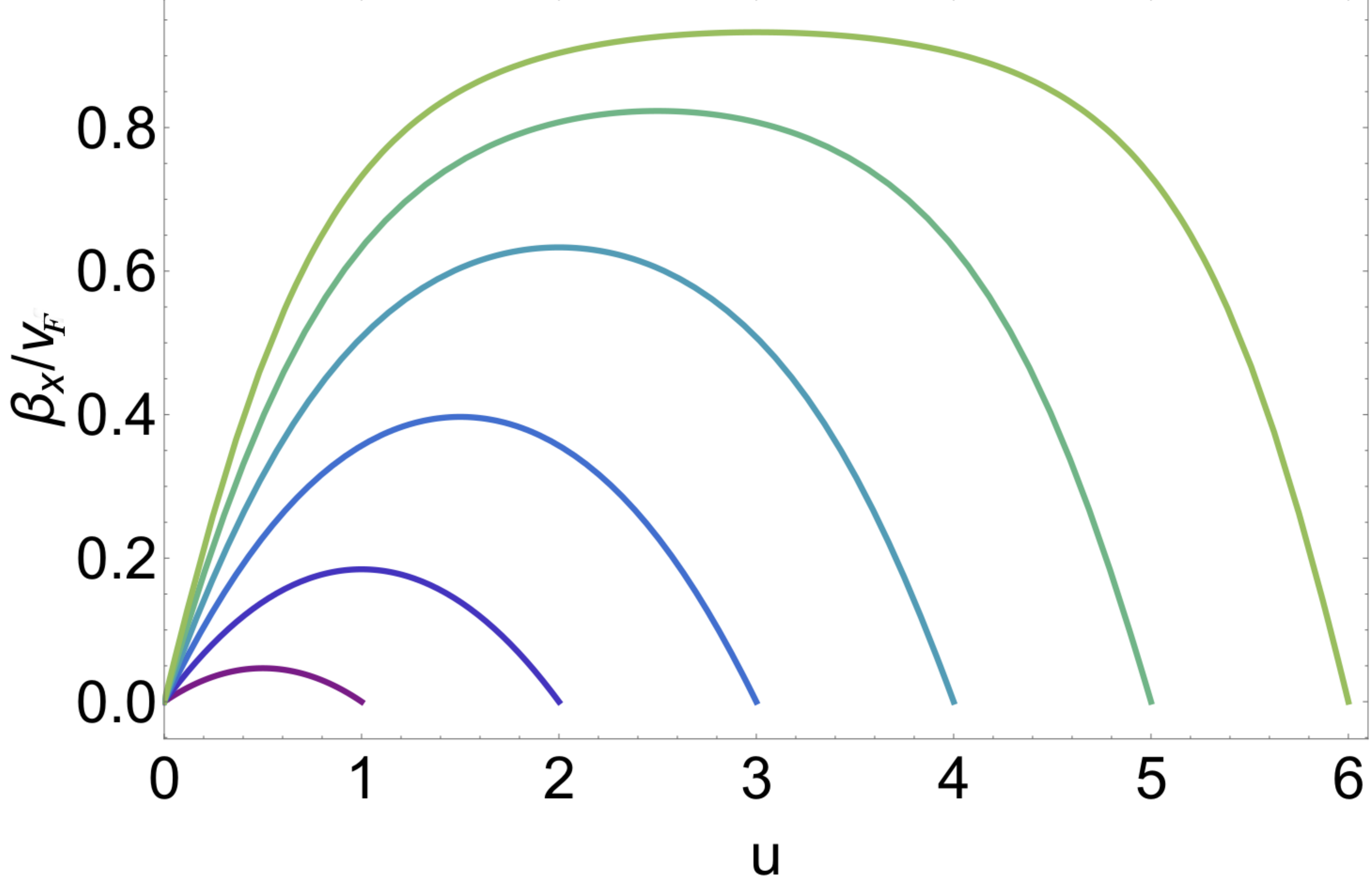}
		\caption{Velocity of the fluid $\beta_x$ for the Poiseuille flow as a function of $u$ at different widths $W,$ keeping fixed $\mathcal{E}^x=1$, $\mu/k_B T=1$. From top to bottom, $w=6,5,4,3,2,1$. \label{fig:poisw}}
	\end{center}
\end{figure}

	\begin{figure}[tp]
		\begin{center}
		\includegraphics[scale=0.20]{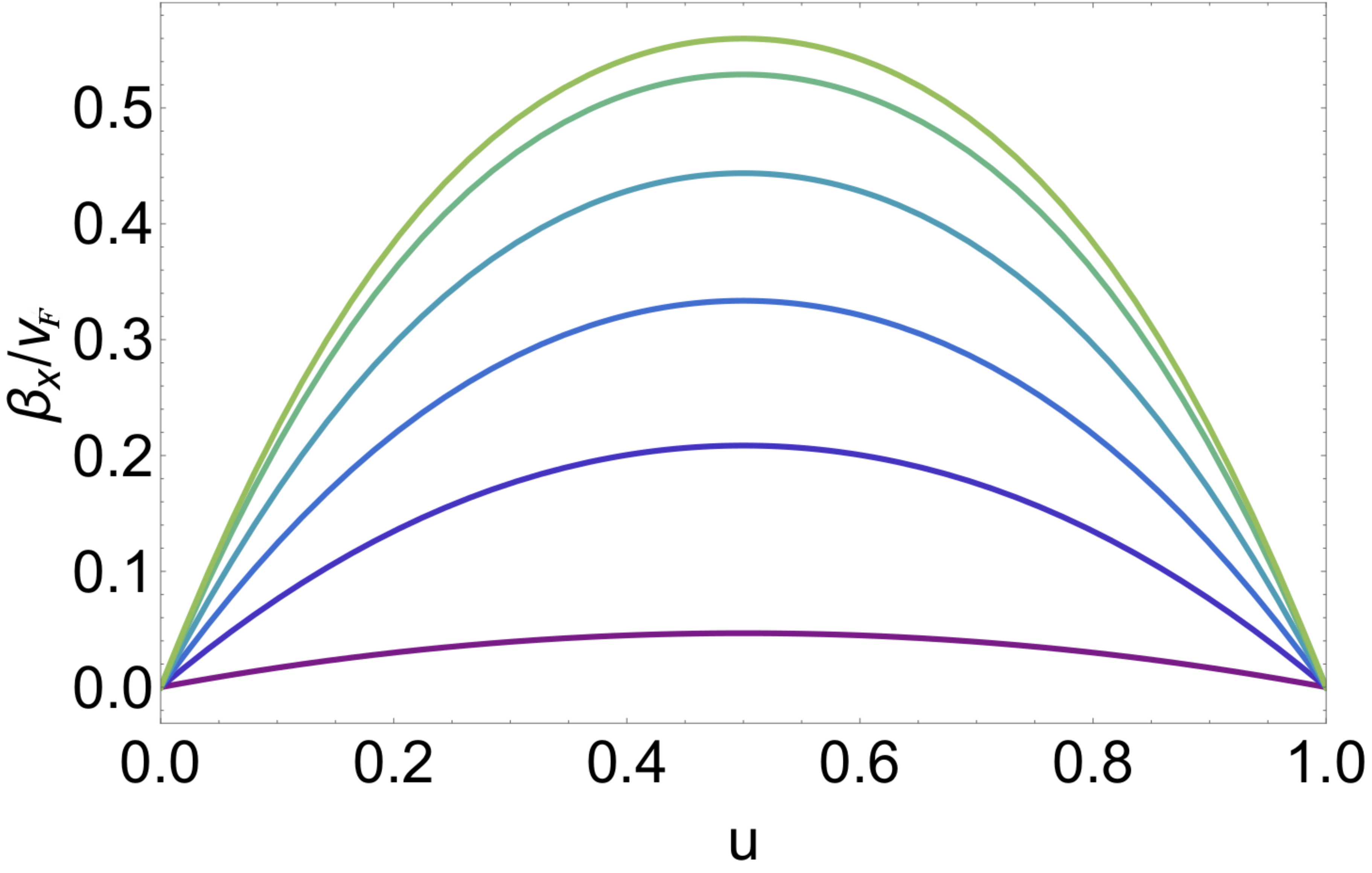}
		\caption{Velocity of the fluid $\beta_x$ for the
                  Poiseuille flow as a function of the dimensionless
                  variable $u$ at $\mathcal{E}^x= 1,w=1$ and different
                  values of $\mu/k_B T $. From top to bottom, $\mu/k_B
                  T=  100,50,20,10,5,1$. If we assume constant $T$, as the
                  density increases, so does the velocity of the fluid.\label{fig:poisx}}
	\end{center}
\end{figure}

\begin{figure}[tp]
	\begin{center} 
		\includegraphics[scale=0.20]{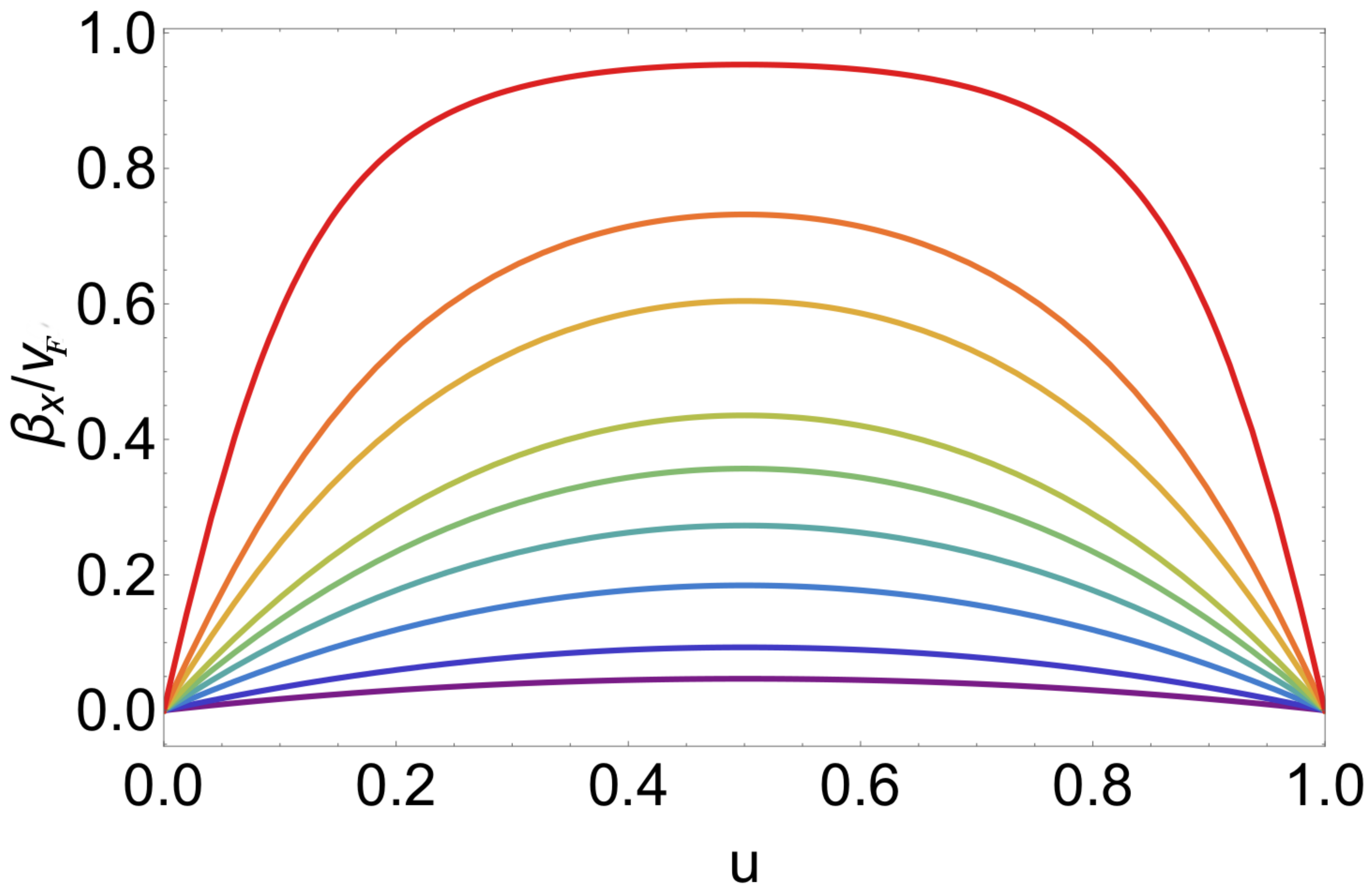}
		\caption{Velocity profile $\beta_x$ for the Poiseuille flow as a
                  function of $u$ at different values of $
                  \mathcal{E}^x$. From top to bottom: $\mathcal{E}^x =
                  50,20,15,10,8,6,4,2,1$. For illustrative purposes,
                  we have taken $w=1$ and $\mu/k_B T=1$. Increasing
                  the electric field implies that the fluid velocity increases.\label{fig:poise}} 
	\end{center}
\end{figure}

\begin{figure}[tp] 
	\begin{center}
		\includegraphics[scale=0.20]{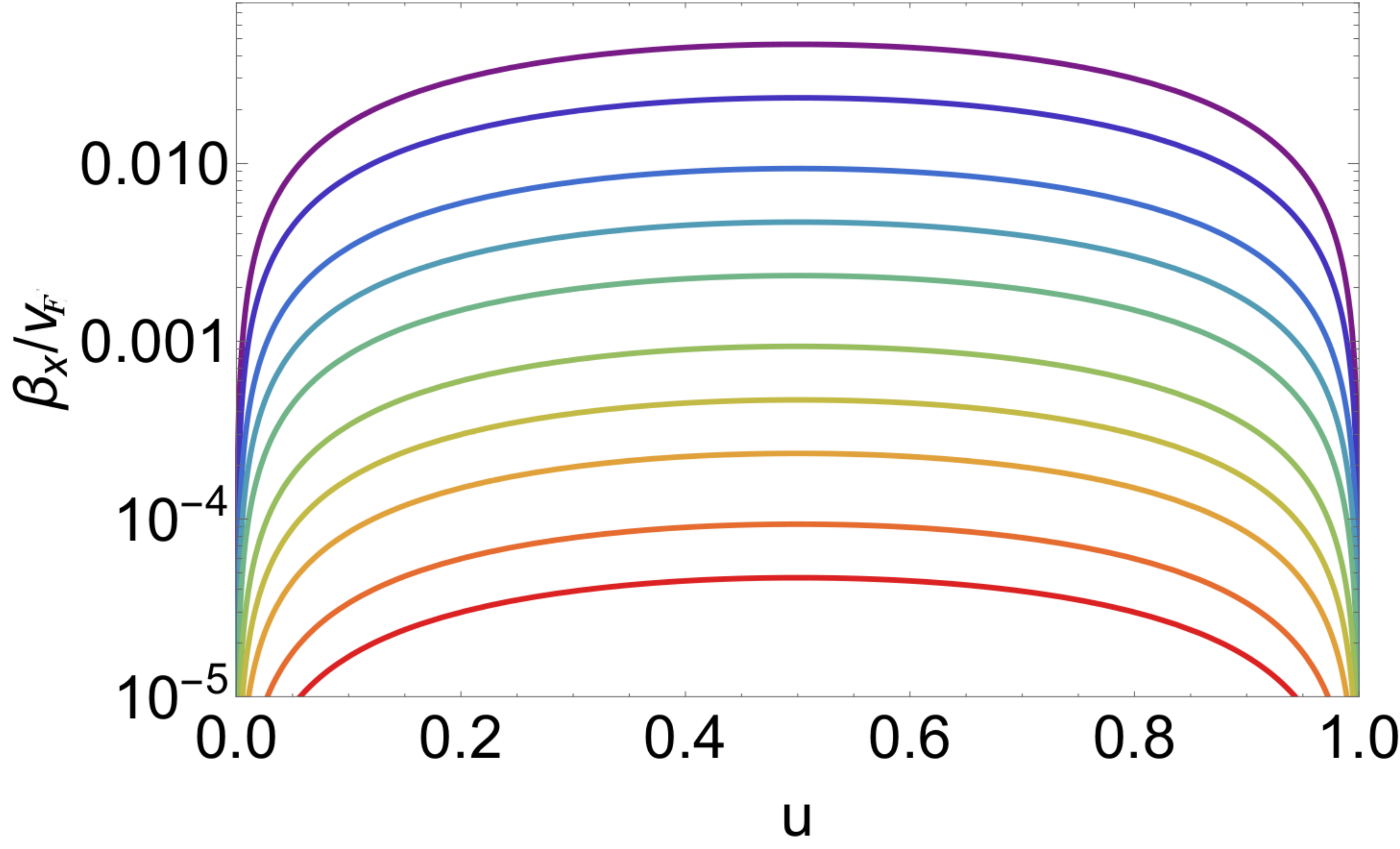}
		\caption{Velocity of the fluid $\beta_x$ for the Poiseuille flow as a function of $u$ at different values of the ratio $\eta/s$ while keeping fixed $\mathcal{E}^x=1$, $\mu/k_B T=1$ and $w=1$. From top to bottom, $4\pi k_B \eta/s\hbar =1, 2, 5, 10, 20, 50, 100, 200, 500, 1000$. In order to be able to plot all possibilities, we use a logarithmic scale in the $y$-axis. \label{fig:poisetas}}
	\end{center} 
\end{figure}

\begin{itemize} 
\item First, from figure \ref{fig:poisw} we infer that the fluid attains a higher maximal velocity in the middle of the channel as the width $w$ is increased. The reason is that for wider channels, the fluid has more space between the middle of the channel and the boundary, and can transfer a larger total momentum to the walls: In the absence of impurities, the only way for the fluid to lose momentum is to transport it to the boundary via the viscous force between fluid layers. The rate of this momentum transfer between adjacent layers is constant and controlled by $\eta/s$. Hence, integrating \eqref{eq:betaeq}, losing a larger amount of momentum will need a wider channel. Note that the applicability of our analysis is restricted by the following argument: As in e.g.~\cite{Molenkamp:1994ii,Molenkamp:1994kb,deJong:1995bn}, a cross-over to a ballistic regime is expected  if the width of the channel becomes shorter than the electron-electron mean free path $\ell_{\rm ee} = \tau_{\rm ee} v_F$. On the other hand, if the width becomes large enough, the maximal velocity in the middle of the channel will be large enough to trigger the onset of turbulent behavior (c.f.~ appendix ~\ref{app:reynolds} for a discussion of the Reynolds number for relativistic flows). Hence the allowed values of $w$ will be restricted to a certain window by these two limits.

\item Second, we observe in figure \ref{fig:poisx} that increasing the chemical potential at fixed temperature, or equivalently decreasing the temperature at fixed chemical potential, leads to an increase of the fluid velocity. 
This is due to the increase in $\rho/s$ in \eqref{eq:betaeq}, which governs the coupling between the external electric field and the momentum density. Hence the momentum transfer from the electric field is increased as $\mu/(k_B T)$ increases. 

\item Third, increasing the external electric field $\mathcal{E}^x$ increases the velocity of the fluid (see figure \ref{fig:poise}), since a stronger electric field transfers more momentum to the system.

\item Finally, from \Eq{eq:betaeq}, we see that if $\eta/s$ increases, the kinetic term needs to become smaller in order for the equation to be satisfied at a fixed applied electric field. We confirm this from figure \ref{fig:poisetas}, where we display the velocity profile as function of increasing $\eta/s$. We find that strongly coupled fluids move faster than their counterparts at weaker coupling, i.e. at larger $\eta/s$. This is one of the main results of our work.

\end{itemize}

\begin{figure}
	\begin{center}
		\includegraphics[scale=0.20]{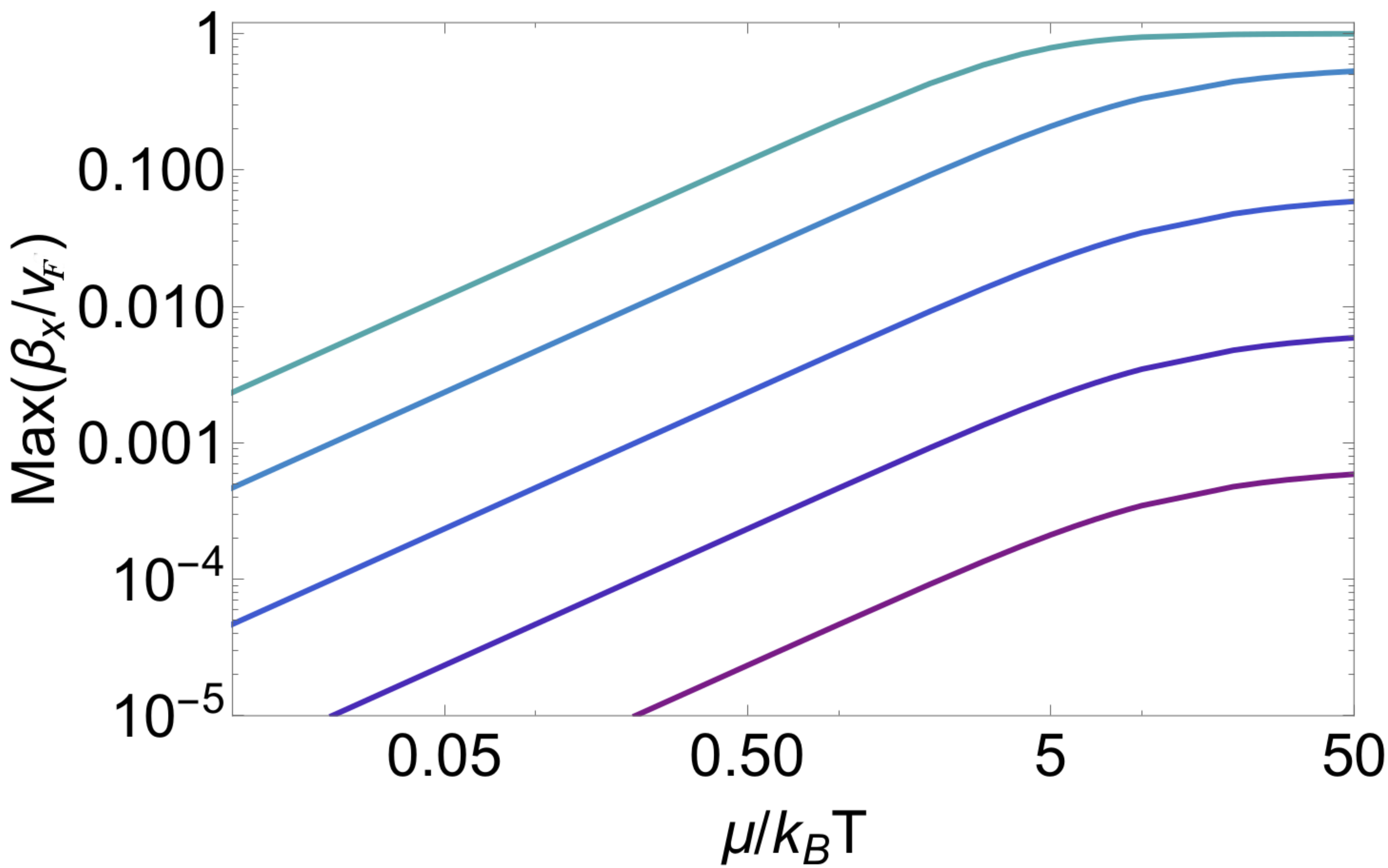}
		\caption{Keeping the ratio $\eta/s$ fixed to
                  $\hbar/4\pi k_B$, we plot the maximum speed of the
                  fluid $\beta_x$ as a function of the ratio $\mu/k_B
                  T$ at different values of the electric field ${\mathcal E}^x
                  = 5,1,0.1,0.01,10^{-3}$. For illustrative purposes,
                  we have set $w=1$. In order to present all possible
                  curves into a single plot, we have employed a
                  logarithmic scale, both on the $x$-axis and
                  $y$-axis. Note also that $\beta_x$ never reaches the Fermi velocity $v_F$. 
                  \label{fig:maxbeta}}
	\end{center}
\end{figure}

We now determine the range of values for the parameters $\lbrace \eta/s,\mathcal{E}^x,\mu/k_B T\rbrace$ for which the fluid flows relativistically, i.e. for which the velocity is not small any more compared to the Fermi velocity. From the final point above, we expect the fluid to flow fastest in the strong coupling limit, and hence we take $\eta/s$ to be the holographic value \eqref{etasAdSCFT}. From figure 
\ref{fig:maxbeta}, we see that at $\mathcal{E}^x \simeq 5$ (corresponding to $E_x \approx 7.5$ mV/$\mu$m), 
$\mu/k_B T\simeq 0.5$, and the width $w=1$, the velocity becomes about 10\% of the
Fermi velocity. Hence, for holographic fluids satisfying \eqref{etasAdSCFT}, the relativistic regime can be reached for small electric fields and chemical potentials. 

\subsection{Momentum relaxation}

The wall relaxation time $\tau_w$ introduced in sec.~\ref{sec:momentumrelaxation} characterizes the rate of momentum outflow through the boundaries. It is expected to be finite due to the finite size $w$ of the channel.  Nevertheless, the magnitude of this time-scale relative to the other scales in the problem (c.f. \eqref{hydroregime}) is \emph{a priori} unclear. In this section we compute $\tau_w$ as a function of the input parameters $\mu/k_B T,\mathcal{E}_x,w$ and $\eta/s$ in the absence of impurities\footnote{We recall that $\tau_w$ is {\it a priori} physically independent of the electron-electron scattering rate $\tau_{\rm ee}$.}.  Again, we resort to numerics in order to evaluate \Eq{eq:momentum2} on the profile solution $\beta_x(y)$. 

\begin{figure*}[tp]
\begin{tabular}{cc}
		\includegraphics[scale=0.19]{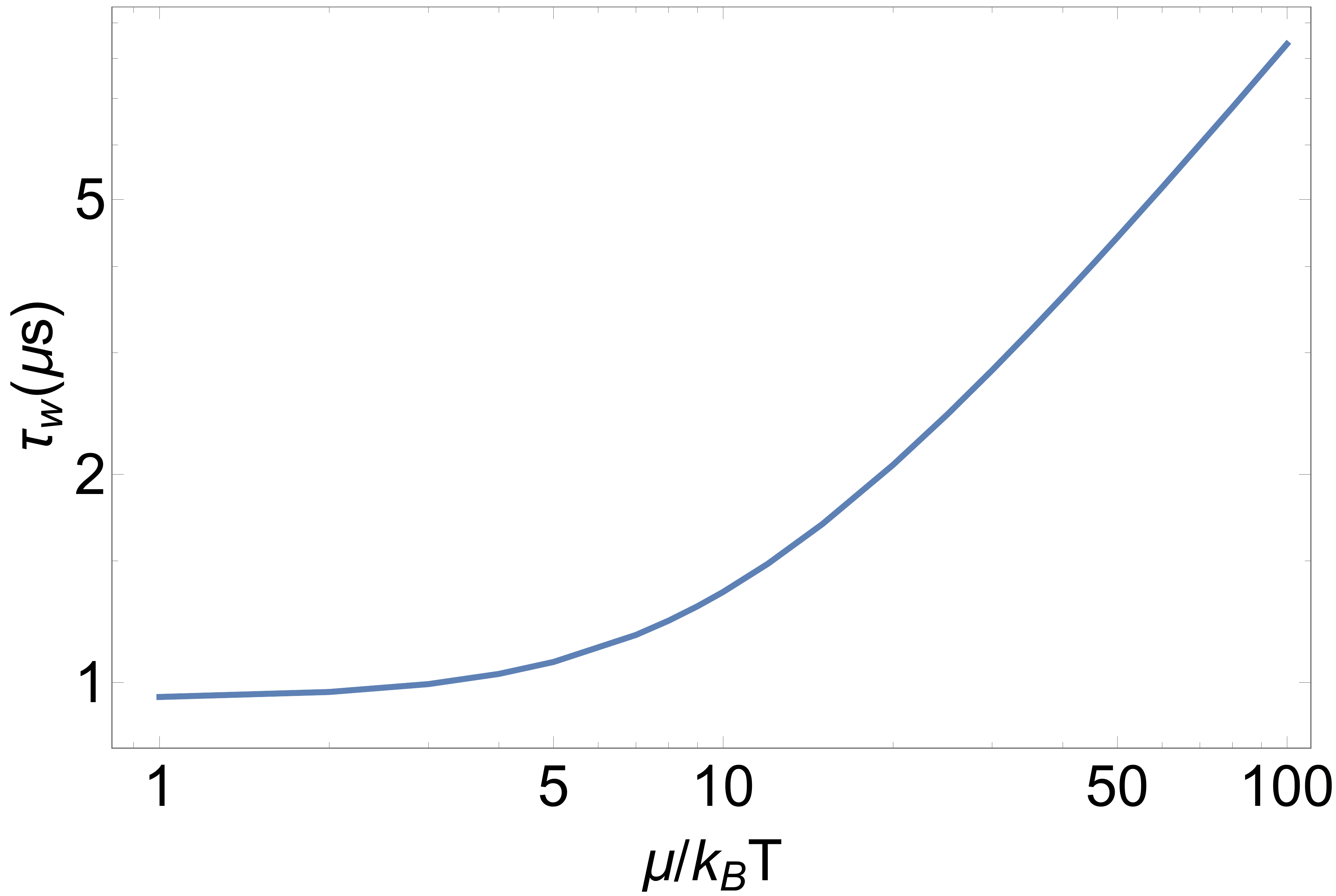}&
		\includegraphics[scale=0.2]{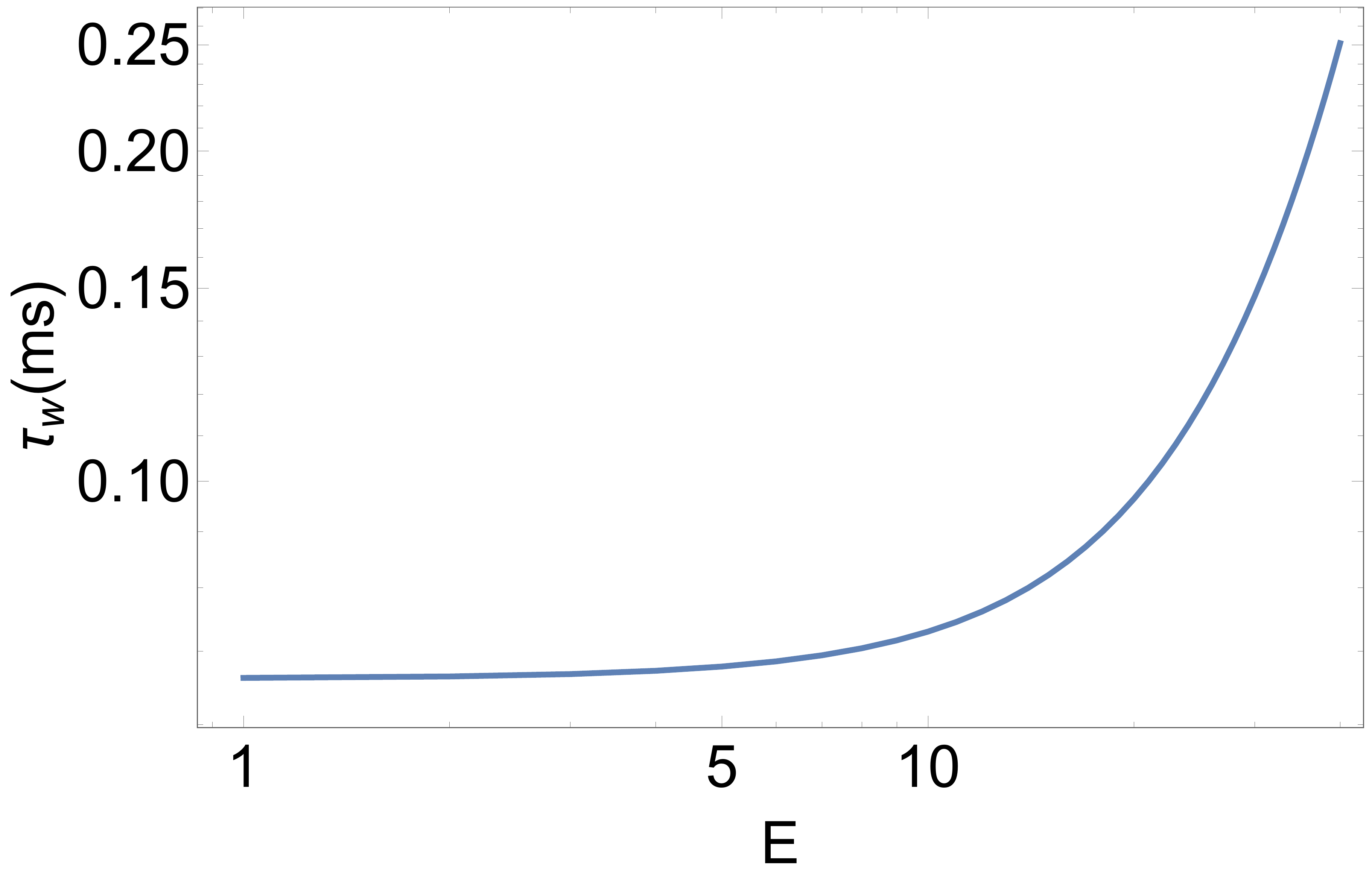}\\
		\includegraphics[scale=0.2]{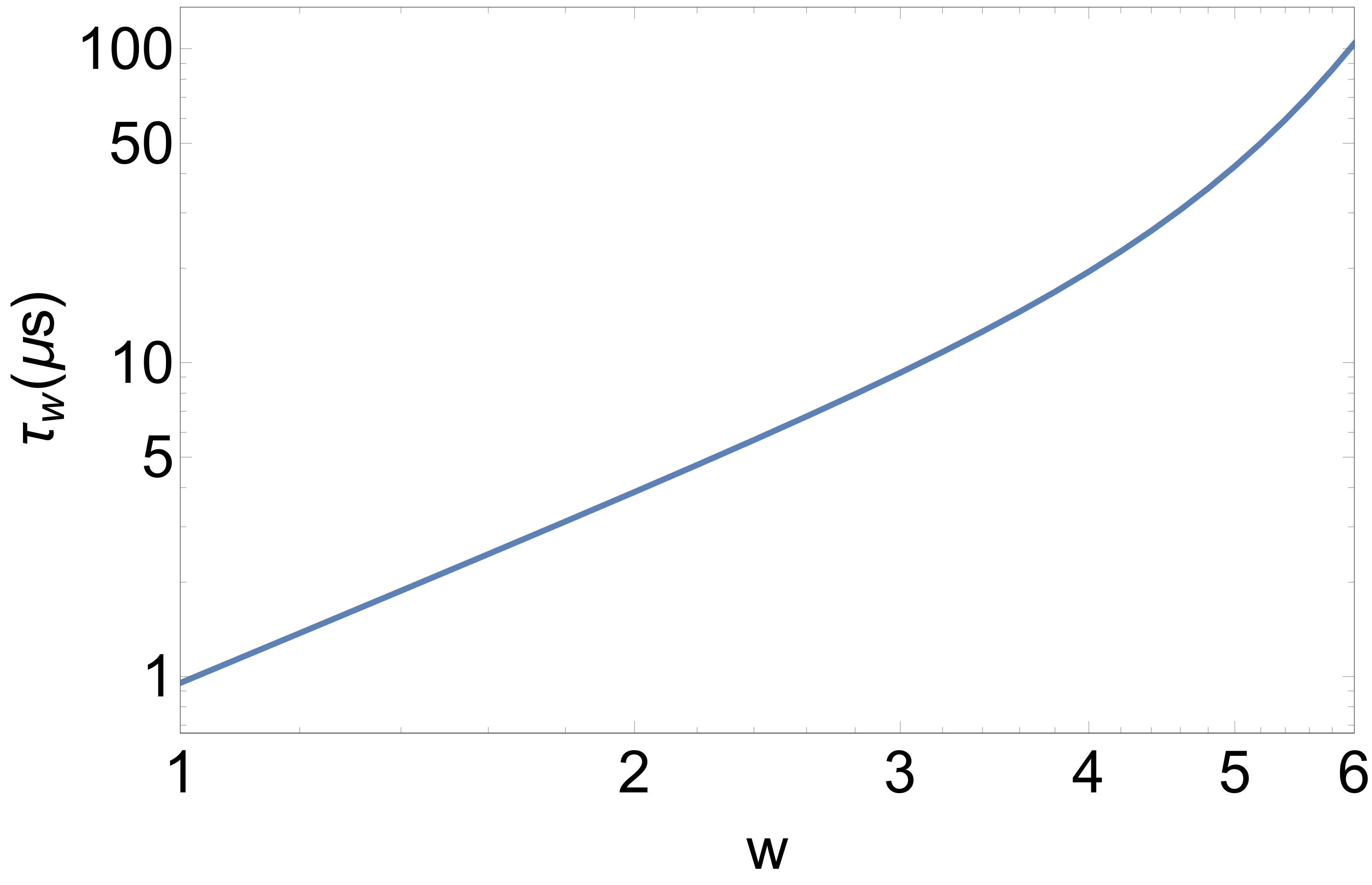}&
		\includegraphics[scale=0.2]{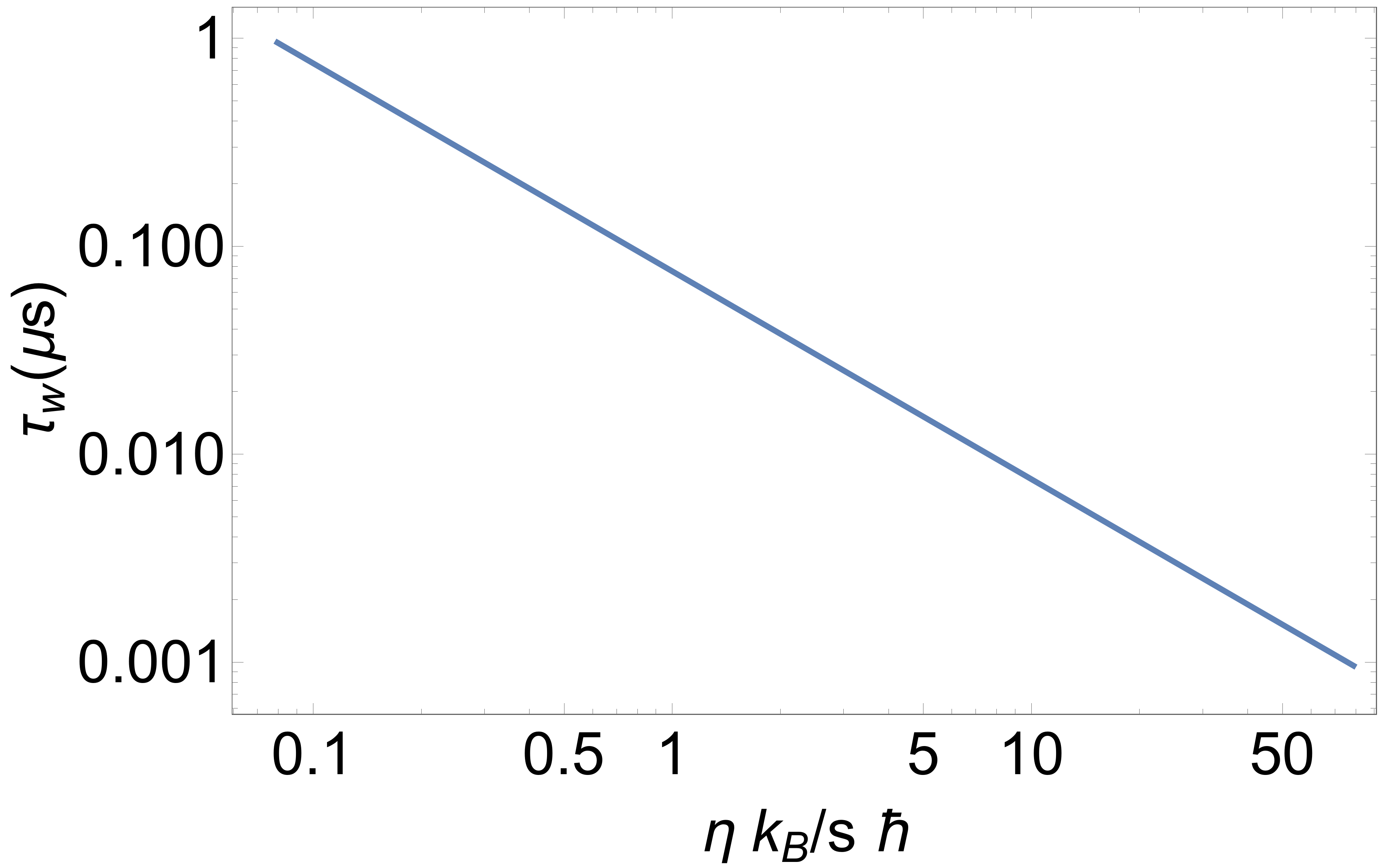}\\
\end{tabular}
		\caption{Plot $\tau_w$ at different input parameter configurations. Excluding the varying parameter, which varies from plot to plot, we have taken $\mu/k_B T =1, \eta/s = \hbar/4\pi k_B$, $w=1$, $\mathcal{E}_x =1$.  \label{fig:tauw}}
\end{figure*}

In figure \ref{fig:tauw} we plot $\tau_w$ as a respective function of one of the input parameters $\mu/k_B T,\mathcal{E}_x,w$ and $\eta/s$, while keeping fixed all other input parameters. From these plots, we infer that $\tau_w$ is closely related to the speed of the hydrodynamic fluid. Combinations of input parameters that lead to  higher velocities will also unavoidably lead to a larger $\tau_w$. The underlying reason is that $\tau_w$ is a measure for the time needed to lose all the momentum in the flow through the walls. This process naturally takes longer if there is more total momentum in the flow, i.e. if the velocity $\beta_x(y)$ is larger. By looking at the slope $\tau_w$ as a function of $\eta/s$ in the lower right plot of fig.~\ref{fig:tauw}, we find a scaling law of the wall relaxation time as a function of $\eta/s$,
\be 
\tau_w\propto \left( \eta/s \right)^{-1}.
\ee 
From \ref{fig:tauw}, we see that the typical window of values for
 $\tau_w$ is 
 \be \label{eq:tauwvalues}
 \tau_w \in \left[10^{-9}-10^{-4} \right] \, \rm{s},
 \ee
We should also point out that what we computed in \Eq{eq:tauwapprox} is the instantaneous rate of momentum loss after instantaneously switching off the electric field $E_x$. In other words, we assumed that the profile $\beta_x(y)$ after switching off the electric field is still the steady state Poiseuille flow. It is expected that the process of momentum loss will slow down as the flow profile $\beta_x(y,t)$ evolves non-linearly as a function of time. Hence we expect that the obtained $\tau_w$ is a lower bound to the $\tau_w$ that would be calculated from solving the time-dependent hydrodynamic equations \Eq{eq:eq1} and \Eq{eq:eq2}.

\subsection{Differential wire resistance}

In this section we  determine the differential wire resistance 
$\mathcal{R}$, defined by
 the inverse of the derivative of the current $I$ through the wire with respect to the voltage $V$,
\be \label{eq:cond}
\mathcal{R}(I) = \frac{d V}{d I}.
\ee 
Using \eqref{eq:j}, the total current $I$ that flows across the channel is given by
\be \label{eq:intcurrent}
I 
= \int^W_0 j^x(y) dy = \int^W_0 \left[ j^x_{\rm (fluid)} + j^x_{\rm (Q)} \right] dy,
\ee 
with
\be \label{eq:currents}
j^x_{\rm (fluid)} = e\rho u^x, \qquad j^x_{\rm (Q)} = \sigma_Q \frac{h}{e^3 c} u_\alpha D^{x\alpha}.
\ee
From \Eq{eq:intcurrent} and \eqref{eq:jcomps}, we observe that there are two different types of contributions to $I$: The first contribution, $j^x_{\rm (fluid)}\propto u^x$, is the flow of the particles along a transverse section of the channel. The second contribution, labeled as $j^x_{\rm (Q)}$, is related to the quantum critical conductivity $\sigma_Q$ and it accounts for the counterflow of electrons and holes. For the AdS/CFT charged black brane model \eqref{eq:bulkaction}, $\sigma_Q$ is given by \cite{Hartnoll:2007ih,Hartnoll:2007ip}
\be \label{eq:sigmaQ} 
\sigma_Q = \left( \frac{s T}{\varepsilon + p } \right)^2 \frac{L^2}{G_4}\frac{e^2}{2h}.
\ee 
For obtaining physical values for the currents, we fix the overall ratio $L^2/G_4$, where $L$ is the AdS radius $L$ and the Newton constant $G_4$, using the AdS/CFT approach described in appendix \ref{app:holoren}. The ratio is fixed to be $L^2/G_4 = 64\sqrt{3}$.

Since we apply a constant electric field along the $x$ direction, the
voltage is proportional to the applied electric field, $V = E^x \,l$,
with $l$ the length of the channel, assumed much larger
than the width $W$. Both Eqs.\Eq{eq:intcurrent} and \Eq{eq:cond} have
to be evaluated numerically, since the velocity profile is obtained from a numerical integration. We recast both the $I$ 
and the differential resistance $\mathcal{R}$ in terms of the reduced 
variables \Eq{eq:changes},  
\begin{align}
\label{eq:intiu}
I &= \int\limits^w_0 du \gamma(u)\left[ e \rho \beta_x(u) +  \sigma_Q \mathcal{E}^x \frac{T^2 k_B^2}{e h v_F} \right]\,,    \\ 
\mathcal{R} &= \frac{ek^2_B}{h^2\epsilon_0}\frac{l}{v_F^2\ep_r} T^2\left(\frac{d I }{d\mathcal{E}^x}\right)^{-1}\nonumber .
\end{align}

In sec.~\ref{sec:flow}, we focused on deriving the qualitative features of the Poiseuille flow at strong coupling. The aim of this subsection is to obtain a realistic prediction for the value of the resistance. Therefore, we need to use a physically viable combination of input parameters. In experiments, the temperatures are in the range  of a few Kelvin \cite{Molenkamp:1994kb,Molenkamp:1994ii,deJong:1995bn}, whereas the width and length are of order of a few micrometers. We will hereafter assume 
\be\label{eq:input2}
T= 2 {\rm K},\quad W=4\mu{\rm m} \quad {\rm and}\quad l=20\mu{\rm m}\,. 
\ee
The chemical potential can be found from the number density $\rho$ after solving \Eq{eq:thermoads} for $\mu$.  Taking for instance the typical value of the density in experiments, $\rho \simeq 10^{11}\, {\rm cm}^{-2}$, and a spin degeneracy of $g=2$ as well as \eqref{eq:input2}, we find 
\be \label{eq:inputmu}
\mu \simeq 4.5\, \text{meV}.
\ee 
The Fermi velocity and relative permittivity are set according to \Eq{eq:input1}. 

\begin{figure}[tp]
	\begin{center}
		\includegraphics[scale=0.15]{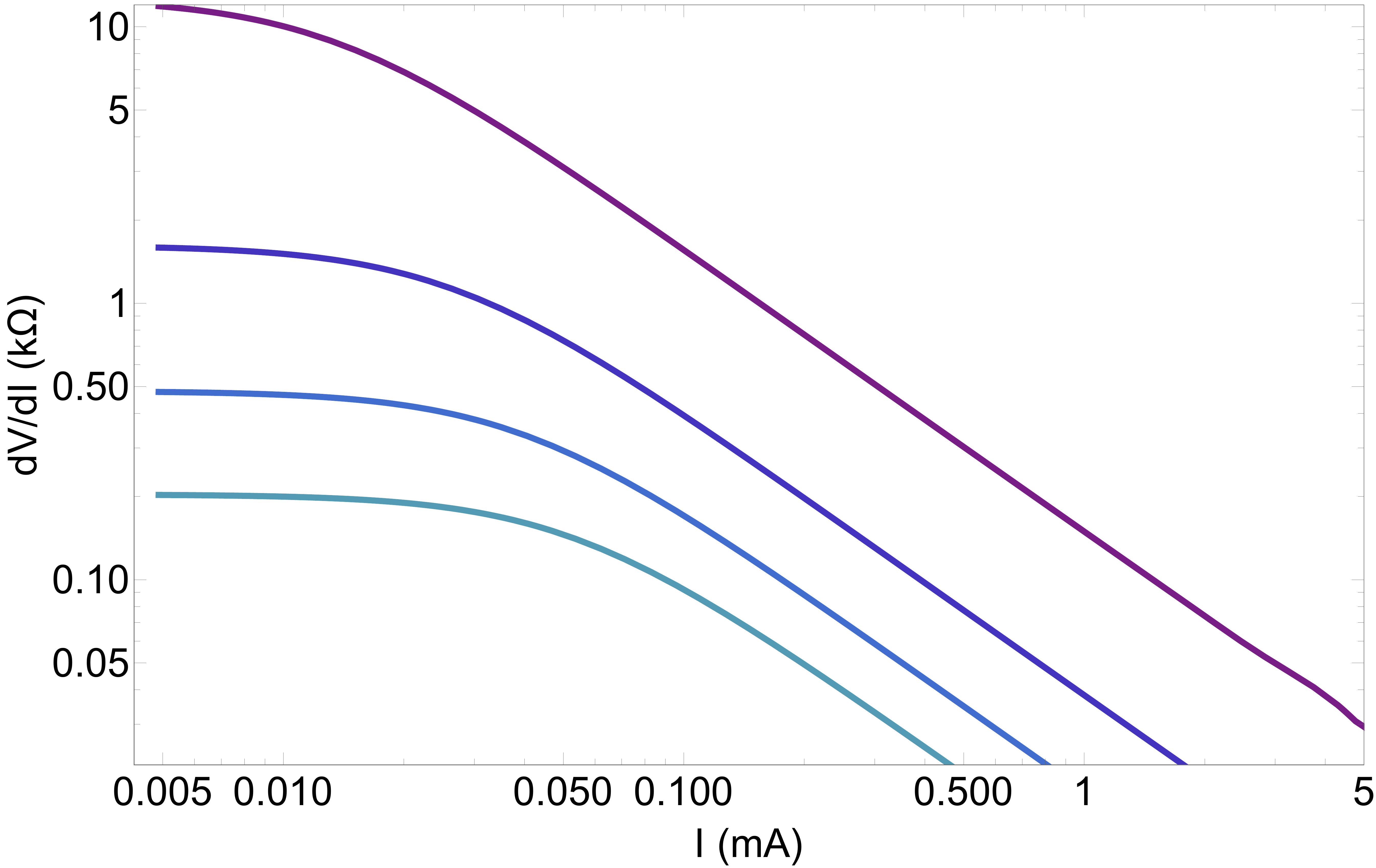}
		\caption{Differential resistance $\mathcal{R} = \frac{dV}{dI}$ as a function of $I$ in absence of impurities and at different widths. From top to bottom (in micrometers), $W=4, 3 ,2,1 $. In particular, we have taken $\eta/s = \hbar/4\pi k_B$. \label{fig:gurzhi1}}
	\end{center}
\end{figure}

In fig.~\ref{fig:gurzhi1} we display the differential resistance as a function of $I$ at different values of the width $W$. It starts with a finite value at $I=0$, and decreases with increasing current until it asymptotes to zero.  Increasing the width of the channel leads to  a decrease on the differential resistance, as expected for a Poiseuille flow. One might expect that at low $I$, or equivalently $\beta_x \ll v_F$, the differential resistance should coincide with the quantum critical resistance\footnote{The quantum critical resistance is defined via Ohm's law, $R = l/(\sigma_Q A)$, with $A$ the cross-sectional area of the wire. Here $A=W$.}, $\mathcal{R}_{\rm Q} \sim l/(W \sigma_Q)$, since at low $I$ there is no macroscopic charge transfer and the wire should act as an Ohmic resistor with an intrinsic conductivity $\sigma_Q$ given by \Eq{eq:sigmaQ}. However, this expectation does not bear out. Consider the solution for $\beta_x$ as obtained from \Eq{eq:eqbeta} restricted to small values of $I$. In this limit the applied electric field must be small and $\beta_x\ll v_F$. The convection term that appears in the right hand side of \Eq{eq:eqbeta} can be neglected and we obtain a linear differential equation
\be 
0 = h \epsilon_0 \frac{v_F E_x \epsilon_r  \rho}{e \eta }+\beta_x''(y),
\ee 
whose solution, consistent also with the boundary conditions \Eq{eq:bound}, is given by the usual parabolic profile of the Poiseuille flow
\be \label{eq:betasmall}
\beta_x = \frac{\epsilon_0 h}{e} \frac{E_x \epsilon_r v_F \rho  }{2\eta } y  (W-y).
\ee 
Inserting \Eq{eq:betasmall} into \Eq{eq:intcurrent} and integrating  over $y$, we obtain the differential resistance 
\be \label{eq:R1}
\frac{1}{\mathcal{R}} =h\epsilon_0 \epsilon_r v_F \frac{W}{l} \left(\frac{\sigma_Q}{e^2}+\frac{W^2 \rho^2}{12 \eta} \right) = \frac{1}{\mathcal{R}_{\rm Q}} + \frac{1}{\mathcal{R}_{\eta }}\,.
\ee
Equation \Eq{eq:R1} contains two contributions of different physical origin: the
first comes from the \emph{quantum critical resistance} $\mathcal{R}_{\rm Q}$, while the second, $\mathcal{R}_{\eta}$, is the \emph{viscous resistance} 
 generated by viscous effects alone. At fixed chemical potential and temperature, the quantum critical conductance \Eq{eq:sigmaQ} as well as the number density $\rho$ defined in \Eq{eq:thermoads} are both fixed and
\be 
\mathcal{R}^{-1}_{\rm Q} \propto W, \qquad \mathcal{R}^{-1}_{ \eta}\propto \frac{W^3}{\eta}.
\ee 
Therefore, $\mathcal{R} \simeq \mathcal{R}_Q$ only if $W$ is small, or $\eta$ large. Finally, we see that the two contributions to \eqref{eq:R1} follow an inverse Matthiessen's rule, i.e. 
\begin{gather}
\mathcal{R}^{-1}  = \sum\limits_{i=1}^2 \frac{1}{\mathcal{R}_i} \, ,
\end{gather}
where the $\mathcal{R}_i$ are resistances arranged in parallel. 
The role of $\mathcal{R}_1^{-1}$ is played by $\mathcal{R}_Q$ and the
role of $\mathcal{R}_2^{-1}$ is played by $\mathcal{R}_\eta$.

The asymptotic form of $\mathcal{R}(I)$ in Fig.\ref{fig:gurzhi1} is related to the asymptotic behavior of $\beta_x$ when approaching the Fermi velocity. For large enough electric field, the velocity profile approaches the Fermi velocity, and the first piece of \Eq{eq:intcurrent} asymptotes to a
constant value, while the second contribution is still linear in $E_x
\propto V$. Taking the derivative with respect to $V$, only the
second contribution survives, and the resistance asymptotes to
zero due to the $\gamma$ factor. This asymptotic behavior is universal, in the sense that it is not affected by the choice of input parameters within the regime of applicability of hydrodynamics.

\begin{figure}
	\begin{center}
		\includegraphics[scale=0.18]{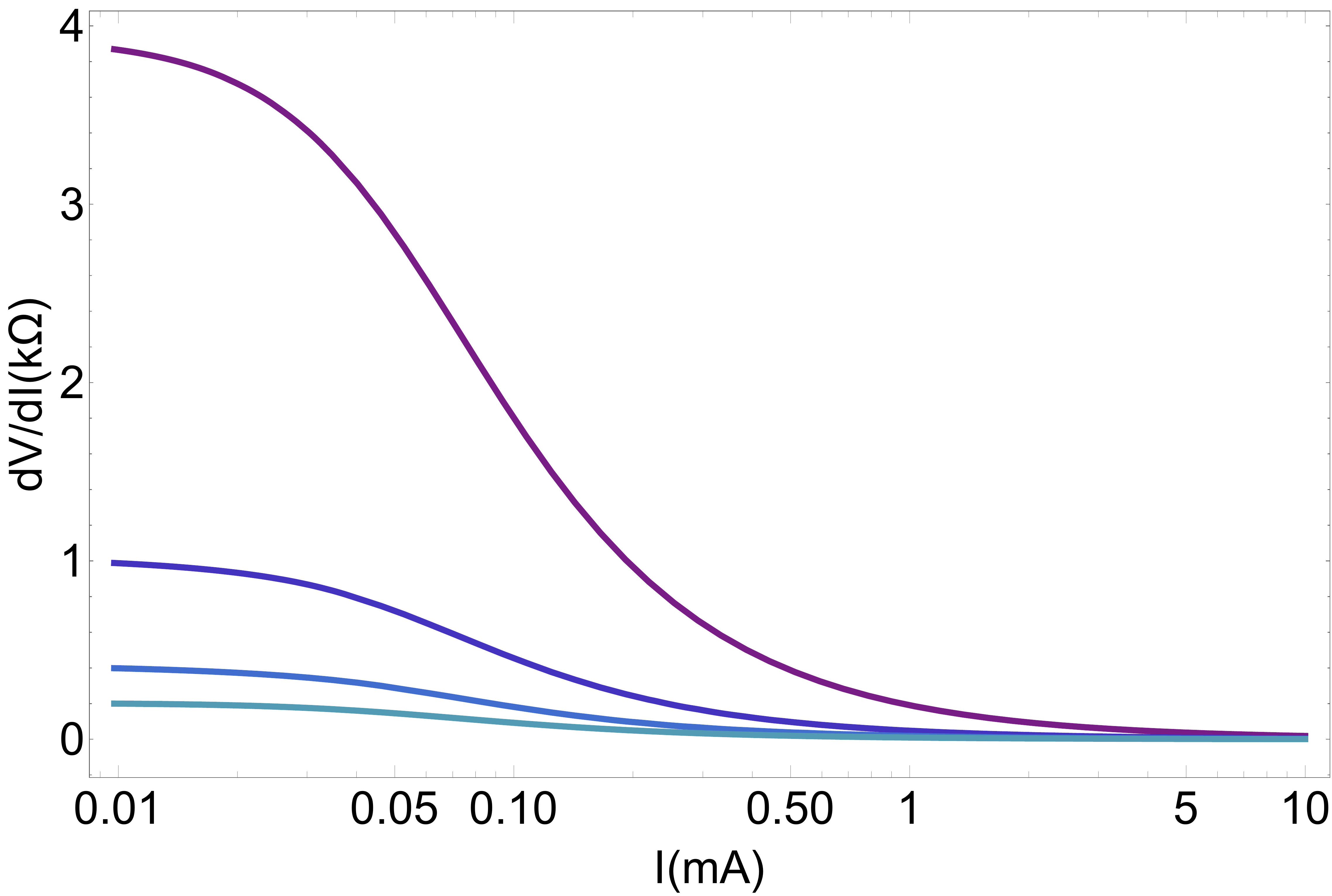}
		\caption{Differential resistance $\mathcal{R} = \frac{dV}{dI}$ as a function of $I$ in absence of impurities. From top to bottom, $k_B \eta/s \hbar = 5/\pi,5/4\pi,1/2\pi,1/4\pi$. \label{fig:gurzhi2}}
	\end{center}
\end{figure}

We now analyze the dependence of the resistance $\mathcal{R}$ on the
viscosity over entropy density ratio $\eta/s$.  This provides important 
information for two reasons:  First, as explained in sec.~\ref{sec:introconc}, $\eta/s$ can
be understood as a measure for the coupling strength 
\cite{Fritz:2008go,Muller:2009cy}. Second, if the remaining 
input parameters are known and two different samples are well in the hydrodynamic regime, it will be possible to infer the relative value of $\eta/s$ through a measurement of the wire resistance at small current. This will be 
of interest from the experimental point of view. 
In figure 
\ref{fig:gurzhi2} we plot the wire resistance $\mathcal{R}$ as a function
of the integrated current $I$ at different ratios of $\eta/s$. We find
that $dV/dI$ increases with increasing $\eta/s$. This behaviour is
related to the decrease of the maximal velocity $\beta_x$, and hence of the
integral \Eq{eq:intcurrent}, as $\eta/s$ increases.

\section{Flow and differential resistance in the presence of impurities}\label{sec:impurities}

We now examine the effect of impurities on the results of
section \ref{sec:flowdVdI}, as is
essential in view of comparison with experiments. In this case, the AdS/CFT
results \eqref{eq:thermoads} are no longer directly applicable since
the pressure becomes $y$-dependent, as given by \eqref{eq:eqp} for
finite $\tau_{\mathrm{imp}}$. In the non-relativistic limit, however,
where $\gamma \rightarrow 1$, this $y$-dependence drops out again from
all the thermodynamic variables. 

We thus consider the non-relativistic limit in this section. This
restricts us to small values of the applied electric field. On the
other hand, we are able to obtain analytical results in this regime,
and to perform a general
hydrodynamical analysis applicable to any value of the
electron-electron coupling.

\subsection{Velocity profiles}\label{sec:velimp} 

\begin{figure}
	\begin{center}
		\includegraphics[scale=0.20]{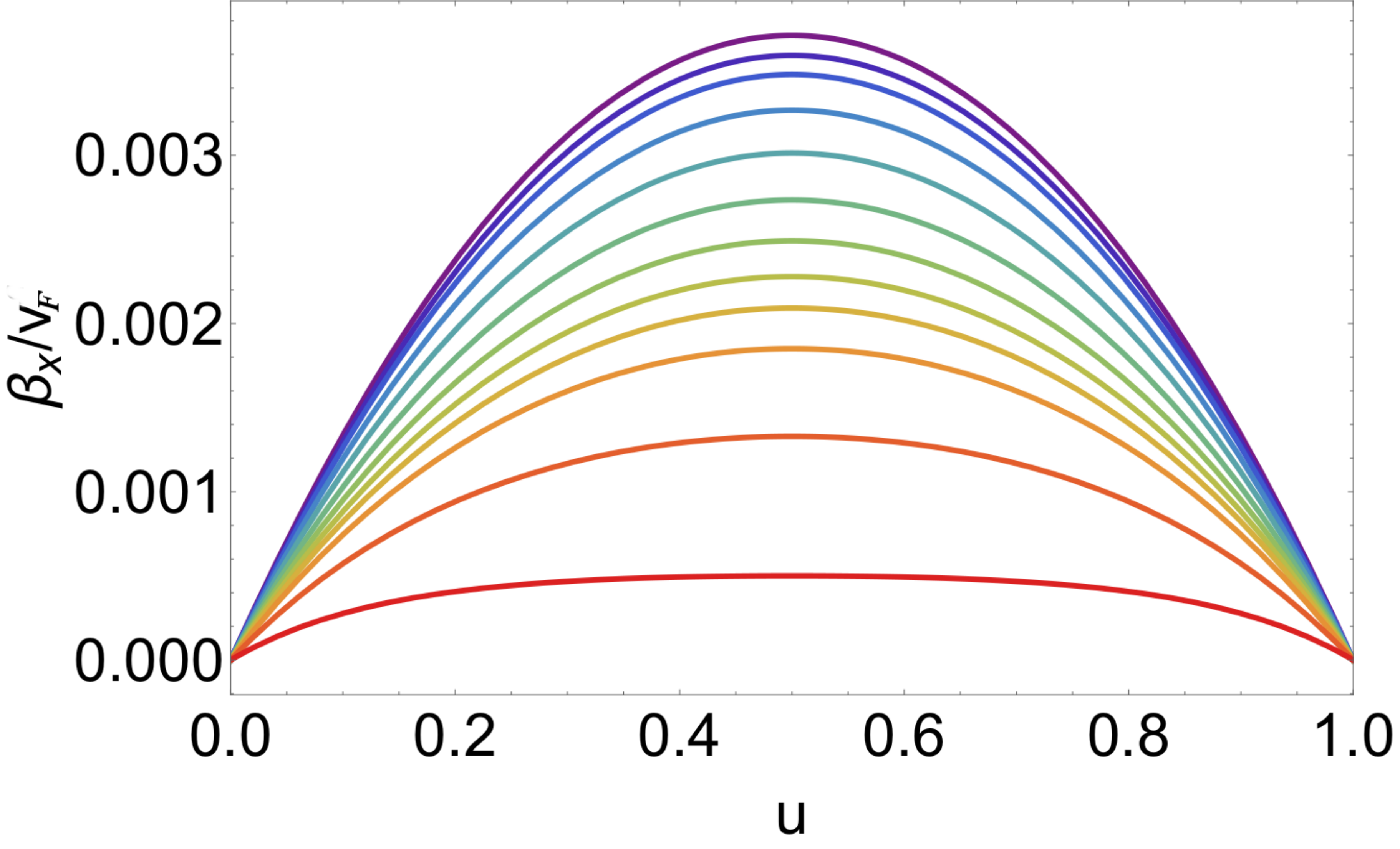}
		\caption{Velocity profile $\beta_x$ along the $x$ direction, $\beta_x(u)$, at different mean free time $\tau_{\rm imp}$. From top to bottom: $T\tau_{\rm imp} k_B/h =10^4,20,10,5,3,2,1.5,1.2,1,0.8,0.5,0.2$. It is remarkable to notice that for the lowest curve, the profile is almost zero in the interior. For all the curves, we have set $\mu/k_B T =1$, $\eta/s=\hbar/4\pi k_B$ and $\mathcal{E}^x = w = 1$. \label{fig:poistau}}
	\end{center}
\end{figure}

We begin by examining the velocity profile of the fluid. 
As discussed above, we consider the
 non-relativistic limit,  in which the non-linear terms that enter
the Navier-Stokes equations \Eq{eq:eqbeta} are suppressed and it is possible to find an
analytic solution for $\beta_x$. 
Linearizing \Eq{eq:eqbeta} in $\beta_x$ gives
\be \label{eq:betataueq}
\frac{\eta}{s}\beta_x'' + E_x \epsilon_r v_F\frac{\rho}{s} \left(\frac{h  \epsilon_0}{e}\right) - \frac{\beta_x}{\tau_{\rm imp}v_F^2}\frac{\varepsilon + p}{s} =0.
\ee
Imposing $\beta_x(0)=\beta_x(W)=0$, we obtain the profile as
a function of the $y$ coordinate,

\begin{align}
 \label{eq:betatau}
\beta_x(y) &= v_F \epsilon_r \frac{E_x h \epsilon_0 }{e}\frac{\rho}{\eta \mathcal{G}^2} \nonumber\\
 \times& \left[ 1- \cosh\left( \mathcal{G} y \right)+ \sinh\left(\mathcal{G}y \right)\tanh\left( \frac{\mathcal{G} W}{2}\right)  \right],
\end{align}
with 
\be \label{eq:G}
\mathcal{G} = \frac{1}{v_F}\sqrt{\frac{\varepsilon + p}{\eta \tau_{\rm imp}} } ,\qquad  \left[ \mathcal{G} \right] = {\rm m}^{-1}.
\ee
Note that the result for the velocity profile is analytical in the
limit considered here, while the relativistic result of the preceding
section is numerical. 
We also note that   $\beta_x$ as given by \Eq{eq:betatau} reduces to
the parabolic flow \Eq{eq:betasmall} in the limit $\tau_{\rm imp}\to
\infty$, as expected.
In Fig.~\ref{fig:poistau}, we display the velocity profile of
\Eq{eq:betatau}  as function
of the impurity scattering time.  We observe that when increasing the impurity density
while keeping all other input parameters fixed, the velocity
decreases. 

If the density of impurities is such that $\tau_{\rm imp}$
becomes the shortest time scale of the system, shorter than the
electron-electron scattering time, then impurity collision effects are
dominant. In this case,  the
Poiseuille hydrodynamic behavior is suppressed and a standard Ohmic
conductivity law is expected instead. Indeed, the Poiseuille flow
connects smoothly to an Ohmic regime as may be seen as follows.
We consider
the non-relativistic limit of  \Eq{eq:eqbeta} and assume that
$\eta/s \ll T\tau_{\rm imp} $.  Then
\be 
\beta_x = \frac{v_F^3 \tau_{\rm imp}\ep_r \ep_0 h}{e(\varepsilon +
  P)}\rho E_x  \, .
\ee 
Inserting this expression in $j_x=  e\rho \beta_x + \sigma_Q E_x $, we obtain in the non-relativistic limit
\be \label{eq:jx2}\hspace{-0.25truecm}
j_x = \left( \sigma_Q + \rho^2 \frac{v_F^3 \tau_{\rm imp}\ep_r
    \ep_0 h}{\varepsilon + P}\right) E_x = \left( \sigma_Q +
  \sigma_D\right) E_x\, . 
\ee
The overall conductivity $\sigma$  consists of the sum of the quantum critical
conductivity $\sigma_Q$ and the Drude conductivity $\sigma_D$, given by
\be
\label{Drude sigma}
\sigma_D = \rho^2 \frac{v_F^3 \tau_{\rm imp}\ep_r \ep_0
  h}{\varepsilon + P} \, .
\ee
From the quantum critical conductivity defined in \Eq{eq:sigmaQ} and for a typical value of $\tau_{\rm imp} =
10^{-12}$s~\cite{deJong:1995bn}, we find that 
\be \label{conductivityratios}
\frac{\sigma_D}{\sigma_Q} = \frac{\rho_Q}{\rho_D} \simeq 0.84 
\ee 
for the input parameters given around \Eq{eq:input1} and \Eq{eq:inputmu}.
This implies that the quantum critical conductivity  plays a more significant
role in the current $j^x$ of \Eq{eq:jx2}, although both are of the same order of
magnitude. This shows that at strong coupling, the electron-hole
scattering effect that gives rise to the quantum critical conductivity
\cite{Fritz:2008go} cannot be neglected compared to impurity
scattering: At weak coupling, \cite{Fritz:2008go} the quantum critical
resistivity drops to zero as $\alpha^2(T)$ and the Drude resistivity
is expected to approach a constant. At strong coupling on the other hand \cite{Fritz:2008go}, the quantum critical resistivity is expected to saturate as well, to a universal value $\sim \frac{h}{e^2}$. Hence, while at weak coupling the ratio \eqref{conductivityratios} is expected to approach zero, it is expected to asymptote to a constant for strongly interacting electron systems. We will further elaborate on the differences of strong and weakly coupled electron systems in sec.~\ref{sec:disc_outlook}.

In conclusion, we observe that when increasing the impurity density
while keeping all other input parameters fixed, the fluid velocity decreases.
Moreover, if $\tau_{\rm imp}<\tau_{\rm ee}$, the flow will no longer
be of Poiseuille form, since the impurities absorb momentum. If the density of impurities is large enough, impurity effects dominate and the electron flow becomes Ohmic. Since the transition between the Poiseuille and Ohmic flows is determined by the interplay of two scales, the electron-electron scattering rate and the impurity scattering rate, this  transition is a cross-over. \\

\subsection{Momentum relaxation}
\label{subsec:momimp}

Here we calculate the wall relaxation time $\tau_w$ within
hydrodynamics for the Poiseuille flow \ref{eq:betatau}. The result is
valid for any value of the electron-electron coupling.  This is
particularly interesting from the AdS/CFT point of view since wall
momentum relaxation has not yet been analyzed in this context.
Moreover, we aim at retrieving the physics associated to the interplay of
impurity and finite size effects, 
both of relevance in  experiments.

Since the velocity profile  $\beta_x$ of \Eq{eq:betatau}  is known
analytically,  $\tau_w$ may also be computed 
analytically. We expand \Eq{eq:momentum2} at small $\beta_x$,
\be \label{eq:tauwnonrel}\hspace{-0.25truecm}
\int^W_0 dy\, \frac{\beta_x}{v_F} \left(p + \varepsilon\right) \left(\frac{1}{\tau_w}+\frac{1}{\tau_{\rm imp}}\right) = \eta\left.\beta'_x(y)\right|_{y=W}^{y=0},
\ee 
 and insert the solution \Eq{eq:betatau} into \eqref{eq:tauwnonrel} to find
\be \label{eq:taunonrelw1}
\tau_w = \frac{\tau_{\rm imp}}{2}\left[ \frac{W \mathcal{G} }{4\tanh \left( \frac{W\mathcal{G}}{2} \right)-W\mathcal{G}  } -1\right] .
\ee 
As we  see from \Eq{eq:G}, $\mathcal{G}$ is an involved function of
the thermodynamic variables, the shear viscosity $\eta$ and the
impurity scattering rate $\tau_{\rm imp}$. 

For a fluid propagating through the channel, we expect that the wall momentum
relaxation time is larger in presence than in absence of impurities,
$\tau_w^{(\tau_{\rm imp}< \infty)} > \tau_{w}^{(\tau_{\rm imp}\to
  \infty)}$. This is due to impurity momentum relaxation
competing with wall momentum absorption. Indeed, if we consider a small density of impurities or equivalently large $\tau_{\rm imp}$, we find from expanding \Eq{eq:taunonrelw1} that the wall relaxation time in the presence of impurities is always longer than in a completely clean sample,
\begin{align} 
\tau_w^{(\tau_{\rm imp}< \infty)} &\simeq \tau_{w}^{(\tau_{\rm imp} \to \infty)} \nonumber \\
&+ \frac{W^2}{180 \tau_{\rm imp} v_F^4} \left[\frac{W^2 (p+\epsilon )^2}{\eta ^2}+15
	v_F^2\right]\,, 
\end{align}
with 
\be 
\tau_{w}^{(\tau_{\rm imp} \to \infty)} = \frac{W^2 (p+\epsilon )}{12 \eta \, v_F^2}.
\ee
In particular, after using the Gibbs-Duhem relation \Eq{eq:GibbsDuhem}, 
\be 
\tau_{w}^{(\tau_{\rm imp} \to \infty)} = \frac{W^2}{12 \eta \, v_F^2} \left( \mu \frac{\rho}{\eta} + T \frac{s}{\eta}\right),
\ee 
we can confirm analytically the $(\eta/s)^{-1}$ scaling dependence in the relaxation time through the walls, as it was firstly predicted in Sec.\ref{subsec:momimp}. Notice however that no physical reasons \emph{a priori} state why this scaling dependence is preserved in the ultra-relativistic regime ($\beta_x \simeq v_F$), rendering thus the fully-fledged analysis of the Navier-Stokes equation \Eq{eq:betaeq} of great relevance.

In addition, from  the balance equation
\Eq{eq:momentum2}, we see that the r.h.s.~is determined by
the velocity profile at fixed $\eta$, whereas the l.h.s.~is
proportional to the sum of $\tau_w^{-1}$ and $\tau_{\rm imp}^{-1}
$. Therefore, for a fixed velocity profile $\beta_x$, $\tau_w$ 
becomes larger  if $\tau_{\rm imp}$ decreases. Equivalently,
momentum diffusion through the boundaries is impeded by the presence
of impurities, even though both mechanisms lead to momentum absorption.

\subsection{Differential resistance} 

Finally we compute the differential resistance of our channel in the
presence of impurities. We use again the same formulae for the current
and differential resistance, \Eq{eq:intcurrent}  and \Eq{eq:cond}. As
before, for a constant energy density even in the presence of
impurities, we have to consider the non-relativistic limit. This in
turn leads to analytical results.

We insert the velocity profile \Eq{eq:betatau} into the current \Eq{eq:intcurrent} and find 
\be \label{eq:nonrelr}
I= \frac{V}{l} v_F \epsilon_r \left( h \epsilon_0\right) \Bigg \lbrace
\frac{W}{e^2}\sigma_Q + \frac{\rho^2}{\mathcal{G}^3 \eta } \left[
  W\mathcal{G} -2 \tanh\left( \frac{W \mathcal{G}}{2}\right)  \right]
\Bigg\rbrace \, .
\ee
Here, $W$ is the dimensionful width of the channel, $l$ is its length,  $\mathcal{G}$ is
given by \Eq{eq:G}, $\sigma_Q$ is the quantum critical conductivity, $\rho$ is
the number density and $\eta$ the shear viscosity. 
\Eq{eq:nonrelr}
 is linear in $V$, so that
 \begin{align} \label{eq:Rimp} 
\mathcal{R}^{-1} = \frac{I}{V} &=  v_F h \epsilon_0 \epsilon_r \frac{1}{l} \times \nonumber\\
\times \Bigg \lbrace  \frac{W}{e^2}\sigma_Q & + \frac{\rho^2}{\mathcal{G}^3 \eta } \left[ W\mathcal{G} -2 \tanh\left( \frac{W \mathcal{G}}{2}\right)  \right] \Bigg\rbrace.
\end{align}

The differential resistance $\mathcal{R}$ given by \Eq{eq:Rimp} is
independent of the current $I$. This follows from the non-relativistic
limit in which $I$ is linear in $V \propto E_x$. 
Terms of higher order in 
$E^x$ contributing to  the equations of motion \Eq{eq:betataueq} and to
the velocity  profile solution \Eq{eq:betatau} would signal relativistic effects, as we
may see from the analysis of  the velocity profile when
impurities are absent  (Fig.\ref{fig:maxbeta}). This means that the relation between
the current $I$ and the quantities on the r.h.s.~of
\Eq{eq:nonrelr}
is valid only for small current $I$, in consistency with the
approximation of small $E_x$. Nevertheless,
\Eq{eq:nonrelr} can be used to predict the value of the maximum of
the differential resistance in a hydrodynamic Poiseuille flow once the input parameters and thermodynamic variables are fixed.

Viscous and impurity effects cannot be disentangled in the
expression for $\mathcal{R}$ (recall that $\mathcal{G}$, given by
\Eq{eq:G}, in \Eq{eq:Rimp}, depends both on $\eta$ and $\tau_{\rm
  imp}$). This follows from the fact that the $j^x_{\rm (fluid)}$
contribution to the current density \Eq{eq:currents} inherits a
non-trivial dependence on $\eta$ and $\tau_{\rm imp}$ from the
velocity profile \Eq{eq:betatau}. 
Viscous and impurity effects may not even be disentangled in the limit
of vanishing $\tau_\mathrm{imp}\rightarrow \infty$. This may be seen
  by expanding \Eq{eq:Rimp} at large $\tau_{\rm imp}$,
\bea 
\mathcal{R}^{-1}_{(\tau_{\rm imp}<\infty)} &\sim & \mathcal{R}^{-1}_{(\tau_{\rm imp}\to \infty)} -\frac{W^5 }{120 \eta ^2 \tau_{\rm imp} }\frac{h}{v_F\, l} \epsilon_0\epsilon_r \rho ^2(p+\varepsilon )\nonumber\\
& \sim & \mathcal{R}^{-1}_{(\tau_{\rm imp}\to \infty)} \nonumber\\
& -& \frac{W^5 }{120 \eta \tau_{\rm imp} } \frac{h}{v_F\, l} \epsilon_0 \epsilon_r \rho ^2\left(T\frac{s}{\eta} + \mu \frac{\rho}{\eta} \right), \label{eq:Rimplowtau}
\eea
with $\mathcal{R}^{-1}_{(\tau_{\rm imp}\to \infty)}$ as defined in \Eq{eq:R1}. The second piece on the left-hand side of \Eq{eq:Rimplowtau} depends not only $\eta$ and $\tau_{\rm imp}$, but also on the ratio $\eta/s$. Nevertheless, laboratory samples usually have a significant impurity density, implying that $\tau_{\rm imp}$ can be rather small. Expanding \Eq{eq:Rimp} at low $\ell_{\rm imp}$ compared to the width of the channel $W$\footnote{We assume that $\ell_{ee}\ll \ell_{\rm imp}$ holds, such that we are in the hydrodynamic regime.}, we find
\be \label{eq:Rimptau0}
\mathcal{R}^{-1} \sim \mathcal{R}^{-1}_{Q} + \mathcal{R}^{-1}_{\eta},
\ee 
with $\mathcal{R}^{-1}_{Q}$ the quantum critical resistance as defined in \ref{eq:R1}, and this time, $\mathcal{R}^{-1}_{\eta}$ defined by
\be 
\mathcal{R}^{-1}_{\eta} = \sigma_D - \mathcal{R}^{-1}_{(1)} + \cdots ,
\ee 
with $\sigma_D$ the Drude conductivity, defined in \ref{Drude sigma} and
\be 
\mathcal{R}^{-1}_{(1)} =  2 E_x h \rho ^2 v_F^4 \epsilon_r \epsilon_0 \tau_{\rm imp}\left[\frac{\tanh\left(\frac{\mathcal{G} W}{2}\right)}{\varepsilon + p }\right] \sqrt{\frac{\tau_{\rm imp}  \eta}{\mu  \rho +s T}},
\ee
which in particular depends on the ratio $\eta/s$. This last term, despite of being subleading with respect to the Drude conductivity, its contribution can be significant enough to be accounted for. Indeed, for the set of input parameters \ref{eq:input1}, \ref{eq:input2} and \ref{eq:inputmu}, the relative difference $\xi$ between $\mathcal{R}^{-1}_{(1)}$ and $\sigma_D$ is found to be
\be 
\xi = \Big\vert\frac{\sigma_D-\mathcal{R}^{-1}_{(1)}}{\sigma_D}\Bigg\vert \sim 0.9,
\ee
for $\tau_{\rm imp}\sim 10^{-11}$s. This relative difference becomes even smaller ($\xi\to 0$) for cleaner samples. Given the fact that $\mathcal{R}^{-1}_{(1)}$ depends significantly on $\eta/s$ and hence on the coupling, one can state that for cleaner samples, it is expected to experimentally observe significant departures from the hydrodynamical differential resistance with respect to the $\mathcal{R}_{\eta} \propto \eta$ law found in the literature\cite{}. This prediction is complementary to the analysis at zero impurities carried out in section \ref{sec:flowdVdI}, where we showed that the differential resistance highly depends on $\eta/s$. 

\section{Reynolds Number for Channel Flows}\label{sec:turbulence}

 One of the main results of this work is that strongly coupled fluids flow fastest. For sufficiently large fluid velocities, the fluid can become turbulent. Between the laminar regime and fully developed turbulence, there is a transitional preturbulent regime in which both laminar and turbulent flows coexist. The Reynolds number $Re$ is used to quantitatively determine the transition between laminar and turbulent flows. In appendix \ref{app:reynolds}, we provide a derivation of the Reynolds number, both for ultra-relativistic and non-relativistic viscous fluids. The non-relativistic Reynolds number is shown in \eqref{8}. 
For Newtonian fluids moving through a 2+1-dimensional channel, a Reynolds number $Re<100$ is associated with clearly laminar flows, while highly turbulent flows occur for $Re\gg 100$ \cite{hanks1963laminar,hanks1969theory}.
Using the KSS bound for $\eta/s$ \Eq{etasAdSCFT2},  
an upper bound on the Reynolds number \Eq{8},
\be \label{eq:reynoldsholo}
Re \leq \frac{4\pi k_B}{\hbar}\frac{\left(\varepsilon + p\right)W}{s v_{\rm max}}\left( \frac{v_{\rm max} }{v_F} \right)^2,
\ee
can be derived. Based on this, for the same maximal flow velocity $v_{\rm max}$ and width $W$, it is reasonable to assume that a strongly coupled fluid is more likely to become turbulent than its weakly coupled counterpart. From \Eq{eq:thermoads}, in the near Fermi liquid limit we obtain
\be 
\frac{\varepsilon + p}{s}\Bigg\vert_{\mu \gg k_B T} = \frac{\sqrt{3} }{16\pi }\frac{\mu}{k_B}.
\ee 
Hence, large chemical potentials enhance the transition to the turbulent regime. Estimating the right hand side of \eqref{eq:reynoldsholo} with input parameters \eqref{eq:input2}-\eqref{eq:inputmu}, we find $ Re \leq 50$. Therefore, for our channel widths, the flow is clearly laminar \cite{hanks1963laminar,hanks1969theory}, but if wider channels could be synthesized in the hydrodynamic regime, higher Reynolds numbers can be achieved since $Re\propto W$. Also, following \cite{Mendoza:2011ka}, inserting an obstacle into the channel or considering flows through a constriction may allow for preturbulent physics such as vortex shedding. 

However, impurities should be accounted for in this analysis. We can accomplish this by introducing the Gurzhi length $\lambda_G$, \cite{Lucas:2017wa} defined as  
\be \label{eq:lambdag}
\lambda_G = v_F \sqrt{\kappa \tau_{\rm imp}} , \qquad \kappa = \frac{\eta}{\varepsilon + p}.
\ee 
where $\kappa$ is the kinematic viscosity of the fluid\footnote{We define the kinematic viscosity $\kappa$ as the ratio of viscosity and energy density. There exists another kinematic viscosity, $\nu$, defined as the ratio of viscosity and mass density. The two viscosities are related by $\nu= v_F^2 \kappa$.}.Using $\lambda_G$ we may define two inequivalent regimes: First, if $W < \lambda_G$, it is likely that the fluid will propagate according to a Poiseuille flow, with a large speed in the middle of the channel. By using the definition of Reynolds number in the ultra-relativistic limit \Eq{9'} derived in appendix \ref{app:reynolds}, we find
\be 
Re_{(W< \lambda_G)} = \frac{\tau_{\rm imp}v_F}{W}\left( \frac{W}{\lambda_G}\right)^2\left(\frac{v_F}{v_{\rm max}}\right)^3.
\ee
Notice that the condition $W< \lambda_G$ introduces a bound on $\tau_{\rm imp}$. This bound is given by
\be\label{eq:nonreltaubound}
\tau_{\rm imp} > \frac{W^2}{v_F^2 \kappa}.
\ee 
Since $v_{\rm max} \leq v_F$, even though they are of similar magnitude, using \eqref{eq:nonreltaubound} we find that the Reynolds number is bounded from below by
\be 
Re > \frac{W}{v_F \kappa}\left( \frac{W}{\lambda_G}\right)^2. 
\ee 

Second, when $W > \lambda_G$, the fluid motion signals the onset of an Ohmic flow although the no-slip conditions at the boundary are still satisfied. Qualitatively speaking, instead of a Poiseuille-type velocity profile, one should expect a plateau around the middle of the channel. We can confirm this behavior from Fig.\ref{fig:poistau}, wherein we plotted $\beta_x$ at different $\lambda_G$ (mind that $\eta$, the temperature and the chemical potential were fixed, which fixes $\kappa$). We expect the fluid velocity not to be large, $v_{\rm max}\ll v_F$, in the $W>\lambda_G$ regime. The condition $W> \lambda_G$ is translated then into an upper bound for the impurity scattering rate 
\be \label{eq:tauimp2}
\tau_{\rm imp} < \frac{W^2}{v_F^2 \kappa}.
\ee
The (non-relativistic) Reynolds number in this case is given by 
\be \label{eq:reynoldsnrel}
Re_{(W > \lambda_G )} = \frac{2}{3}\frac{v_F \tau_{\rm imp}}{W}\left( \frac{W}{\lambda_G}\right)^2\frac{v_{\rm max}}{v_F}.
\ee 
Inserting \Eq{eq:tauimp2} and \Eq{eq:lambdag} into \Eq{eq:reynoldsnrel}, we see that the Reynolds number is bounded from above. The bound is given by
\be 
Re < \frac{2}{3}\frac{W}{v_F\kappa}\left( \frac{W}{\lambda_G}\right)^2\frac{v_{\rm max}}{v_F}.
\ee
It is also worthwhile to mention that 
\be 
\label{eq:reycomp}
Re_{(W<\lambda_G)} > Re_{(W> \lambda_G)},
\ee 

since $\tau_{\rm imp}$ when $W< \lambda_G$ needs to be large for consistency. Based on this last result \Eq{eq:reycomp}, we can conclude that the onset of turbulence is more likely to appear in clean samples. 

\section{Discussion and outlook}\label{sec:disc_outlook}

In this work, we considered the motion of relativistic strongly coupled electron
fluids propagating along a channel under an applied electric
field. Due to the incompressibility of the flow as well as the
expected near-conformality of relativistic electron systems such as
graphene \cite{novoselov2005two} or HgTe \cite{konig2007quantum},
transport is characterized by a single transport coefficient, the
ratio of shear viscosity to entropy density $\eta/s$. We analyzed the qualitative dependence of the fluid flow on $\eta/s$ in relativistic clean systems in sec.~\ref{sec:flowdVdI}, and the non-relativistic  case with momentum relaxation in sec.~\ref{sec:impurities}. In a phenomenological approach, we varied $\eta/s$ from its strong coupling value \eqref{etasAdSCFT} predicted by AdS/CFT  towards the intermediate coupling regime in which $\eta/s$ is larger \cite{Muller:2009cy}. We generically find that keeping the other input parameters such as e.g. electric field and impurities fixed, strongly coupled holographic fluids satisfying \eqref{etasAdSCFT} flow fastest. 

 A very important observable for hydrodynamic behavior in such channels, also called high-mobility wires, is the differential resistance. It is sensitive to the ballistic-to-hydrodynamic crossover \cite{Molenkamp:1994kb,Molenkamp:1994ii,deJong:1995bn}. Our hydrodynamic simulations show that the differential resistance has the form expected for a Poiseuille flow, and depends sensitively on the value of $\eta/s$. In particular, we found that the differential resistance becomes minimal for strongly coupled fluids satisfying \eqref{etasAdSCFT}. This is due to the dependence of the current defined in \Eq{eq:intcurrent} on the velocity profile, which, as can be seen from Fig.~\ref{fig:poisetas}, itself depends strongly on $\eta/s$. Thus, keeping all other parameters fixed, we expect the holographic fluids to exhibit the smallest resistance. From the qualitative behavior of the channel resistance $dV/dI$ as a function of $\eta/s$ obtained from our hydrodynamic simulations (c.f.~Fig.~\ref{fig:gurzhi2}) we conclude that if the equilibration length associated with \eqref{taueq} is of the order of the channel width $w$, the position of the Gurzhi maximum indicative of the Knudsen-Poiseuille crossover strongly depends on $\eta/s$. 

Moreover, we calculated the wall momentum-relaxation timescale
$\tau_w$, which describes how fast momentum is lost through the
walls. We found it to be largest for holographic fluids. We interpret
this as follows: The shear viscosity $\eta$ is a measure of the
momentum transfer between adjacent fluid layers. The entropy density
$s$ is a measure of the number of degrees of freedom in each layer, at
a given temperature and chemical potential. Hence, $\eta/s$ can be
interpreted as the rate of momentum transfer between adjacent fluid
layers per effective degree of freedom. This implies that the momentum
transfer between layers is less efficient for small $\eta/s$ and hence
$\tau_w$ is larger in this case.
Note also that $\eta/s$ is related to the relativistic
analogue of the kinematic viscosity by means of the Gibbs-Duhem
relation \eqref{eq:GibbsDuhem}.

Furthermore, we found that boundary-induced momentum relaxation is not
independent of momentum loss through impurities: Decreasing the impurity density
leads to an increased total momentum of the flow, which in turn
decreases wall momentum relaxation (c.f.~\eqref{eq:tauwnonrel}). 
 We derived the exact relationship between wall momentum relaxation
 $\tau_w$ and impurity momentum relaxation $\tau_{\rm imp}$ for
 non-relativistic flows in \eqref{eq:taunonrelw1}. Experimentally verifying the consequences of
 wall momentum relaxation requires to synthesize clean enough samples
 with small $\eta/s$ such
that $\tau_w$ is comparable to $\tau_{\rm imp}$ (c.f.~\eqref{eq:tauwvalues}).

In the nonrelativistic limit we derived a closed form for the differential resistance $\mathcal{R}$, given in \eqref{eq:Rimp}. For all other input parameters held fixed, $\mathcal{R}$ is a monotonically decreasing function of $\tau_{\rm imp}$, as expected. We stress that the expression \eqref{eq:Rimp} for $\mathcal{R}$ is only valid for small values of the current $I$. Therefore, provided that the system behaves hydrodynamically, \Eq{eq:Rimp} gives the maximal resistance of the channel. In the present work, we have assumed that the system is well in the hydrodynamic regime \eqref{hydroregime}. 
In addition, we found that $\mathcal{R}$ satisfies an inverse
Matthiessen's rule, which in the absence of impurities is given by
\eqref{eq:R1}. In the presence of impurities, $\mathcal{R}_\eta^{-1}$ in \eqref{eq:R1} 
is replaced by the second term in \eqref{eq:Rimp}. From \eqref{eq:R1} and its expansion around the clean limit \eqref{eq:Rimplowtau}, quantum critical conductance effects are separable from viscous and
impurity effects in the way how they enter \eqref{eq:Rimp}.\footnote{A measurement of $\sigma_Q$ is more readily performed in a bulk sample where viscous effects are negligible, by comparing samples of different impurity content and extrapolating to the clean limit.} The underlying reason is that in the constitutive relation for
the current \Eq{eq:j}, they enter as two independent terms. This will
also hold for flows with other boundary
conditions.\cite{Kiselev:2018wj} This is reminiscent of situations in
AdS/CFT models of strange metallic physics, where the quantum critical
part and the Drude part of the conductivity also follow an inverse
Matthiessen rule \cite{Gouteraux:2014hca},  or even more nonlinear relations. \cite{Karch:2007pd} 
We emphasize that the expression \eqref{eq:Rimp} for the wire
resistance was derived without any particular assumption about the equation of state or the value of $\eta/s$. It is hence valid at weak coupling as well, as long as the conditions for hydrodynamics \eqref{hydroregime} apply.

Throughout our calculations, we used the  expression \eqref{eq:sigmaQ}
for the quantum critical conductivity $\sigma_Q$. This choice is most
natural at strong coupling for the following reason:  The quantum critical conductivity can be
calculated at weak coupling \cite{Fritz:2008go} to leading order in $\alpha(T)$. The result is similar to \eqref{etasFritzetal} for $\eta/s$, 
\be\label{sigmaQFritz}
\sigma_Q = 0.76 \frac{e^2}{h} \frac{1}{\alpha^2(T)}.
\ee
Following the discussion of $\eta/s$ in sec.~\ref{sec:introconc} and
extrapolating this result to strong coupling, the natural value for
the quantum critical conductivity at strong coupling is  of order 
\be\label{sigmaQFritz1}
\sigma_Q \sim \frac{e^2}{h}\,.
\ee
From  \Eq{conductivityratios} we see that at strong coupling, 
 the quantum critical
resistivity is of the same order as the Drude resistivity for
realistic values of the impurity density.  Hence, the quantum critical
part of the resistivity cannot be neglected,  contrary to the situation at weak coupling.
The normalization of the $F^2$ term in our holographic model \eqref{eq:bulkaction} was chosen to bear out this expectation\cite{Hartnoll:2007ih,Hartnoll:2007ip} to yield \eqref{eq:sigmaQ}.\footnote{In particular in the Dirac limit $\mu/(k_B T) \ll 1$, $\varepsilon+p \approx sT$, and the dependence on thermodynamic variables drops out of \eqref{eq:sigmaQ}.}

In view of the discussion on  the onset of turbulence in
sec.~\ref{sec:turbulence}, it will be highly interesting to perform
fully space-time dependent hydrodynamic simulations in 2+1 dimensional
charged fluids. This is also interesting from the point of view of the
viscosity and impurity induced pressure drop \eqref{eq:nr-pre} found as a 
solution to the Hagen-Poiseuille equation \eqref{eq:eqp}. The
classical Hagen-Poiseuille law relates the pressure drop along a
laminar flow in a pipe, in our case the wire, to the viscosity of the
flowing fluid. This effect is present as long as the viscosity is
non-vanishing, independently of the presence of impurities. If we allow for $x$-dependence in our hydrodynamic simulations, we will recover that well-known effect, as well as \eqref{eq:nr-pre}. In any fluid, pressure gradients translate into gradients in energy density via the equation of state $\epsilon(p)$, and hence into a spatially dependent temperature. It will be very interesting to investigate this and other space-time dependent effects further.
 
Another interesting question is the physical significance of the
relative ordering of the channel with  $W$ versus the impurity mean
free path $\ell_{\rm imp}$ in \eqref{hydroregime}. The relevant
length scale to compare $W$ to is the Gurzhi length $\lambda_G = v_F
\sqrt{\kappa \tau_{\rm imp}}$, with $\kappa$ the
kinematic viscosity. The Gurzhi length is the scale on which the
viscous drag in the channel is efficient. \cite{Lucas:2017wa} In the
Poiseuille regime $W \leq \lambda$, the viscous drag from the
boundaries permeates the entire channel. In the opposite regime $W
\geq \lambda$, the viscous drag is important up to a distance of order
$\lambda_G$ away from the walls, but in the center of the channel, the
physics is effectively 2+1-dimensional, and the flow is
Ohmic.\footnote{In fact, the center flow can be understood as an effective channel flow with
  no-stress boundary conditions. These conditions are imposed at the edge of a layer
  of width $\lambda_G$. These boundary conditions then  lead to Ohmic
  hydrodynamic flow at the center of the channel
  \cite{Lucas:2017wa}.} The latter point is noteworthy as it represents an Ohmic electric response in the regime of applicability of hydrodynamics \eqref{hydroregime}. 
  
The results presented in this work assume the validity of
hydrodynamics \eqref{hydroregime} at all electron temperatures and
input parameters.   The hydrodynamic regime is reached if 
the electron-electron mean free path $\ell_{\rm ee}$ is the smallest
length scale present. This is expected to be the case in particular in
systems with strong electron-electron correlations. Realizing such
strongly correlated electron systems will hence be an important avenue
for further development.
The above considerations about the applicability of hydrodynamics may not however apply at intermediate coupling strengths. There, new non-hydrodynamic modes may appear, invalidating the hydrodynamic expansion, as was explained in footnote 48. One possible approach for investigating whether hydrodynamics remains applicable at intermediate couplings is to study gravitational duals at finite charge density that include coupling corrections.\cite{AnninosPastras}\cite{SachdevKrempaPRB.86.235115}\cite{SachdevKrempaPRB.87.155149}\cite{Grozdanov2017}\cite{GROSS198741}
 
 Finally, it will be interesting to investigate the effect of parity
 and time reversal
 breaking\cite{Jensen:2011xb,Hoyos:2014pba,Pellegrino:2017fd,Delacretaz:2017yia}
 in the setup considered, as well as thermal and thermoelectric transport.

\section*{Acknowledgments}

We thank Hartmut Buhmann, Ewelina Hankiewicz, Carlos Hoyos, Oleksiy
Kashuba, Laurens Molenkamp, Valentin M\"uller and Bj\"orn Trauzettel
for useful discussions.  We gratefully acknowledge support from the
DFG via SFB 1170 `Topological and Correlated Electronics at Surfaces
and Interfaces'. We also thank the referees for constructive comments.

\appendix
	
\section{Holographic set-up}
\label{app:holoren}


Here we review some aspects of the AdS/CFT correspondence relevant to our analysis and fix the overall coefficient of the gravity action.
Since we are interested in planar 2+1 dimensional systems, we will work in 3+1 bulk dimensions, i.e. consider a gravity model with an AdS$_{4}$ ground state. 

The simplest holographic set-up which accommodates matter at finite charge density and temperature is the field theory dual to Einstein-Maxwell theory,
\be 
S = \frac{1}{16\pi G_4} \int \, d^4 x \sqrt{-g}\left[R -2\Lambda - L^2 F_{\mu\nu} F^{\mu\nu}\right]\,.\label{eq:bulkaction}
\ee
Here $R$ is the Ricci curvature scalar, $\Lambda = -\frac{3}{L^2}$ is the negative cosmological constant necessary to generate an Anti de Sitter space-time of curvature radius $L$ as the ground state of \eqref{eq:bulkaction}, and $F_{\mu\nu}=\partial_\mu A_\nu - \partial_\nu A_\mu$ the field strength for the $U(1)$ gauge connection $A_\mu$. Turning on a dynamical $U(1)$ gauge field allows for a QFT at finite density. In particular, we consider
\be 
A_\mu dx^\mu = A_0 dt.
\ee 
The chemical potential is defined as the boundary value of the time component of the bulk gauge $U(1)$ field,
\be 
\mu = A_0(r) \big \vert_{r\to \infty}.
\ee 

In order to describe a QFT at non-zero temperature, we need to consider a black brane\footnote{A black brane is a black hole with planar horizon topology, in this case $\mathbbm{R}^2$.} in the gravity dual. The reason is that the Hawking temperature $T_H$ of the black brane is identified with the temperature of the QFT, $T=T_H$. To this purpose, we consider the line element of the AdS-Reissner-Nordstr\"om black brane, given by
\be \label{eq:adsrnmetric}
ds^2 = \frac{L^2}{r^2 f(r)}dr^2 + \frac{r^2}{L^2}\left[ -f(r) v_F^2 dt^2 + dx^2 \right]\,.
\ee
The black brane encloses a charge density inside the horizon, sourcing a Coulomb potential $A_N dx^N = A_0(r) dx^0$ varying in the holographic direction $r$. 
The emblackening factor $f(r)$ and Coulomb potential $A_0(r)$ are
\bea \nonumber
f(r) &=& 1-\frac{r_H^3}{r^3} - \frac{1}{4}\frac{\mu^2 L^4}{\hbar^2 v_F^2} \frac{r_H}{r^3}\left[1-\frac{r_H}{r} \right]\,,\\\label{eq:funcsads}
A_0 &=& \frac{\mu}{\hbar v_F} \left( 1- \frac{r_H}{r}\right) .
\eea 
The Hawking temperature is obtained from the surface gravity $\kappa_H$, defined as
\be 
2\pi \frac{v_F \, k_B}{\hbar} T_H = \kappa_H =  \sqrt{\Big \vert \left(\nabla_A \zeta_B\right)\left(\nabla^A \zeta^B\right)\Big\vert } \Bigg\vert_{r=r_H}.
\ee
For a static black brane, one Killing vector is simply $\zeta = \left(\zeta^r,\zeta^t,\zeta^x,\zeta^y\right) = (0,1,0,0)$. $\nabla_a$ stands for the covariant derivative, defined from the bulk metric \Eq{eq:adsrnmetric}. With this,
\be 
T_H = \frac{\hbar\, v_F}{k_B} \frac{r_H}{4\pi L^2} f'(r_H) e^{A(r_H)},
\ee 
which, for the functions defined in \Eq{eq:funcsads}, gives
\be 
T_H = \frac{1}{\hbar\, v_F k_B}\frac{\mu ^2 L^2}{16 \pi  r_H}-\frac{\hbar v_F}{k_B}\frac{3 r_H}{4 \pi  L^2}.
\ee 

Notice that due to the horizon, the radial coordinate spans from the horizon position $r_H$ to the boundary, that is $r\in [r_H,\infty]$. The metric \Eq{eq:adsrnmetric} asymptotes to an $\rm{AdS}_4$ space of radius $L$ for $r\to \infty$. Since our aim is to describe a hydrodynamic fluid of relativistic electrons, the Fermi velocity $v_F$ replaces the vacuum speed of light $c$. The position of the event horizon $r_H$ is fixed in terms of the chemical potential and the Hawking temperature,
\be 
\frac{r_H}{L^2} = \frac{1}{6} \frac{k_B T }{\hbar \, v_F} \left(4 \pi+ \sqrt{16 \pi ^2 + \frac{3\mu ^2}{k_B^2 T^2} }  \right).
\ee 

The entropy density of the QFT is identified with the Bekenstein-Hawking entropy density, which is given by
\be \label{eq:beken}
s = \frac{\mathcal{A}_H\, k_B}{4 l_*^2}, \qquad l_*^2 = \frac{\hbar G_4}{v_F^3}.
\ee 
The area density, $\mathcal{A}_H$, is
\be 
\mathcal{A}_H = \sqrt{g_{ xx} g_{ yy}}\big\vert_{r\to r_H} .
\ee 

\noindent For the metric element \Eq{eq:adsrnmetric}, we get
\be 
\mathcal{A}_H = \frac{r_H^2}{L^2}.
\ee

A finite density in the dual QFT is obtained by turning on a dynamical $U(1)$ gauge field,
\be 
A_\mu dx^\mu = A_0 dt.
\ee 

There is a non-trivial relation between the bulk fields $f$ and $A_0$ and the chemical potential and temperature. The chemical potential is defined as the boundary value of the time component of the bulk gauge $U(1)$ field,
\be 
\mu = A_0(r) \big \vert_{r\to \infty}.
\ee 

Having identified the temperature and chemical potential, we now proceed to determine the thermodynamical variables that appear in \eqref{eq:thermoads}, with the exception of the entropy density, already defined in \eqref{eq:beken}. The holographic duality establishes that the partition function of the QFT and of the gravity theory are identified. Taking variations of the on-shell gravity action \eqref{eq:bulkaction} with respect to the metric or the gauge field will give the corresponding $n$-point correlation functions. The on-shell action is divergent and we need to renormalize it. This constitutes the holographic renormalization prescription \cite{deHaro:2000vlm,Papadimitriou:2004rz,Papadimitriou:2005ii}. We introduce a cutoff in the radial direction $r_\Lambda$ such that the divergences of the bulk action \Eq{eq:bulkaction} are regulated. The volume divergence is canceled by adding the counterterm 
\be
S_{c.t.}  =  - \frac{1}{16\pi G_4}\int_{r=r_\Lambda} d^3 x\,\cL_{c.t.} , \quad \cL_{c.t.} = \sqrt{-\gamma} \frac{6}{L} \ .
\ee
Further divergences, subleading with respect to the volume divergence are remove if in addition to $S_{c.t.}$, one adds the Gibbons-Hawking term 
\be
S_{GH}  = \frac{1}{8\pi G_4}\int_{r=r_\Lambda} d^3 x \sqrt{-\gamma} K.
\ee 
The on-shell action together with the counterterms $S_{c.t.}$ and $S_{GH}$ guarantee a well-defined variational principle. We assume that the form of the metric is
\begin{equation}
ds^2=N^2 dr^2+\gamma_{\mu\nu}dx^\mu dx^\nu \ ,
\end{equation}
and define the extrinsic curvature and Brown-York tensors in the usual way
\bea
\quad K_{\m\n} = \frac{1}{2N}\partial_r \gamma_{\m\n} \ , \ K = \gamma^{\m\n} K_{\m\n}\ , \nonumber\\
\ \Pi^{\m\n}_{BY} = \left( K^{\mu\nu}-\gamma^{\mu\nu}K\right) \ .
\eea

In particular, there is no need for adding a counterterm for the gauge field. The renormalized action is then
\begin{equation}
S_{ren}=\lim_{r_\Lambda\to\infty}\left[S_{bulk}+S_{GH}+S_{\rm c.t.}\right]\ .
\end{equation}

Renormalized expectation values of the stress tensor and current are computed from variations of the action with respect to the metric $g_{\mu\nu}$ and gauge field $A_\mu$,
\bea
\vev{T^{\m\n}} &=& \frac{1}{8\pi G_4} \nonumber\\
& \times & \lim_{r_\Lambda \to \infty}\frac{r^2}{L^2}\left[- \sqrt{-\gamma}  \Pi_{BY}^{\m\n} +\frac{\delta \cL_{c.t.} }{\delta \gamma_{\mu\nu}}\right] \Bigg\vert_{r\to r_\Lambda} \,, \label{eq:temren}\nonumber\\
\vev{j^\mu} &=& \frac{1}{16\pi G_4}  \nonumber\\
& \times & \lim_{r_\Lambda \to \infty}   \left[-4L^2\sqrt{-g} g^{rr} g^{\mu \alpha} F_{r\alpha} \right]\big\vert_{r\to r_\Lambda} \ .\label{eq:jren}
\eea 

The stress tensor reads in components $\vev{T^{\mu\nu}} = \text{diag}(\varepsilon,p,p)$ while the current density $\vev{\j^{\mu}} = \left( \rho, 0,0\right)$. Identifying this with \eqref{eq:temren} and \eqref{eq:jren} yields the energy density, pressure and number density as functions of the chemical potential and temperature, as they appear in \Eq{eq:thermoads}.\\

The dimensionless ratio of the AdS$_4$ radius $L$ and the gravity coupling $G_4$ in \Eq{eq:thermoads} is related to the rank of the gauge group $N$ in the dual quantum field theory,\cite{Erdmenger:2016wyp}
\be \label{eq:N}
\frac{L^2}{G_4}  = \frac{1}{3} \left(2 N\right)^{3/2}.
\ee

Evaluating observables such as the channel resistance and ultimately comparing with other theoretical approaches as well as with experiment, it is necessary to fix the rank $N$. To do so, we adopt the following phenomenological procedure. We match the value of the charge density of our charged black brane \eqref{eq:TDvariables} to a typical charge density found in condensed matter experiments.\cite{Molenkamp:1994ii,Molenkamp:1994kb,deJong:1995bn} These  experiments are performed in the limit of the chemical potential being much larger than the temperature, $\mu\gg k_B T$, which is also the regime we work in.\footnote{The renormalization of the Fermi velocity $v_F$ is negligible at most temperatures except exponentially small ones \protect\cite{Lucas:2017wa}.}.\footnote{We in particular do
	not claim that Einstein-Maxwell holography
	\protect\eqref{eq:bulkaction} can describe a weakly interacting
	Fermi liquid regime such as the one present in e.g. graphene at
	$\mu/T\gg 1$.}

Moreover, we remark that the particular IR fixed point of \eqref{eq:adsrnmetric}, $AdS_2\times \mathbbm{R}^2$ quantum  criticality, does not influence the dependence of our hydrodynamic solutions on $\eta/s$: Besides the simple Einstein-Maxwell theory \eqref{eq:bulkaction}, other holographic models with different IR fixed points such as Schr\"odinger symmetry\cite{Balasubramanian:2008dm} or  hyperscaling violating Lifshitz scaling\cite{Charmousis:2010zz} exist. These models describe strongly coupled fluids with very different IR symmetries, some closer to that of the Fermi liquid, but all have a hydrodynamic description with \eqref{etasAdSCFT} in the strong coupling limit. The dependence of hydrodynamic observables such as the channel resistance on $\eta/s$ hence will also be the same in these models. Finally, since relativistic hydrodynamics captures the universal finite temperature dynamics of any holographic dual, our matching procedure also applies to top-down string theoretic constructions.\cite{Maldacena:1997re,Aharony:2008ug}

Since adequate strong-coupling expressions are not available,
for simplicity we assume this charge density to be given by the Fermi-Dirac distribution,
\bea \label{eq:FermiDirac}\nonumber
& \rho_{\rm Dirac} = n_{e^-} -n_{h^+} & \\
= & \frac{g}{(2\pi)^2}\int d^2 p \left[ \frac{1}{e^{\left(E-\mu\right)/k_B T} +1} - \frac{1}{e^{E/k_B T} +1} \right],&
\eea
where $g$ stands for the degeneracy. For massless fermions, $E= v_F \vert p\vert = \hbar v_F \vert k\vert$. Integrating \eqref{eq:FermiDirac} yields 
\be \label{eq:diracrho}
\rho_{\rm Dirac} = -g\left(\frac{k_B T}{\hbar \, v_F}\right)^2 \left[ \frac{1}{2\pi} {\rm Li}_2\left(-e^{\mu/k_B T}\right) + \frac{\pi  }{24}\right] ,
\ee 
with ${\rm Li}_s(z) = \sum^{\infty}_{n=1}\frac{z^n}{n^s}$. At leading order in $\mu/k_B T\gg 1$ we obtain
\be \label{eq:rhodirac}
\rho_{\rm Dirac} \Big\vert_{\mu \gg k_B T} = \frac{g}{4\pi}\frac{\mu^2}{\hbar^2 v_F^2}.
\ee
In the same limit, the number density \Eq{eq:thermoads} becomes 
\be \label{eq:rhoads}
\rho_{\rm AdS}\Big\vert_{\mu \gg k_B T} =  \frac{1}{128\pi \sqrt{3}}\frac{L^2}{G_4}\frac{\mu^2}{\hbar^2 v_F^2}.
\ee 
Identifying \Eq{eq:rhodirac} with \Eq{eq:rhoads} in the limit $\mu/k_B
T\gg 1$, the ratio \eqref{eq:N} is fixed to
\be \label{eq:coupval}
\frac{L^2}{G_4} = 32\sqrt{3} g.
\ee

\begin{figure}[tbp]
	\begin{center}
		\includegraphics[scale=0.23]{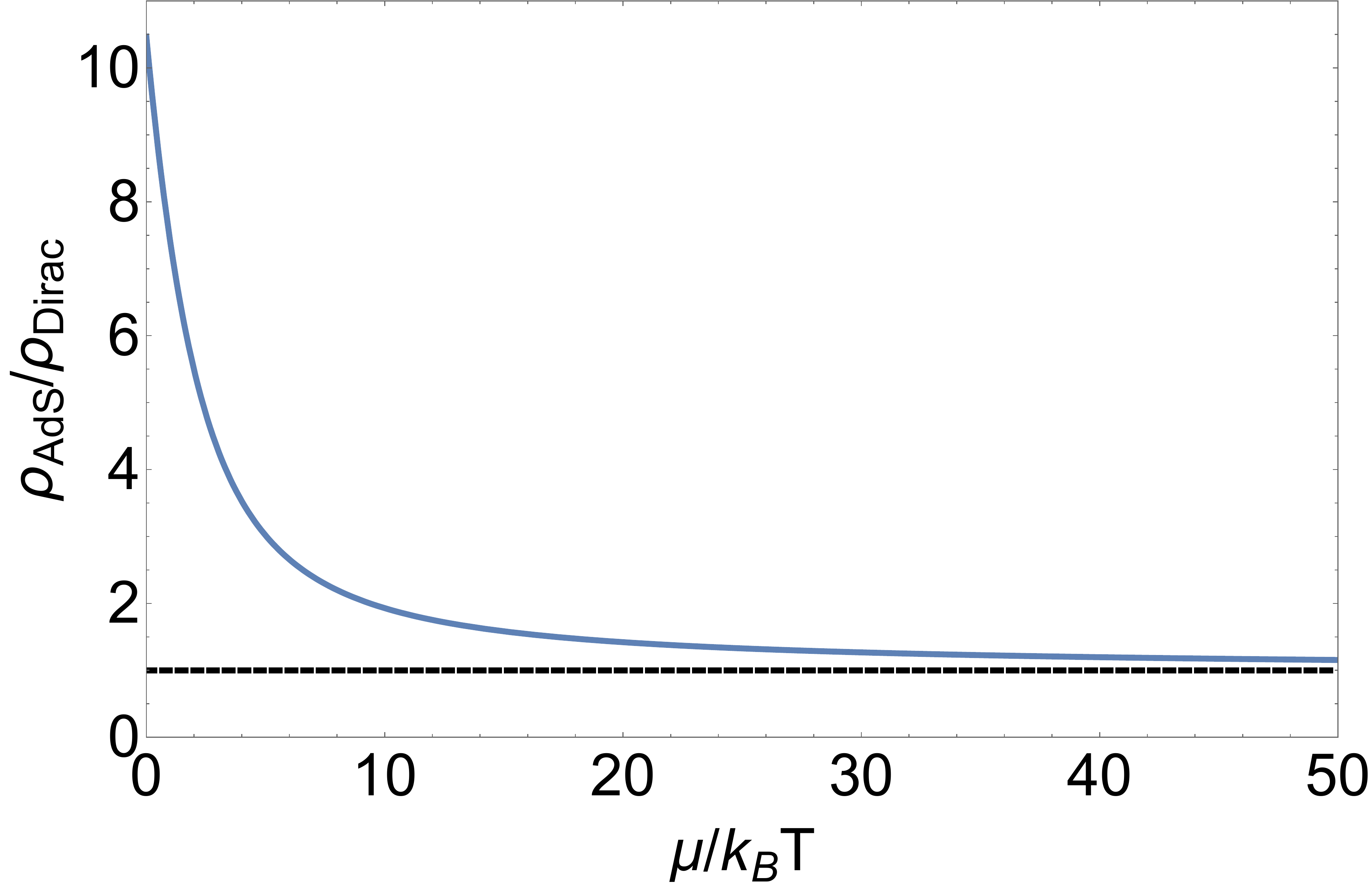}
	\end{center}
	\caption{Ratio of the Dirac and holographic charge densities $\rho_{\rm AdS}/\rho_{\rm Dirac}$ as a function of $\mu/k_B T$. The black dashed line denotes $\rho_{\rm AdS}/\rho_{\rm Dirac} = 1$.
		\label{fig:ratio}} 
	
\end{figure}

For $g=2$, the phenomenological approach described yields $N\simeq 24$. We can therefore state that the approach we chose here of matching charge densities is self-consistent as it leads to a large value of $N\gg 1$, which is large enough to also suppress quantum gravitational $1/N^2$ corrections. We will use this value of $N$ for numerical estimates. Figure \ref{fig:ratio} shows the ratio between the charge densities from the Fermi-Dirac distribution and from holography. The difference is small for $\mu/k_B T > 10$, and not too large for values closer to the charge neutrality point. This is due to the coinciding $\mu/k_B T$ dependence of both densities  for $\mu> k_B T$. The dependence on $\mu/(k_B T)$ also drops out in this ratio for $\mu/(k_B T)\ll 1$. We hence could have matched the two distributions at the charge neutrality point as well, obtaining a slightly smaller value of $N\simeq 5$.

Phenomenological extrapolations from strongly coupled models have been widely employed in QCD.\cite{Hoyos:2016zke,CasalderreySolana:2011us} Experience with these models of QCD tells that $N=3$ is already large enough to make quantitative statements at the $10\%$ level,\cite{CasalderreySolana:2011us} even at 't Hooft coupling constants that are not  far from $1$.  

Finally, we comment on the quantum critical conductivity \eqref{eq:sigmaQ} from AdS/CFT models such as \eqref{eq:bulkaction}. One reason for the naturalness of $\sigma_Q = e^2/h$ is the electromagnetic self-duality of the holographic action \eqref{eq:bulkaction} if the Maxwell term is normalized canonically to $-\frac{1}{4} F_{\mu\nu}^2$.\cite{Herzog:2007ij} This emergent discrete symmetry fixes $\sigma_Q = e^2/h$ in the Dirac regime $\mu/(k_B T) \ll 1$. Finite coupling corrections which break the self-duality can be calculated from universal higher derivative terms on the gravity side of the AdS/CFT correspondence.\cite{Myers:2010pk} In summary, there is ample evidence that $\sigma_Q = e^2/h$ is the strong coupling limit of the perturbative quantum critical conductance $\sigma_Q = e^2/h$, which motivates our choice of normalization of the $F^2$ term in \eqref{eq:bulkaction}.

\section{Reynolds number for relativistic hydrodynamics}
\label{app:reynolds}

In the present appendix we derive the Reynolds number, $Re$, for relativistic flows. In non-relativistic hydrodynamics, $Re$ appears in the dimensionless\footnote{For example, $\vec{v} = \vec{u}/v_c$ where $\vec{u}, v_c$ is the fluid and characteristic velocity, respectively.} Navier-Stokes equations as

\be
\label{1}
{D\vec{v} \over Dt} = -\nabla P + {1 \over Re_{NS}} \triangledown^2 \vec{v}
\ee

\noindent where $ {D \over Dt} \equiv \pa_t + \vec{v}\cdot\nabla$ is the convective derivative.  \\

A similar definition for $Re$ does not exist for relativistic hydrodynamics; we amend this below.

Our first derivation of $Re$ mimics the derivation of $Re_{NS}$; we'll bring the relativistic equations of motion (EOM), $\partial_\mu T^{\mu\nu} = 0$, into a form similar to the Navier-Stokes equations, and read off $Re$. Thus, we have\footnote{Without loss of generality, we have set the external fields to zero and taken  $\pa_\m \e = 0 =\pa_\m u^\m $.}

\be
\label{4}
\begin{split}
{1 \over v^2_F}{ D\vec{u} \over D\tau} &= - \left( {\vec{u} \over v^2_F} {D \over D\tau} + \nabla\right)\ln (\varepsilon + p) - {\h \over \varepsilon + p }\vec{\Sigma} 
\\
&\vec{\Sigma}^i \equiv \pa_\m \s^{\m i} \spa i=1,2
\spa
{D \over D\tau} \equiv u^\m \pa_\m
\end{split}
\ee

\noindent Equation \refeq{4} is clearly the relativistic counterpart of the Navier-Stokes equations, \refeq{1}. We bring \refeq{4} into a dimensionless form by applying the transformation $\vec{u} \rightarrow {\vec{u}/ v_{\rm max}}  \spa x^\mu \rightarrow {x^\m / W}$, where $v_{\rm max}\equiv |\vec{u}|(y=W/2), W$ are the characteristic velocity and length scales of our flows (see Figure \ref{fig:pois}.). The result is

\begin{align}
\label{7}
{D\vec{u} \over D\tau} &= - \left[ \vec{u}  {D \over D\tau} +\left({v_F \over v_{\rm max}}\right)^2 \nabla\right]\ln (\varepsilon + p) \nonumber\\
 &- \left({v_F \over v_{\rm max}}\right)^2{\h v_{\rm max} \over W(\varepsilon + p) }\vec{\tilde{\Sigma}}, 
\end{align}
\noindent where $\vec{\tilde{\Sigma}}$ is the rescaled version of $\vec{\Sigma}$.

Now we can use \eqref{7} in order to read off $Re$. Unfortunately, there is not a single dimensionless number multiplying the second term in the r.h.s. of \eqref{7}, because of $\vec{\tilde{\Sigma}}$. 
\footnote{
$\vec{\tilde{\Sigma}}$ after rescaling contains terms proportional to $\left({v_{\rm max} \over v_F}\right)^a \spa a=0,2,4$.
}
Thus there is no natural definition of $Re$ in the relativistic case. We can bypass this problem by considering the non-relativistic (NR) and the ultra-relativistic (UR) limits of \eqref{7}.

\begin{itemize}
\item \textit{non-relativistic limit:} In this case, $v_F /v_{\rm max} \gg 1$, and the non-relativistic $Re$ is 

\be
\label{8}
Re_{NR} = {(\varepsilon + p )W \over \h v_{\rm max}} \left({v_{\rm max} \over v_F}\right)^2 .
\ee

$Re_{NR}$ is proportional to the Navier-Stokes $Re$. We see this if we substitute $\varepsilon = \rho v^2_F$ and $p={1 \over 2} \e$ \footnote{Recall that the fluid in particular is conformal. This is in contrast to common non-relativistic fluids, for which $p\ll\varepsilon$. In this case, $Re_{NR} = Re_{NS}$.}, where $\r$ is the fluid mass density, into \eqref{8} i.e.

\be
\label{8'} Re_{NR} = {3 \over 2}Re_{NS}= {3 \over 2}{\r v_{\rm max} W \over \h} .
\ee

\item \textit{ultra-relativistic limit:} In this case, we keep the terms proportional to $({v_{\rm max} / v_F})^4$ in $\vec{\tilde{\Sigma}}$ and obtain

\be
\label{9'} 
Re_{UR} = {(\varepsilon + p )W \over \h v_{\rm max}} \left({v_F \over v_{\rm max}}\right)^2 . 
\ee

\end{itemize}

Our second derivation of $Re$ is based on the physical meaning of $Re$, which is defined as the ratio of inertial forces to the ratio of viscous forces. As usual, the inertial and viscous forces are given by

\be
\label{11} F^i_{in}= \pa_t T^{0i}  \spa F^i_{v} = -\h \pa_j\sigma^{ji} \spa i=1,2
\ee

Therefore $Re$ can be defined through the ratio $\vec{F}_{in}/ \vec{F}_{v}$. Computing this ratio leads us, as expected, to equations \eqref{8'} and \eqref{9'}.

Equations \eqref{8'} and \eqref{9'} above summarize our derivation of $Re$ for non-relativistic and ultra-relativistic fluids.

\bibliography{biblio2}

\begin{thebibliography}{120}%
\makeatletter
\providecommand \@ifxundefined [1]{%
 \@ifx{#1\undefined}
}%
\providecommand \@ifnum [1]{%
 \ifnum #1\expandafter \@firstoftwo
 \else \expandafter \@secondoftwo
 \fi
}%
\providecommand \@ifx [1]{%
 \ifx #1\expandafter \@firstoftwo
 \else \expandafter \@secondoftwo
 \fi
}%
\providecommand \natexlab [1]{#1}%
\providecommand \enquote  [1]{``#1''}%
\providecommand \bibnamefont  [1]{#1}%
\providecommand \bibfnamefont [1]{#1}%
\providecommand \citenamefont [1]{#1}%
\providecommand \href@noop [0]{\@secondoftwo}%
\providecommand \href [0]{\begingroup \@sanitize@url \@href}%
\providecommand \@href[1]{\@@startlink{#1}\@@href}%
\providecommand \@@href[1]{\endgroup#1\@@endlink}%
\providecommand \@sanitize@url [0]{\catcode `\\12\catcode `\$12\catcode
  `\&12\catcode `\#12\catcode `\^12\catcode `\_12\catcode `\%12\relax}%
\providecommand \@@startlink[1]{}%
\providecommand \@@endlink[0]{}%
\providecommand \url  [0]{\begingroup\@sanitize@url \@url }%
\providecommand \@url [1]{\endgroup\@href {#1}{\urlprefix }}%
\providecommand \urlprefix  [0]{URL }%
\providecommand \Eprint [0]{\href }%
\providecommand \doibase [0]{http://dx.doi.org/}%
\providecommand \selectlanguage [0]{\@gobble}%
\providecommand \bibinfo  [0]{\@secondoftwo}%
\providecommand \bibfield  [0]{\@secondoftwo}%
\providecommand \translation [1]{[#1]}%
\providecommand \BibitemOpen [0]{}%
\providecommand \bibitemStop [0]{}%
\providecommand \bibitemNoStop [0]{.\EOS\space}%
\providecommand \EOS [0]{\spacefactor3000\relax}%
\providecommand \BibitemShut  [1]{\csname bibitem#1\endcsname}%
\let\auto@bib@innerbib\@empty
\bibitem [{\citenamefont {Narozhny}\ \emph {et~al.}(2017)\citenamefont
  {Narozhny}, \citenamefont {Gornyi}, \citenamefont {Mirlin},\ and\
  \citenamefont {Schmalian}}]{Narozhny:2017vc}%
  \BibitemOpen
  \bibfield  {author} {\bibinfo {author} {\bibfnamefont {B.~N.}\ \bibnamefont
  {Narozhny}}, \bibinfo {author} {\bibfnamefont {I.~V.}\ \bibnamefont
  {Gornyi}}, \bibinfo {author} {\bibfnamefont {A.~D.}\ \bibnamefont {Mirlin}},
  \ and\ \bibinfo {author} {\bibfnamefont {J.}~\bibnamefont {Schmalian}},\
  }\href@noop {} {\  (\bibinfo {year} {2017})},\ \Eprint
  {http://arxiv.org/abs/1704.03494} {1704.03494} \BibitemShut {NoStop}%
\bibitem [{\citenamefont {Lucas}\ and\ \citenamefont
  {Fong}(2017)}]{Lucas:2017wa}%
  \BibitemOpen
  \bibfield  {author} {\bibinfo {author} {\bibfnamefont {A.}~\bibnamefont
  {Lucas}}\ and\ \bibinfo {author} {\bibfnamefont {K.~C.}\ \bibnamefont
  {Fong}},\ }\href@noop {} {\  (\bibinfo {year} {2017})},\ \Eprint
  {http://arxiv.org/abs/1710.08425} {1710.08425} \BibitemShut {NoStop}%
\bibitem [{\citenamefont {Molenkamp}\ and\ \citenamefont
  {de~Jong}(1994{\natexlab{a}})}]{Molenkamp:1994kb}%
  \BibitemOpen
  \bibfield  {author} {\bibinfo {author} {\bibfnamefont {L.~W.}\ \bibnamefont
  {Molenkamp}}\ and\ \bibinfo {author} {\bibfnamefont {M.~J.~M.}\ \bibnamefont
  {de~Jong}},\ }\href@noop {} {\bibfield  {journal} {\bibinfo  {journal}
  {Solid-State Electronics}\ }\textbf {\bibinfo {volume} {37}},\ \bibinfo
  {pages} {551} (\bibinfo {year} {1994}{\natexlab{a}})}\BibitemShut {NoStop}%
\bibitem [{\citenamefont {Molenkamp}\ and\ \citenamefont
  {de~Jong}(1994{\natexlab{b}})}]{Molenkamp:1994ii}%
  \BibitemOpen
  \bibfield  {author} {\bibinfo {author} {\bibfnamefont {L.~W.}\ \bibnamefont
  {Molenkamp}}\ and\ \bibinfo {author} {\bibfnamefont {M.~J.~M.}\ \bibnamefont
  {de~Jong}},\ }\href@noop {} {\bibfield  {journal} {\bibinfo  {journal}
  {Physical Review B}\ }\textbf {\bibinfo {volume} {49}},\ \bibinfo {pages}
  {5038} (\bibinfo {year} {1994}{\natexlab{b}})}\BibitemShut {NoStop}%
\bibitem [{\citenamefont {{de Jong}}\ and\ \citenamefont
  {{Molenkamp}}(1995)}]{deJong:1995bn}%
  \BibitemOpen
  \bibfield  {author} {\bibinfo {author} {\bibfnamefont {M.~J.~M.}\
  \bibnamefont {{de Jong}}}\ and\ \bibinfo {author} {\bibfnamefont {L.~W.}\
  \bibnamefont {{Molenkamp}}},\ }\href {\doibase 10.1103/PhysRevB.51.13389}
  {\bibfield  {journal} {\bibinfo  {journal} {Physical Review B}\ }\textbf
  {\bibinfo {volume} {51}},\ \bibinfo {pages} {13389} (\bibinfo {year}
  {1995})},\ \Eprint {http://arxiv.org/abs/cond-mat/9411067} {cond-mat/9411067}
  \BibitemShut {NoStop}%
\bibitem [{\citenamefont {Crossno}\ \emph {et~al.}(2015)\citenamefont
  {Crossno}, \citenamefont {Shi}, \citenamefont {Wang}, \citenamefont {Liu},
  \citenamefont {Harzheim}, \citenamefont {Lucas}, \citenamefont {Sachdev},
  \citenamefont {Kim}, \citenamefont {Taniguchi}, \citenamefont {Watanabe},
  \citenamefont {Ohki},\ and\ \citenamefont {Fong}}]{Crossno:2015iy}%
  \BibitemOpen
  \bibfield  {author} {\bibinfo {author} {\bibfnamefont {J.}~\bibnamefont
  {Crossno}}, \bibinfo {author} {\bibfnamefont {J.~K.}\ \bibnamefont {Shi}},
  \bibinfo {author} {\bibfnamefont {K.}~\bibnamefont {Wang}}, \bibinfo {author}
  {\bibfnamefont {X.}~\bibnamefont {Liu}}, \bibinfo {author} {\bibfnamefont
  {A.}~\bibnamefont {Harzheim}}, \bibinfo {author} {\bibfnamefont
  {A.}~\bibnamefont {Lucas}}, \bibinfo {author} {\bibfnamefont
  {S.}~\bibnamefont {Sachdev}}, \bibinfo {author} {\bibfnamefont
  {P.}~\bibnamefont {Kim}}, \bibinfo {author} {\bibfnamefont {T.}~\bibnamefont
  {Taniguchi}}, \bibinfo {author} {\bibfnamefont {K.}~\bibnamefont {Watanabe}},
  \bibinfo {author} {\bibfnamefont {T.~A.}\ \bibnamefont {Ohki}}, \ and\
  \bibinfo {author} {\bibfnamefont {K.~C.}\ \bibnamefont {Fong}},\ }\href@noop
  {} {\bibfield  {journal} {\bibinfo  {journal} {Science}\ ,\ \bibinfo {pages}
  {1058}} (\bibinfo {year} {2015})},\ \Eprint {http://arxiv.org/abs/1509.04713}
  {1509.04713} \BibitemShut {NoStop}%
\bibitem [{\citenamefont {{Moll}}\ \emph {et~al.}(2016)\citenamefont {{Moll}},
  \citenamefont {{Kushwaha}}, \citenamefont {{Nandi}}, \citenamefont
  {{Schmidt}},\ and\ \citenamefont {{Mackenzie}}}]{Moll:2016ju}%
  \BibitemOpen
  \bibfield  {author} {\bibinfo {author} {\bibfnamefont {P.~J.~W.}\
  \bibnamefont {{Moll}}}, \bibinfo {author} {\bibfnamefont {P.}~\bibnamefont
  {{Kushwaha}}}, \bibinfo {author} {\bibfnamefont {N.}~\bibnamefont {{Nandi}}},
  \bibinfo {author} {\bibfnamefont {B.}~\bibnamefont {{Schmidt}}}, \ and\
  \bibinfo {author} {\bibfnamefont {A.~P.}\ \bibnamefont {{Mackenzie}}},\
  }\href {\doibase 10.1126/science.aac8385} {\bibfield  {journal} {\bibinfo
  {journal} {Science}\ }\textbf {\bibinfo {volume} {351}},\ \bibinfo {pages}
  {1061} (\bibinfo {year} {2016})},\ \Eprint {http://arxiv.org/abs/1509.05691}
  {arXiv:1509.05691 [cond-mat.str-el]} \BibitemShut {NoStop}%
\bibitem [{\citenamefont {Bandurin}\ \emph {et~al.}(2016)\citenamefont
  {Bandurin}, \citenamefont {Torre}, \citenamefont {Kumar}, \citenamefont
  {Shalom}, \citenamefont {Tomadin}, \citenamefont {Principi}, \citenamefont
  {Auton}, \citenamefont {Khestanova}, \citenamefont {Novoselov}, \citenamefont
  {Grigorieva} \emph {et~al.}}]{bandurin2016negative}%
  \BibitemOpen
  \bibfield  {author} {\bibinfo {author} {\bibfnamefont {D.}~\bibnamefont
  {Bandurin}}, \bibinfo {author} {\bibfnamefont {I.}~\bibnamefont {Torre}},
  \bibinfo {author} {\bibfnamefont {R.~K.}\ \bibnamefont {Kumar}}, \bibinfo
  {author} {\bibfnamefont {M.~B.}\ \bibnamefont {Shalom}}, \bibinfo {author}
  {\bibfnamefont {A.}~\bibnamefont {Tomadin}}, \bibinfo {author} {\bibfnamefont
  {A.}~\bibnamefont {Principi}}, \bibinfo {author} {\bibfnamefont
  {G.}~\bibnamefont {Auton}}, \bibinfo {author} {\bibfnamefont
  {E.}~\bibnamefont {Khestanova}}, \bibinfo {author} {\bibfnamefont
  {K.}~\bibnamefont {Novoselov}}, \bibinfo {author} {\bibfnamefont
  {I.}~\bibnamefont {Grigorieva}},  \emph {et~al.},\ }\href@noop {} {\bibfield
  {journal} {\bibinfo  {journal} {Science}\ }\textbf {\bibinfo {volume}
  {351}},\ \bibinfo {pages} {1055} (\bibinfo {year} {2016})}\BibitemShut
  {NoStop}%
\bibitem [{\citenamefont {Nam}\ \emph {et~al.}(2017)\citenamefont {Nam},
  \citenamefont {Ki}, \citenamefont {Soler-Delgado},\ and\ \citenamefont
  {Morpurgo}}]{nam2017electron}%
  \BibitemOpen
  \bibfield  {author} {\bibinfo {author} {\bibfnamefont {Y.}~\bibnamefont
  {Nam}}, \bibinfo {author} {\bibfnamefont {D.-K.}\ \bibnamefont {Ki}},
  \bibinfo {author} {\bibfnamefont {D.}~\bibnamefont {Soler-Delgado}}, \ and\
  \bibinfo {author} {\bibfnamefont {A.~F.}\ \bibnamefont {Morpurgo}},\
  }\href@noop {} {\bibfield  {journal} {\bibinfo  {journal} {Nature Physics}\
  }\textbf {\bibinfo {volume} {13}},\ \bibinfo {pages} {1207} (\bibinfo {year}
  {2017})}\BibitemShut {NoStop}%
\bibitem [{\citenamefont {Gurzhi}(1968)}]{Gurzhi:1968ji}%
  \BibitemOpen
  \bibfield  {author} {\bibinfo {author} {\bibfnamefont {R.~N.}\ \bibnamefont
  {Gurzhi}},\ }\href@noop {} {\bibfield  {journal} {\bibinfo  {journal} {Soviet
  Physics Uspekhi}\ }\textbf {\bibinfo {volume} {11}},\ \bibinfo {pages} {255}
  (\bibinfo {year} {1968})}\BibitemShut {NoStop}%
\bibitem [{Note1()}]{Note1}%
  \BibitemOpen
  \bibinfo {note} {In order to meaningfully extract the hydrodynamic shear
  viscosity, the conditions for hydrodynamics need to apply as a low-energy
  effective theory. These conditions are e.g. spelled out in sec.~\ref {sec21}.
  Otherwise the result for the viscosity from e.g. the Kubo formula for the
  viscosity in quantum field theory or from effective approaches such as
  kinetic theory will not have a meaningful hydrodynamic interpretation as the
  viscosity of a fluid in local thermal equilibrium.}\BibitemShut {Stop}%
\bibitem [{\citenamefont {Levitov}\ and\ \citenamefont
  {Falkovich}(2016)}]{levitov2016electron}%
  \BibitemOpen
  \bibfield  {author} {\bibinfo {author} {\bibfnamefont {L.}~\bibnamefont
  {Levitov}}\ and\ \bibinfo {author} {\bibfnamefont {G.}~\bibnamefont
  {Falkovich}},\ }\href@noop {} {\bibfield  {journal} {\bibinfo  {journal}
  {Nature Physics}\ }\textbf {\bibinfo {volume} {12}},\ \bibinfo {pages} {672}
  (\bibinfo {year} {2016})}\BibitemShut {NoStop}%
\bibitem [{\citenamefont {Pellegrino}\ \emph {et~al.}(2016)\citenamefont
  {Pellegrino}, \citenamefont {Torre}, \citenamefont {Geim},\ and\
  \citenamefont {Polini}}]{Pellegrino:2016kp}%
  \BibitemOpen
  \bibfield  {author} {\bibinfo {author} {\bibfnamefont {F.~M.}\ \bibnamefont
  {Pellegrino}}, \bibinfo {author} {\bibfnamefont {I.}~\bibnamefont {Torre}},
  \bibinfo {author} {\bibfnamefont {A.~K.}\ \bibnamefont {Geim}}, \ and\
  \bibinfo {author} {\bibfnamefont {M.}~\bibnamefont {Polini}},\ }\href@noop {}
  {\bibfield  {journal} {\bibinfo  {journal} {Physical Review B}\ }\textbf
  {\bibinfo {volume} {94}},\ \bibinfo {pages} {155414} (\bibinfo {year}
  {2016})}\BibitemShut {NoStop}%
\bibitem [{\citenamefont {{Torre}}\ \emph {et~al.}(2015)\citenamefont
  {{Torre}}, \citenamefont {{Tomadin}}, \citenamefont {{Geim}},\ and\
  \citenamefont {{Polini}}}]{Torre:2015eja}%
  \BibitemOpen
  \bibfield  {author} {\bibinfo {author} {\bibfnamefont {I.}~\bibnamefont
  {{Torre}}}, \bibinfo {author} {\bibfnamefont {A.}~\bibnamefont {{Tomadin}}},
  \bibinfo {author} {\bibfnamefont {A.~K.}\ \bibnamefont {{Geim}}}, \ and\
  \bibinfo {author} {\bibfnamefont {M.}~\bibnamefont {{Polini}}},\ }\href
  {\doibase 10.1103/PhysRevB.92.165433} {\bibfield  {journal} {\bibinfo
  {journal} {Physical Review B}\ }\textbf {\bibinfo {volume} {92}},\ \bibinfo
  {eid} {165433} (\bibinfo {year} {2015})},\ \Eprint
  {http://arxiv.org/abs/1508.00363} {arXiv:1508.00363 [cond-mat.mes-hall]}
  \BibitemShut {NoStop}%
\bibitem [{\citenamefont {Tomadin}\ \emph {et~al.}(2014)\citenamefont
  {Tomadin}, \citenamefont {Vignale},\ and\ \citenamefont
  {Polini}}]{Tomadin:2014by}%
  \BibitemOpen
  \bibfield  {author} {\bibinfo {author} {\bibfnamefont {A.}~\bibnamefont
  {Tomadin}}, \bibinfo {author} {\bibfnamefont {G.}~\bibnamefont {Vignale}}, \
  and\ \bibinfo {author} {\bibfnamefont {M.}~\bibnamefont {Polini}},\
  }\href@noop {} {\bibfield  {journal} {\bibinfo  {journal} {Physical Review
  Letters}\ }\textbf {\bibinfo {volume} {113}},\ \bibinfo {pages} {235901}
  (\bibinfo {year} {2014})}\BibitemShut {NoStop}%
\bibitem [{\citenamefont {Moessner}\ \emph {et~al.}(2018)\citenamefont
  {Moessner}, \citenamefont {Sur\'owka},\ and\ \citenamefont
  {Witkowski}}]{PhysRevB.97.161112}%
  \BibitemOpen
  \bibfield  {author} {\bibinfo {author} {\bibfnamefont {R.}~\bibnamefont
  {Moessner}}, \bibinfo {author} {\bibfnamefont {P.}~\bibnamefont {Sur\'owka}},
  \ and\ \bibinfo {author} {\bibfnamefont {P.}~\bibnamefont {Witkowski}},\
  }\href {\doibase 10.1103/PhysRevB.97.161112} {\bibfield  {journal} {\bibinfo
  {journal} {Phys. Rev. B}\ }\textbf {\bibinfo {volume} {97}},\ \bibinfo
  {pages} {161112} (\bibinfo {year} {2018})}\BibitemShut {NoStop}%
\bibitem [{\citenamefont {Kashuba}\ \emph {et~al.}(2018)\citenamefont
  {Kashuba}, \citenamefont {Trauzettel},\ and\ \citenamefont
  {Molenkamp}}]{2018arXiv180502987K}%
  \BibitemOpen
  \bibfield  {author} {\bibinfo {author} {\bibfnamefont {O.}~\bibnamefont
  {Kashuba}}, \bibinfo {author} {\bibfnamefont {B.}~\bibnamefont {Trauzettel}},
  \ and\ \bibinfo {author} {\bibfnamefont {L.~W.}\ \bibnamefont {Molenkamp}},\
  }\href@noop {} {\bibfield  {journal} {\bibinfo  {journal} {Physical Review
  B}\ }\textbf {\bibinfo {volume} {97}},\ \bibinfo {pages} {205129} (\bibinfo
  {year} {2018})}\BibitemShut {NoStop}%
\bibitem [{\citenamefont {Maldacena}(1998)}]{Maldacena:1997re}%
  \BibitemOpen
  \bibfield  {author} {\bibinfo {author} {\bibfnamefont {J.~M.}\ \bibnamefont
  {Maldacena}},\ }\href {\doibase 10.1023/A:1026654312961} {\bibfield
  {journal} {\bibinfo  {journal} {Adv. Theor. Math. Phys.}\ }\textbf {\bibinfo
  {volume} {2}},\ \bibinfo {pages} {231} (\bibinfo {year} {1998})},\ \Eprint
  {http://arxiv.org/abs/arxiv:9711200} {arxiv:9711200 [hep-th]} \BibitemShut
  {NoStop}%
\bibitem [{\citenamefont {Gubser}\ \emph {et~al.}(1998)\citenamefont {Gubser},
  \citenamefont {Klebanov},\ and\ \citenamefont {Polyakov}}]{Gubser:1998bc}%
  \BibitemOpen
  \bibfield  {author} {\bibinfo {author} {\bibfnamefont {S.~S.}\ \bibnamefont
  {Gubser}}, \bibinfo {author} {\bibfnamefont {I.~R.}\ \bibnamefont
  {Klebanov}}, \ and\ \bibinfo {author} {\bibfnamefont {A.~M.}\ \bibnamefont
  {Polyakov}},\ }\href {\doibase 10.1016/S0370-2693(98)00377-3} {\bibfield
  {journal} {\bibinfo  {journal} {Phys.Lett.}\ }\textbf {\bibinfo {volume}
  {B428}},\ \bibinfo {pages} {105} (\bibinfo {year} {1998})},\ \Eprint
  {http://arxiv.org/abs/hep-th/9802109} {arXiv:hep-th/9802109 [hep-th]}
  \BibitemShut {NoStop}%
\bibitem [{\citenamefont {Witten}(1998)}]{Witten:1998qj}%
  \BibitemOpen
  \bibfield  {author} {\bibinfo {author} {\bibfnamefont {E.}~\bibnamefont
  {Witten}},\ }\href@noop {} {\bibfield  {journal} {\bibinfo  {journal}
  {Adv.Theor.Math.Phys.}\ }\textbf {\bibinfo {volume} {2}},\ \bibinfo {pages}
  {253} (\bibinfo {year} {1998})},\ \Eprint
  {http://arxiv.org/abs/hep-th/9802150} {arXiv:hep-th/9802150 [hep-th]}
  \BibitemShut {NoStop}%
\bibitem [{\citenamefont {Casalderrey-Solana}\ \emph
  {et~al.}(2014)\citenamefont {Casalderrey-Solana}, \citenamefont {Liu},
  \citenamefont {Mateos}, \citenamefont {Rajagopal},\ and\ \citenamefont
  {Wiedemann}}]{CasalderreySolana:2011us}%
  \BibitemOpen
  \bibfield  {author} {\bibinfo {author} {\bibfnamefont {J.}~\bibnamefont
  {Casalderrey-Solana}}, \bibinfo {author} {\bibfnamefont {H.}~\bibnamefont
  {Liu}}, \bibinfo {author} {\bibfnamefont {D.}~\bibnamefont {Mateos}},
  \bibinfo {author} {\bibfnamefont {K.}~\bibnamefont {Rajagopal}}, \ and\
  \bibinfo {author} {\bibfnamefont {U.~A.}\ \bibnamefont {Wiedemann}},\
  }\href@noop {} {\emph {\bibinfo {title} {Gauge/string duality, hot QCD and
  heavy ion collisions}}}\ (\bibinfo  {publisher} {Cambridge University
  Press},\ \bibinfo {year} {2014})\BibitemShut {NoStop}%
\bibitem [{\citenamefont {Ammon}\ and\ \citenamefont
  {Erdmenger}(2015)}]{Ammon:2015}%
  \BibitemOpen
  \bibfield  {author} {\bibinfo {author} {\bibfnamefont {M.}~\bibnamefont
  {Ammon}}\ and\ \bibinfo {author} {\bibfnamefont {J.}~\bibnamefont
  {Erdmenger}},\ }\href@noop {} {\emph {\bibinfo {title} {Gauge/Gravity
  Duality: Foundations and Applications}}}\ (\bibinfo  {publisher} {Cambridge
  University Press},\ \bibinfo {year} {2015})\BibitemShut {NoStop}%
\bibitem [{\citenamefont {Zaanen}\ \emph {et~al.}(2015)\citenamefont {Zaanen},
  \citenamefont {Liu}, \citenamefont {Sun},\ and\ \citenamefont
  {Schalm}}]{zaanen2015holographic}%
  \BibitemOpen
  \bibfield  {author} {\bibinfo {author} {\bibfnamefont {J.}~\bibnamefont
  {Zaanen}}, \bibinfo {author} {\bibfnamefont {Y.}~\bibnamefont {Liu}},
  \bibinfo {author} {\bibfnamefont {Y.-W.}\ \bibnamefont {Sun}}, \ and\
  \bibinfo {author} {\bibfnamefont {K.}~\bibnamefont {Schalm}},\ }\href@noop {}
  {\emph {\bibinfo {title} {Holographic duality in condensed matter physics}}}\
  (\bibinfo  {publisher} {Cambridge University Press},\ \bibinfo {year}
  {2015})\BibitemShut {NoStop}%
\bibitem [{\citenamefont {Hartnoll}\ \emph {et~al.}(2016)\citenamefont
  {Hartnoll}, \citenamefont {Lucas},\ and\ \citenamefont
  {Sachdev}}]{Hartnoll:2016apf}%
  \BibitemOpen
  \bibfield  {author} {\bibinfo {author} {\bibfnamefont {S.~A.}\ \bibnamefont
  {Hartnoll}}, \bibinfo {author} {\bibfnamefont {A.}~\bibnamefont {Lucas}}, \
  and\ \bibinfo {author} {\bibfnamefont {S.}~\bibnamefont {Sachdev}},\
  }\href@noop {} {\  (\bibinfo {year} {2016})},\ \Eprint
  {http://arxiv.org/abs/1612.07324} {arXiv:1612.07324 [hep-th]} \BibitemShut
  {NoStop}%
\bibitem [{Note2()}]{Note2}%
  \BibitemOpen
  \bibinfo {note} {There are examples where hidden conserved quantities
  together with a long-range disorder potential invalidate the hydrodynamic
  approximation~\cite {Lucas:2017vlc}.}\BibitemShut {Stop}%
\bibitem [{\citenamefont {Zaanen}(2004)}]{zaanen2004superconductivity}%
  \BibitemOpen
  \bibfield  {author} {\bibinfo {author} {\bibfnamefont {J.}~\bibnamefont
  {Zaanen}},\ }\href@noop {} {\bibfield  {journal} {\bibinfo  {journal}
  {Nature}\ }\textbf {\bibinfo {volume} {430}},\ \bibinfo {pages} {512}
  (\bibinfo {year} {2004})}\BibitemShut {NoStop}%
\bibitem [{Note3()}]{Note3}%
  \BibitemOpen
  \bibinfo {note} {For an explanation of this fact, c.f.~e.g.~sec.~3.5~of~\cite
  {Hartnoll:2016apf}.}\BibitemShut {Stop}%
\bibitem [{\citenamefont {Bhattacharyya}\ \emph
  {et~al.}(2008{\natexlab{a}})\citenamefont {Bhattacharyya}, \citenamefont
  {Hubeny}, \citenamefont {Minwalla},\ and\ \citenamefont
  {Rangamani}}]{Bhattacharyya:2008jc}%
  \BibitemOpen
  \bibfield  {author} {\bibinfo {author} {\bibfnamefont {S.}~\bibnamefont
  {Bhattacharyya}}, \bibinfo {author} {\bibfnamefont {V.~E.}\ \bibnamefont
  {Hubeny}}, \bibinfo {author} {\bibfnamefont {S.}~\bibnamefont {Minwalla}}, \
  and\ \bibinfo {author} {\bibfnamefont {M.}~\bibnamefont {Rangamani}},\ }\href
  {\doibase 10.1088/1126-6708/2008/02/045} {\bibfield  {journal} {\bibinfo
  {journal} {JHEP}\ }\textbf {\bibinfo {volume} {02}},\ \bibinfo {pages} {045}
  (\bibinfo {year} {2008}{\natexlab{a}})},\ \Eprint
  {http://arxiv.org/abs/0712.2456} {arXiv:0712.2456 [hep-th]} \BibitemShut
  {NoStop}%
\bibitem [{\citenamefont {Bhattacharyya}\ \emph
  {et~al.}(2008{\natexlab{b}})\citenamefont {Bhattacharyya}, \citenamefont
  {Hubeny}, \citenamefont {Loganayagam}, \citenamefont {Mandal}, \citenamefont
  {Minwalla}, \citenamefont {Morita}, \citenamefont {Rangamani},\ and\
  \citenamefont {Reall}}]{Bhattacharyya:2008xc}%
  \BibitemOpen
  \bibfield  {author} {\bibinfo {author} {\bibfnamefont {S.}~\bibnamefont
  {Bhattacharyya}}, \bibinfo {author} {\bibfnamefont {V.~E.}\ \bibnamefont
  {Hubeny}}, \bibinfo {author} {\bibfnamefont {R.}~\bibnamefont {Loganayagam}},
  \bibinfo {author} {\bibfnamefont {G.}~\bibnamefont {Mandal}}, \bibinfo
  {author} {\bibfnamefont {S.}~\bibnamefont {Minwalla}}, \bibinfo {author}
  {\bibfnamefont {T.}~\bibnamefont {Morita}}, \bibinfo {author} {\bibfnamefont
  {M.}~\bibnamefont {Rangamani}}, \ and\ \bibinfo {author} {\bibfnamefont
  {H.~S.}\ \bibnamefont {Reall}},\ }\href {\doibase
  10.1088/1126-6708/2008/06/055} {\bibfield  {journal} {\bibinfo  {journal}
  {JHEP}\ }\textbf {\bibinfo {volume} {06}},\ \bibinfo {pages} {055} (\bibinfo
  {year} {2008}{\natexlab{b}})},\ \Eprint {http://arxiv.org/abs/0803.2526}
  {arXiv:0803.2526 [hep-th]} \BibitemShut {NoStop}%
\bibitem [{\citenamefont {Bhattacharyya}\ \emph {et~al.}(2009)\citenamefont
  {Bhattacharyya}, \citenamefont {Loganayagam}, \citenamefont {Minwalla},
  \citenamefont {Nampuri}, \citenamefont {Trivedi},\ and\ \citenamefont
  {Wadia}}]{Bhattacharyya:2008ji}%
  \BibitemOpen
  \bibfield  {author} {\bibinfo {author} {\bibfnamefont {S.}~\bibnamefont
  {Bhattacharyya}}, \bibinfo {author} {\bibfnamefont {R.}~\bibnamefont
  {Loganayagam}}, \bibinfo {author} {\bibfnamefont {S.}~\bibnamefont
  {Minwalla}}, \bibinfo {author} {\bibfnamefont {S.}~\bibnamefont {Nampuri}},
  \bibinfo {author} {\bibfnamefont {S.~P.}\ \bibnamefont {Trivedi}}, \ and\
  \bibinfo {author} {\bibfnamefont {S.~R.}\ \bibnamefont {Wadia}},\ }\href
  {\doibase 10.1088/1126-6708/2009/02/018} {\bibfield  {journal} {\bibinfo
  {journal} {JHEP}\ }\textbf {\bibinfo {volume} {02}},\ \bibinfo {pages} {018}
  (\bibinfo {year} {2009})},\ \Eprint {http://arxiv.org/abs/0806.0006}
  {arXiv:0806.0006 [hep-th]} \BibitemShut {NoStop}%
\bibitem [{\citenamefont {Blake}(2015)}]{Blake:2015epa}%
  \BibitemOpen
  \bibfield  {author} {\bibinfo {author} {\bibfnamefont {M.}~\bibnamefont
  {Blake}},\ }\href {\doibase 10.1007/JHEP09(2015)010} {\bibfield  {journal}
  {\bibinfo  {journal} {JHEP}\ }\textbf {\bibinfo {volume} {09}},\ \bibinfo
  {pages} {010} (\bibinfo {year} {2015})},\ \Eprint
  {http://arxiv.org/abs/1505.06992} {arXiv:1505.06992 [hep-th]} \BibitemShut
  {NoStop}%
\bibitem [{\citenamefont {Policastro}\ \emph {et~al.}(2001)\citenamefont
  {Policastro}, \citenamefont {Son},\ and\ \citenamefont
  {Starinets}}]{Policastro:2001yc}%
  \BibitemOpen
  \bibfield  {author} {\bibinfo {author} {\bibfnamefont {G.}~\bibnamefont
  {Policastro}}, \bibinfo {author} {\bibfnamefont {D.~T.}\ \bibnamefont {Son}},
  \ and\ \bibinfo {author} {\bibfnamefont {A.~O.}\ \bibnamefont {Starinets}},\
  }\href {\doibase 10.1103/PhysRevLett.87.081601} {\bibfield  {journal}
  {\bibinfo  {journal} {Phys. Rev. Lett.}\ }\textbf {\bibinfo {volume} {87}},\
  \bibinfo {pages} {081601} (\bibinfo {year} {2001})},\ \Eprint
  {http://arxiv.org/abs/hep-th/0104066} {arXiv:hep-th/0104066 [hep-th]}
  \BibitemShut {NoStop}%
\bibitem [{\citenamefont {Kovtun}\ \emph {et~al.}(2005)\citenamefont {Kovtun},
  \citenamefont {Son},\ and\ \citenamefont {Starinets}}]{Kovtun:2004de}%
  \BibitemOpen
  \bibfield  {author} {\bibinfo {author} {\bibfnamefont {P.}~\bibnamefont
  {Kovtun}}, \bibinfo {author} {\bibfnamefont {D.~T.}\ \bibnamefont {Son}}, \
  and\ \bibinfo {author} {\bibfnamefont {A.~O.}\ \bibnamefont {Starinets}},\
  }\href {\doibase 10.1103/PhysRevLett.94.111601} {\bibfield  {journal}
  {\bibinfo  {journal} {Phys. Rev. Lett.}\ }\textbf {\bibinfo {volume} {94}},\
  \bibinfo {pages} {111601} (\bibinfo {year} {2005})},\ \Eprint
  {http://arxiv.org/abs/hep-th/0405231} {arXiv:hep-th/0405231 [hep-th]}
  \BibitemShut {NoStop}%
\bibitem [{Note4()}]{Note4}%
  \BibitemOpen
  \bibinfo {note} {For rotationally non-invariant systems, there can be
  violations of the bound,~\cite {Mateos:2011ix,Rebhan:2011vd}. However, all
  such violations that have been found so far to leading order in the inverse
  string tension $\alpha '$ (dual to the coupling constant) or the gauge group
  rank $N$ (controlling the large $N$ limit) occur in unstable ground
  state,~\cite {Mateos:2011ix,Mateos:2011tv}. On the other hand, in systems
  with rotationally non-invariant but obviously stable ground states, the KSS
  bound \protect \eqref {etasAdSCFT2} seems to be observed for all components
  of the viscosity tensor \cite {Erdmenger:2012zu}. To subleading order in
  $\alpha '$ (corresponding to finite coupling corrections), violations of
  \protect \eqref {etasAdSCFT2} induce causality violations in the dual field
  theory \cite {Brigante:2008gz,Camanho:2014apa}. Presumably only string
  theories including all order $\alpha '$ corrections is causal \cite
  {Camanho:2014apa}, but $\eta /s$ has never bee calculated in such a setup.
  Similar remarks apply to finite $N$ corrections \cite {Kats:2007mq} which
  also correspond to higher derivative terms in the dual gravitational theory.
  In summary, there is good evidence\cite {Cremonini:2011iq} that \protect
  \textup {\hbox {\mathsurround \z@ \protect \normalfont (\ignorespaces \ref
  {etasAdSCFT}\unskip \@@italiccorr )}} holds in the stable ground states of
  all consistent models.}\BibitemShut {Stop}%
\bibitem [{\citenamefont {Delacr{\'e}taz}\ \emph {et~al.}(2018)\citenamefont
  {Delacr{\'e}taz}, \citenamefont {Hartman}, \citenamefont {Hartnoll},\ and\
  \citenamefont {Lewkowycz}}]{Delacretaz:2018cfk}%
  \BibitemOpen
  \bibfield  {author} {\bibinfo {author} {\bibfnamefont {L.~V.}\ \bibnamefont
  {Delacr{\'e}taz}}, \bibinfo {author} {\bibfnamefont {T.}~\bibnamefont
  {Hartman}}, \bibinfo {author} {\bibfnamefont {S.~A.}\ \bibnamefont
  {Hartnoll}}, \ and\ \bibinfo {author} {\bibfnamefont {A.}~\bibnamefont
  {Lewkowycz}},\ }\href@noop {} {\  (\bibinfo {year} {2018})},\ \Eprint
  {http://arxiv.org/abs/1805.04194} {arXiv:1805.04194 [hep-th]} \BibitemShut
  {NoStop}%
\bibitem [{\citenamefont {Hartman}\ \emph {et~al.}(2017)\citenamefont
  {Hartman}, \citenamefont {Hartnoll},\ and\ \citenamefont
  {Mahajan}}]{Hartman:2017hhp}%
  \BibitemOpen
  \bibfield  {author} {\bibinfo {author} {\bibfnamefont {T.}~\bibnamefont
  {Hartman}}, \bibinfo {author} {\bibfnamefont {S.~A.}\ \bibnamefont
  {Hartnoll}}, \ and\ \bibinfo {author} {\bibfnamefont {R.}~\bibnamefont
  {Mahajan}},\ }\href {\doibase 10.1103/PhysRevLett.119.141601} {\bibfield
  {journal} {\bibinfo  {journal} {Phys. Rev. Lett.}\ }\textbf {\bibinfo
  {volume} {119}},\ \bibinfo {pages} {141601} (\bibinfo {year} {2017})},\
  \Eprint {http://arxiv.org/abs/1706.00019} {arXiv:1706.00019 [hep-th]}
  \BibitemShut {NoStop}%
\bibitem [{\citenamefont {Song}\ \emph {et~al.}(2011)\citenamefont {Song},
  \citenamefont {Bass}, \citenamefont {Heinz}, \citenamefont {Hirano},\ and\
  \citenamefont {Shen}}]{song2011200}%
  \BibitemOpen
  \bibfield  {author} {\bibinfo {author} {\bibfnamefont {H.}~\bibnamefont
  {Song}}, \bibinfo {author} {\bibfnamefont {S.~A.}\ \bibnamefont {Bass}},
  \bibinfo {author} {\bibfnamefont {U.}~\bibnamefont {Heinz}}, \bibinfo
  {author} {\bibfnamefont {T.}~\bibnamefont {Hirano}}, \ and\ \bibinfo {author}
  {\bibfnamefont {C.}~\bibnamefont {Shen}},\ }\href@noop {} {\bibfield
  {journal} {\bibinfo  {journal} {Physical Review Letters}\ }\textbf {\bibinfo
  {volume} {106}},\ \bibinfo {pages} {192301} (\bibinfo {year}
  {2011})}\BibitemShut {NoStop}%
\bibitem [{Note5()}]{Note5}%
  \BibitemOpen
  \bibinfo {note} {For electrons in solids we have direct access to the
  hydrodynamic regime. In QCD, we need to reconstruct the hydrodynamic regime
  from the data collected by particle detectors back through the hadronization
  crossover.}\BibitemShut {Stop}%
\bibitem [{\citenamefont {Novoselov}\ \emph {et~al.}(2005)\citenamefont
  {Novoselov}, \citenamefont {Geim}, \citenamefont {Morozov}, \citenamefont
  {Jiang}, \citenamefont {Katsnelson}, \citenamefont {Grigorieva},
  \citenamefont {Dubonos}, \citenamefont {Firsov},\ and\ \citenamefont
  {AA}}]{novoselov2005two}%
  \BibitemOpen
  \bibfield  {author} {\bibinfo {author} {\bibfnamefont {K.~S.}\ \bibnamefont
  {Novoselov}}, \bibinfo {author} {\bibfnamefont {A.~K.}\ \bibnamefont {Geim}},
  \bibinfo {author} {\bibfnamefont {S.}~\bibnamefont {Morozov}}, \bibinfo
  {author} {\bibfnamefont {D.}~\bibnamefont {Jiang}}, \bibinfo {author}
  {\bibfnamefont {M.}~\bibnamefont {Katsnelson}}, \bibinfo {author}
  {\bibfnamefont {I.}~\bibnamefont {Grigorieva}}, \bibinfo {author}
  {\bibfnamefont {S.}~\bibnamefont {Dubonos}}, \bibinfo {author} {\bibnamefont
  {Firsov}}, \ and\ \bibinfo {author} {\bibnamefont {AA}},\ }\href@noop {}
  {\bibfield  {journal} {\bibinfo  {journal} {nature}\ }\textbf {\bibinfo
  {volume} {438}},\ \bibinfo {pages} {197} (\bibinfo {year}
  {2005})}\BibitemShut {NoStop}%
\bibitem [{\citenamefont {K{\"o}nig}\ \emph {et~al.}(2007)\citenamefont
  {K{\"o}nig}, \citenamefont {Wiedmann}, \citenamefont {Br{\"u}ne},
  \citenamefont {Roth}, \citenamefont {Buhmann}, \citenamefont {Molenkamp},
  \citenamefont {Qi},\ and\ \citenamefont {Zhang}}]{konig2007quantum}%
  \BibitemOpen
  \bibfield  {author} {\bibinfo {author} {\bibfnamefont {M.}~\bibnamefont
  {K{\"o}nig}}, \bibinfo {author} {\bibfnamefont {S.}~\bibnamefont {Wiedmann}},
  \bibinfo {author} {\bibfnamefont {C.}~\bibnamefont {Br{\"u}ne}}, \bibinfo
  {author} {\bibfnamefont {A.}~\bibnamefont {Roth}}, \bibinfo {author}
  {\bibfnamefont {H.}~\bibnamefont {Buhmann}}, \bibinfo {author} {\bibfnamefont
  {L.~W.}\ \bibnamefont {Molenkamp}}, \bibinfo {author} {\bibfnamefont {X.-L.}\
  \bibnamefont {Qi}}, \ and\ \bibinfo {author} {\bibfnamefont {S.-C.}\
  \bibnamefont {Zhang}},\ }\href@noop {} {\bibfield  {journal} {\bibinfo
  {journal} {Science}\ }\textbf {\bibinfo {volume} {318}},\ \bibinfo {pages}
  {766} (\bibinfo {year} {2007})}\BibitemShut {NoStop}%
\bibitem [{\citenamefont {Klinkhamer}\ and\ \citenamefont
  {Volovik}(2005)}]{Klinkhamer:2004hg}%
  \BibitemOpen
  \bibfield  {author} {\bibinfo {author} {\bibfnamefont {F.~R.}\ \bibnamefont
  {Klinkhamer}}\ and\ \bibinfo {author} {\bibfnamefont {G.~E.}\ \bibnamefont
  {Volovik}},\ }\href {\doibase 10.1142/S0217751X05020902} {\bibfield
  {journal} {\bibinfo  {journal} {Int. J. Mod. Phys.}\ }\textbf {\bibinfo
  {volume} {A20}},\ \bibinfo {pages} {2795} (\bibinfo {year} {2005})},\ \Eprint
  {http://arxiv.org/abs/hep-th/0403037} {arXiv:hep-th/0403037 [hep-th]}
  \BibitemShut {NoStop}%
\bibitem [{\citenamefont {Volovik}(2006)}]{Volovik:2003fe}%
  \BibitemOpen
  \bibfield  {author} {\bibinfo {author} {\bibfnamefont {G.~E.}\ \bibnamefont
  {Volovik}},\ }\href@noop {} {\bibfield  {journal} {\bibinfo  {journal} {Int.
  Ser. Monogr. Phys.}\ }\textbf {\bibinfo {volume} {117}},\ \bibinfo {pages}
  {1} (\bibinfo {year} {2006})}\BibitemShut {NoStop}%
\bibitem [{\citenamefont {Wan}\ \emph {et~al.}(2011)\citenamefont {Wan},
  \citenamefont {Turner}, \citenamefont {Vishwanath},\ and\ \citenamefont
  {Savrasov}}]{PhysRevB.83.205101}%
  \BibitemOpen
  \bibfield  {author} {\bibinfo {author} {\bibfnamefont {X.}~\bibnamefont
  {Wan}}, \bibinfo {author} {\bibfnamefont {A.~M.}\ \bibnamefont {Turner}},
  \bibinfo {author} {\bibfnamefont {A.}~\bibnamefont {Vishwanath}}, \ and\
  \bibinfo {author} {\bibfnamefont {S.~Y.}\ \bibnamefont {Savrasov}},\ }\href
  {\doibase 10.1103/PhysRevB.83.205101} {\bibfield  {journal} {\bibinfo
  {journal} {Phys. Rev. B}\ }\textbf {\bibinfo {volume} {83}},\ \bibinfo
  {pages} {205101} (\bibinfo {year} {2011})}\BibitemShut {NoStop}%
\bibitem [{\citenamefont {{Fritz}}\ \emph {et~al.}(2008)\citenamefont
  {{Fritz}}, \citenamefont {{Schmalian}}, \citenamefont {{M{\"u}ller}},\ and\
  \citenamefont {{Sachdev}}}]{Fritz:2008go}%
  \BibitemOpen
  \bibfield  {author} {\bibinfo {author} {\bibfnamefont {L.}~\bibnamefont
  {{Fritz}}}, \bibinfo {author} {\bibfnamefont {J.}~\bibnamefont
  {{Schmalian}}}, \bibinfo {author} {\bibfnamefont {M.}~\bibnamefont
  {{M{\"u}ller}}}, \ and\ \bibinfo {author} {\bibfnamefont {S.}~\bibnamefont
  {{Sachdev}}},\ }\href {\doibase 10.1103/PhysRevB.78.085416} {\bibfield
  {journal} {\bibinfo  {journal} {Physical Review B}\ }\textbf {\bibinfo
  {volume} {78}},\ \bibinfo {eid} {085416} (\bibinfo {year} {2008})},\ \Eprint
  {http://arxiv.org/abs/0802.4289} {arXiv:0802.4289} \BibitemShut {NoStop}%
\bibitem [{\citenamefont {{Müller}}\ \emph {et~al.}(2009)\citenamefont
  {{Müller}}, \citenamefont {{Schmalian}},\ and\ \citenamefont
  {{Fritz}}}]{Muller:2009cy}%
  \BibitemOpen
  \bibfield  {author} {\bibinfo {author} {\bibfnamefont {M.}~\bibnamefont
  {{Müller}}}, \bibinfo {author} {\bibfnamefont {J.}~\bibnamefont
  {{Schmalian}}}, \ and\ \bibinfo {author} {\bibfnamefont {L.}~\bibnamefont
  {{Fritz}}},\ }\href@noop {} {\bibfield  {journal} {\bibinfo  {journal}
  {Physical Review Letters}\ }\textbf {\bibinfo {volume} {103}},\ \bibinfo
  {pages} {025301} (\bibinfo {year} {2009})},\ \Eprint
  {http://arxiv.org/abs/0903.4178} {arXiv:0903.4178 [cond-mat.mes-hall]}
  \BibitemShut {NoStop}%
\bibitem [{Note6()}]{Note6}%
  \BibitemOpen
  \bibinfo {note} {In experiment, the channels are typically prepared by
  etching them out of a thin layer of material. During that process, the
  channel edges are sufficiently disordered, such that zero velocity boundary
  conditions are applicable.}\BibitemShut {Stop}%
\bibitem [{Note7()}]{Note7}%
  \BibitemOpen
  \bibinfo {note} {In this work, we assume that we can vary $\eta /s$
  independently of $s$. This amounts to varying the coupling strength
  independent of the temperature, in which case, the entropy density $s$ at
  fixed temperature is a constant in thermal and chemical equilibrium. We leave
  the investigation on the possible experimental realization for the
  future.}\BibitemShut {Stop}%
\bibitem [{Note8()}]{Note8}%
  \BibitemOpen
  \bibinfo {note} {\label {foot:newhydromodes}The appearance of new
  non-hydrodynamic modes at finite coupling\cite {grozdanov2016strong} may
  invalidate the hydrodynamic approximation. However, we note that in the
  present work our analysis is performed at finite density in 2+1 dimensions,
  while these results\cite {grozdanov2016strong} apply to 3+1 dimensions at
  vanishing density. Since the structure of higher derivative terms in 3+1
  dimensional gravity will be very different from the ones in 4+1 dimensional
  gravity, these results\cite {grozdanov2016strong} are not directly applicable
  to our analysis. We note that in the context of applying AdS/CFT to heavy ion
  physics, the physical implications of these terms are still under
  debate.\cite {Strickland:2018exs} At present we therefore consider our
  monotonic coupling interpolation to be in the hydrodynamic regime. We point
  out that a detailed analysis of this issue will be necessary in 2+1
  dimensions and at finite density as well.}\BibitemShut {Stop}%
\bibitem [{\citenamefont {Davison}\ \emph {et~al.}(2014)\citenamefont
  {Davison}, \citenamefont {Schalm},\ and\ \citenamefont
  {Zaanen}}]{davison2014holographic}%
  \BibitemOpen
  \bibfield  {author} {\bibinfo {author} {\bibfnamefont {R.~A.}\ \bibnamefont
  {Davison}}, \bibinfo {author} {\bibfnamefont {K.}~\bibnamefont {Schalm}}, \
  and\ \bibinfo {author} {\bibfnamefont {J.}~\bibnamefont {Zaanen}},\
  }\href@noop {} {\bibfield  {journal} {\bibinfo  {journal} {Physical Review
  B}\ }\textbf {\bibinfo {volume} {89}},\ \bibinfo {pages} {245116} (\bibinfo
  {year} {2014})}\BibitemShut {NoStop}%
\bibitem [{\citenamefont {Landau}\ and\ \citenamefont
  {Lifshitz}(1987)}]{Landau:1987}%
  \BibitemOpen
  \bibfield  {author} {\bibinfo {author} {\bibfnamefont {L.}~\bibnamefont
  {Landau}}\ and\ \bibinfo {author} {\bibfnamefont {E.}~\bibnamefont
  {Lifshitz}},\ }\href@noop {} {\emph {\bibinfo {title} {Fluid Mechanics}}},\
  \bibinfo {edition} {2nd}\ ed.,\ Course of Theoretical Physics S\ (\bibinfo
  {publisher} {Butterworth-Heinemann},\ \bibinfo {year} {1987})\BibitemShut
  {NoStop}%
\bibitem [{\citenamefont {Romatschke}(2010)}]{Romatschke:2010jf}%
  \BibitemOpen
  \bibfield  {author} {\bibinfo {author} {\bibfnamefont {P.}~\bibnamefont
  {Romatschke}},\ }\href@noop {} {\bibfield  {journal} {\bibinfo  {journal}
  {International Journal of Modern Physics E}\ }\textbf {\bibinfo {volume}
  {19}},\ \bibinfo {pages} {1} (\bibinfo {year} {2010})}\BibitemShut {NoStop}%
\bibitem [{\citenamefont {Luciano~Rezzolla}(2013)}]{Rezzolla:2013}%
  \BibitemOpen
  \bibfield  {author} {\bibinfo {author} {\bibfnamefont {O.~Z.}\ \bibnamefont
  {Luciano~Rezzolla}},\ }\href@noop {} {\emph {\bibinfo {title} {Relativistic
  Hydrodynamics}}}\ (\bibinfo  {publisher} {Oxford University Press},\ \bibinfo
  {year} {2013})\BibitemShut {NoStop}%
\bibitem [{Note9()}]{Note9}%
  \BibitemOpen
  \bibinfo {note} {Introducing a nontrivial metric is useful when e.g.
  considering perturbations that correspond to a temperature gradient acting on
  the fluid \cite {1963AnPhy..24..419K}, and can be achieved by replacing
  partial derivatives in \protect \textup {\hbox {\mathsurround \z@ \protect
  \normalfont (\ignorespaces \ref {relhydroeqs}\unskip \@@italiccorr
  )}}-\protect \textup {\hbox {\mathsurround \z@ \protect \normalfont
  (\ignorespaces \ref {relhydroeqs2}\unskip \@@italiccorr )}} by the covariant
  derivative with an appropriately chosen connection, which most of the time
  will be the metric compatible Christoffel connection.}\BibitemShut {Stop}%
\bibitem [{Note10()}]{Note10}%
  \BibitemOpen
  \bibinfo {note} {We normalized the electric field by means of the fundamental
  constants $h,e$ and $c$ and in such a way that $\left [ E^\mu \right
  ]=\protect \text {V/m}, \protect \tmspace +\thinmuskip {.1667em} \left [j^\mu
  \right ]=\protect \text {A/m}$ and $\left [ D^{\mu \nu }\right ] = {\protect
  \rm A^2 s/m}$.}\BibitemShut {Stop}%
\bibitem [{Note11()}]{Note11}%
  \BibitemOpen
  \bibinfo {note} {Another common frame is Eckart frame, in which the fluid is
  comoving w.r.t. the current flow, amounting to $j^\mu =0$ in \protect \textup
  {\hbox {\mathsurround \z@ \protect \normalfont (\ignorespaces \ref
  {eq:J1}\unskip \@@italiccorr )}}. Another example can be found in \cite
  {Jensen:2011xb}, where the $M B$ in the presence of a constant magnetic field
  can either be considered zeroth order in derivatives due to the constancy of
  $B$ and incorporated into the pressure, or, if $B$ is small, to be of first
  order due to the fact that the magnetic field is the derivative of the vector
  potential. It is possible to switch between both frames through field
  transformations,~\cite {Kovtun:2012rj}.}\BibitemShut {Stop}%
\bibitem [{Note12()}]{Note12}%
  \BibitemOpen
  \bibinfo {note} {Scale invariance implies vanishing stress-energy tensor
  trace, $T_\mu {}^\mu $=0, and hence $\xi =0$. Furthermore, due to
  renormalization effects the Fermi velocity runs logarithmically at low
  temperatures $v_F=v_F(T)$ \cite {Lucas:2017wa}. Another reason for neglecting
  the bulk viscosity term is that we are working with incompressible fluid
  flows, i.e. $\partial _\mu u^\mu = 0$}\BibitemShut {NoStop}%
\bibitem [{Note13()}]{Note13}%
  \BibitemOpen
  \bibinfo {note} {The conservation equation for the current \protect \textup
  {\hbox {\mathsurround \z@ \protect \normalfont (\ignorespaces \ref
  {relhydroeqs2}\unskip \@@italiccorr )}} is trivially satisfied for the flow
  we will introduce in sec.~\ref {sec:FSE}, and hence we do not discuss it
  hereafter.}\BibitemShut {Stop}%
\bibitem [{\citenamefont {Bruin}\ \emph {et~al.}(2013)\citenamefont {Bruin},
  \citenamefont {Sakai}, \citenamefont {Perry},\ and\ \citenamefont
  {Mackenzie}}]{bruin2013similarity}%
  \BibitemOpen
  \bibfield  {author} {\bibinfo {author} {\bibfnamefont {J.}~\bibnamefont
  {Bruin}}, \bibinfo {author} {\bibfnamefont {H.}~\bibnamefont {Sakai}},
  \bibinfo {author} {\bibfnamefont {R.}~\bibnamefont {Perry}}, \ and\ \bibinfo
  {author} {\bibfnamefont {A.}~\bibnamefont {Mackenzie}},\ }\href@noop {}
  {\bibfield  {journal} {\bibinfo  {journal} {Science}\ }\textbf {\bibinfo
  {volume} {339}},\ \bibinfo {pages} {804} (\bibinfo {year}
  {2013})}\BibitemShut {NoStop}%
\bibitem [{\citenamefont {Gooth}\ \emph {et~al.}(2017)\citenamefont {Gooth},
  \citenamefont {Menges}, \citenamefont {Shekhar}, \citenamefont {S{\"u}{\ss}},
  \citenamefont {Kumar}, \citenamefont {Sun}, \citenamefont {Drechsler},
  \citenamefont {Zierold}, \citenamefont {Felser},\ and\ \citenamefont
  {Gotsmann}}]{gooth2017electrical}%
  \BibitemOpen
  \bibfield  {author} {\bibinfo {author} {\bibfnamefont {J.}~\bibnamefont
  {Gooth}}, \bibinfo {author} {\bibfnamefont {F.}~\bibnamefont {Menges}},
  \bibinfo {author} {\bibfnamefont {C.}~\bibnamefont {Shekhar}}, \bibinfo
  {author} {\bibfnamefont {V.}~\bibnamefont {S{\"u}{\ss}}}, \bibinfo {author}
  {\bibfnamefont {N.}~\bibnamefont {Kumar}}, \bibinfo {author} {\bibfnamefont
  {Y.}~\bibnamefont {Sun}}, \bibinfo {author} {\bibfnamefont {U.}~\bibnamefont
  {Drechsler}}, \bibinfo {author} {\bibfnamefont {R.}~\bibnamefont {Zierold}},
  \bibinfo {author} {\bibfnamefont {C.}~\bibnamefont {Felser}}, \ and\ \bibinfo
  {author} {\bibfnamefont {B.}~\bibnamefont {Gotsmann}},\ }\href@noop {}
  {\bibfield  {journal} {\bibinfo  {journal} {arXiv preprint arXiv:1706.05925}\
  } (\bibinfo {year} {2017})}\BibitemShut {NoStop}%
\bibitem [{Note14()}]{Note14}%
  \BibitemOpen
  \bibinfo {note} {Note that we assume vanishing temperature gradients. If we
  were to relax this assumption, the spacetime dependence of $\eta $ should
  also be taken into account in the equations of motion.}\BibitemShut {Stop}%
\bibitem [{\citenamefont {Vegh}(2013)}]{Vegh:2013sk}%
  \BibitemOpen
  \bibfield  {author} {\bibinfo {author} {\bibfnamefont {D.}~\bibnamefont
  {Vegh}},\ }\href@noop {} {\  (\bibinfo {year} {2013})},\ \Eprint
  {http://arxiv.org/abs/1301.0537} {arXiv:1301.0537 [hep-th]} \BibitemShut
  {NoStop}%
\bibitem [{\citenamefont {Andrade}\ and\ \citenamefont
  {Withers}(2014)}]{Andrade2014}%
  \BibitemOpen
  \bibfield  {author} {\bibinfo {author} {\bibfnamefont {T.}~\bibnamefont
  {Andrade}}\ and\ \bibinfo {author} {\bibfnamefont {B.}~\bibnamefont
  {Withers}},\ }\href {\doibase 10.1007/JHEP05(2014)101} {\bibfield  {journal}
  {\bibinfo  {journal} {Journal of High Energy Physics}\ }\textbf {\bibinfo
  {volume} {2014}},\ \bibinfo {pages} {101} (\bibinfo {year}
  {2014})}\BibitemShut {NoStop}%
\bibitem [{\citenamefont {Davison}(2013)}]{Davison:2013jba}%
  \BibitemOpen
  \bibfield  {author} {\bibinfo {author} {\bibfnamefont {R.~A.}\ \bibnamefont
  {Davison}},\ }\href {\doibase 10.1103/PhysRevD.88.086003} {\bibfield
  {journal} {\bibinfo  {journal} {Phys. Rev.}\ }\textbf {\bibinfo {volume}
  {D88}},\ \bibinfo {pages} {086003} (\bibinfo {year} {2013})},\ \Eprint
  {http://arxiv.org/abs/1306.5792} {arXiv:1306.5792 [hep-th]} \BibitemShut
  {NoStop}%
\bibitem [{\citenamefont {{Br{\"u}ne}}\ \emph {et~al.}(2014)\citenamefont
  {{Br{\"u}ne}}, \citenamefont {{Thienel}}, \citenamefont {{Stuiber}},
  \citenamefont {{B{\"o}ttcher}}, \citenamefont {{Buhmann}}, \citenamefont
  {{Novik}}, \citenamefont {{Liu}}, \citenamefont {{Hankiewicz}},\ and\
  \citenamefont {{Molenkamp}}}]{PhysRevX.4.041045}%
  \BibitemOpen
  \bibfield  {author} {\bibinfo {author} {\bibfnamefont {C.}~\bibnamefont
  {{Br{\"u}ne}}}, \bibinfo {author} {\bibfnamefont {C.}~\bibnamefont
  {{Thienel}}}, \bibinfo {author} {\bibfnamefont {M.}~\bibnamefont
  {{Stuiber}}}, \bibinfo {author} {\bibfnamefont {J.}~\bibnamefont
  {{B{\"o}ttcher}}}, \bibinfo {author} {\bibfnamefont {H.}~\bibnamefont
  {{Buhmann}}}, \bibinfo {author} {\bibfnamefont {E.~G.}\ \bibnamefont
  {{Novik}}}, \bibinfo {author} {\bibfnamefont {C.-X.}\ \bibnamefont {{Liu}}},
  \bibinfo {author} {\bibfnamefont {E.~M.}\ \bibnamefont {{Hankiewicz}}}, \
  and\ \bibinfo {author} {\bibfnamefont {L.~W.}\ \bibnamefont {{Molenkamp}}},\
  }\href {\doibase 10.1103/PhysRevX.4.041045} {\bibfield  {journal} {\bibinfo
  {journal} {Physical Review X}\ }\textbf {\bibinfo {volume} {4}},\ \bibinfo
  {eid} {041045} (\bibinfo {year} {2014})},\ \Eprint
  {http://arxiv.org/abs/1407.6537} {arXiv:1407.6537 [cond-mat.mes-hall]}
  \BibitemShut {NoStop}%
\bibitem [{\citenamefont {Majumder}\ \emph {et~al.}(2007)\citenamefont
  {Majumder}, \citenamefont {Müller},\ and\ \citenamefont
  {Wang}}]{Majumder:2007zh}%
  \BibitemOpen
  \bibfield  {author} {\bibinfo {author} {\bibfnamefont {A.}~\bibnamefont
  {Majumder}}, \bibinfo {author} {\bibfnamefont {B.}~\bibnamefont {Müller}}, \
  and\ \bibinfo {author} {\bibfnamefont {X.-N.}\ \bibnamefont {Wang}},\ }\href
  {\doibase 10.1103/PhysRevLett.99.192301} {\bibfield  {journal} {\bibinfo
  {journal} {Phys. Rev. Lett.}\ }\textbf {\bibinfo {volume} {99}},\ \bibinfo
  {pages} {192301} (\bibinfo {year} {2007})},\ \Eprint
  {http://arxiv.org/abs/hep-ph/0703082} {arXiv:hep-ph/0703082 [hep-ph]}
  \BibitemShut {NoStop}%
\bibitem [{\citenamefont {Buchel}\ \emph {et~al.}(2005)\citenamefont {Buchel},
  \citenamefont {Liu},\ and\ \citenamefont {Starinets}}]{Buchel:2004di}%
  \BibitemOpen
  \bibfield  {author} {\bibinfo {author} {\bibfnamefont {A.}~\bibnamefont
  {Buchel}}, \bibinfo {author} {\bibfnamefont {J.~T.}\ \bibnamefont {Liu}}, \
  and\ \bibinfo {author} {\bibfnamefont {A.~O.}\ \bibnamefont {Starinets}},\
  }\href {\doibase 10.1016/j.nuclphysb.2004.11.055} {\bibfield  {journal}
  {\bibinfo  {journal} {Nucl. Phys.}\ }\textbf {\bibinfo {volume} {B707}},\
  \bibinfo {pages} {56} (\bibinfo {year} {2005})},\ \Eprint
  {http://arxiv.org/abs/hep-th/0406264} {arXiv:hep-th/0406264 [hep-th]}
  \BibitemShut {NoStop}%
\bibitem [{Note15()}]{Note15}%
  \BibitemOpen
  \bibinfo {note} {We recall that $\tau _w$ is {\protect \it a priori}
  physically independent of the electron-electron scattering rate $\tau
  _{\protect \rm ee}$.}\BibitemShut {Stop}%
\bibitem [{\citenamefont {Hartnoll}\ \emph {et~al.}(2007)\citenamefont
  {Hartnoll}, \citenamefont {Kovtun}, \citenamefont {Müller},\ and\
  \citenamefont {Sachdev}}]{Hartnoll:2007ih}%
  \BibitemOpen
  \bibfield  {author} {\bibinfo {author} {\bibfnamefont {S.~A.}\ \bibnamefont
  {Hartnoll}}, \bibinfo {author} {\bibfnamefont {P.~K.}\ \bibnamefont
  {Kovtun}}, \bibinfo {author} {\bibfnamefont {M.}~\bibnamefont {Müller}}, \
  and\ \bibinfo {author} {\bibfnamefont {S.}~\bibnamefont {Sachdev}},\ }\href
  {\doibase 10.1103/PhysRevB.76.144502} {\bibfield  {journal} {\bibinfo
  {journal} {Phys. Rev.}\ }\textbf {\bibinfo {volume} {B76}},\ \bibinfo {pages}
  {144502} (\bibinfo {year} {2007})},\ \Eprint {http://arxiv.org/abs/0706.3215}
  {arXiv:0706.3215 [cond-mat.str-el]} \BibitemShut {NoStop}%
\bibitem [{\citenamefont {Hartnoll}\ and\ \citenamefont
  {Herzog}(2007)}]{Hartnoll:2007ip}%
  \BibitemOpen
  \bibfield  {author} {\bibinfo {author} {\bibfnamefont {S.~A.}\ \bibnamefont
  {Hartnoll}}\ and\ \bibinfo {author} {\bibfnamefont {C.~P.}\ \bibnamefont
  {Herzog}},\ }\href {\doibase 10.1103/PhysRevD.76.106012} {\bibfield
  {journal} {\bibinfo  {journal} {Phys. Rev.}\ }\textbf {\bibinfo {volume}
  {D76}},\ \bibinfo {pages} {106012} (\bibinfo {year} {2007})},\ \Eprint
  {http://arxiv.org/abs/0706.3228} {arXiv:0706.3228 [hep-th]} \BibitemShut
  {NoStop}%
\bibitem [{Note16()}]{Note16}%
  \BibitemOpen
  \bibinfo {note} {The quantum critical resistance is defined via Ohm's law, $R
  = l/(\sigma _Q A)$, with $A$ the cross-sectional area of the wire. Here
  $A=W$.}\BibitemShut {Stop}%
\bibitem [{Note17()}]{Note17}%
  \BibitemOpen
  \bibinfo {note} {We assume that $\ell _{ee}\ll \ell _{\protect \rm imp}$
  holds, such that we are in the hydrodynamic regime.}\BibitemShut {Stop}%
\bibitem [{\citenamefont {Hanks}(1963)}]{hanks1963laminar}%
  \BibitemOpen
  \bibfield  {author} {\bibinfo {author} {\bibfnamefont {R.~W.}\ \bibnamefont
  {Hanks}},\ }\href@noop {} {\bibfield  {journal} {\bibinfo  {journal} {AIChE
  Journal}\ }\textbf {\bibinfo {volume} {9}},\ \bibinfo {pages} {45} (\bibinfo
  {year} {1963})}\BibitemShut {NoStop}%
\bibitem [{\citenamefont {Hanks}(1969)}]{hanks1969theory}%
  \BibitemOpen
  \bibfield  {author} {\bibinfo {author} {\bibfnamefont {R.~W.}\ \bibnamefont
  {Hanks}},\ }\href@noop {} {\bibfield  {journal} {\bibinfo  {journal} {AIChE
  Journal}\ }\textbf {\bibinfo {volume} {15}},\ \bibinfo {pages} {25} (\bibinfo
  {year} {1969})}\BibitemShut {NoStop}%
\bibitem [{\citenamefont {{Mendoza}}\ \emph {et~al.}(2011)\citenamefont
  {{Mendoza}}, \citenamefont {{Herrmann}},\ and\ \citenamefont
  {{Succi}}}]{Mendoza:2011ka}%
  \BibitemOpen
  \bibfield  {author} {\bibinfo {author} {\bibfnamefont {M.}~\bibnamefont
  {{Mendoza}}}, \bibinfo {author} {\bibfnamefont {H.~J.}\ \bibnamefont
  {{Herrmann}}}, \ and\ \bibinfo {author} {\bibfnamefont {S.}~\bibnamefont
  {{Succi}}},\ }\href {\doibase 10.1103/PhysRevLett.106.156601} {\bibfield
  {journal} {\bibinfo  {journal} {Physical Review Letters}\ }\textbf {\bibinfo
  {volume} {106}},\ \bibinfo {eid} {156601} (\bibinfo {year} {2011})},\ \Eprint
  {http://arxiv.org/abs/1201.6590} {arXiv:1201.6590 [cond-mat.mes-hall]}
  \BibitemShut {NoStop}%
\bibitem [{Note18()}]{Note18}%
  \BibitemOpen
  \bibinfo {note} {We define the kinematic viscosity $\kappa $ as the ratio of
  viscosity and energy density. There exists another kinematic viscosity, $\nu
  $, defined as the ratio of viscosity and mass density. The two viscosities
  are related by $\nu = v_F^2 \kappa $.}\BibitemShut {Stop}%
\bibitem [{Note19()}]{Note19}%
  \BibitemOpen
  \bibinfo {note} {A measurement of $\sigma _Q$ is more readily performed in a
  bulk sample where viscous effects are negligible, by comparing samples of
  different impurity content and extrapolating to the clean limit.}\BibitemShut
  {Stop}%
\bibitem [{\citenamefont {Kiselev}\ and\ \citenamefont
  {Schmalian}(2018)}]{Kiselev:2018wj}%
  \BibitemOpen
  \bibfield  {author} {\bibinfo {author} {\bibfnamefont {E.~I.}\ \bibnamefont
  {Kiselev}}\ and\ \bibinfo {author} {\bibfnamefont {J.}~\bibnamefont
  {Schmalian}},\ }\href@noop {} {\bibfield  {journal} {\bibinfo  {journal}
  {arXiv.org}\ } (\bibinfo {year} {2018})},\ \Eprint
  {http://arxiv.org/abs/1806.03933v2} {1806.03933v2} \BibitemShut {NoStop}%
\bibitem [{\citenamefont {Gout\'eraux}(2014)}]{Gouteraux:2014hca}%
  \BibitemOpen
  \bibfield  {author} {\bibinfo {author} {\bibfnamefont {B.}~\bibnamefont
  {Gout\'eraux}},\ }\href {\doibase 10.1007/JHEP04(2014)181} {\bibfield
  {journal} {\bibinfo  {journal} {JHEP}\ }\textbf {\bibinfo {volume} {04}},\
  \bibinfo {pages} {181} (\bibinfo {year} {2014})},\ \Eprint
  {http://arxiv.org/abs/1401.5436} {arXiv:1401.5436 [hep-th]} \BibitemShut
  {NoStop}%
\bibitem [{\citenamefont {Karch}\ and\ \citenamefont
  {O'Bannon}(2007)}]{Karch:2007pd}%
  \BibitemOpen
  \bibfield  {author} {\bibinfo {author} {\bibfnamefont {A.}~\bibnamefont
  {Karch}}\ and\ \bibinfo {author} {\bibfnamefont {A.}~\bibnamefont
  {O'Bannon}},\ }\href {\doibase 10.1088/1126-6708/2007/09/024} {\bibfield
  {journal} {\bibinfo  {journal} {JHEP}\ }\textbf {\bibinfo {volume} {09}},\
  \bibinfo {pages} {024} (\bibinfo {year} {2007})},\ \Eprint
  {http://arxiv.org/abs/0705.3870} {arXiv:0705.3870 [hep-th]} \BibitemShut
  {NoStop}%
\bibitem [{Note20()}]{Note20}%
  \BibitemOpen
  \bibinfo {note} {In particular in the Dirac limit $\mu /(k_B T) \ll 1$,
  $\varepsilon +p \approx sT$, and the dependence on thermodynamic variables
  drops out of \protect \textup {\hbox {\mathsurround \z@ \protect \normalfont
  (\ignorespaces \ref {eq:sigmaQ}\unskip \@@italiccorr )}}.}\BibitemShut
  {Stop}%
\bibitem [{Note21()}]{Note21}%
  \BibitemOpen
  \bibinfo {note} {In fact, the center flow can be understood as an effective
  channel flow with no-stress boundary conditions. These conditions are imposed
  at the edge of a layer of width $\lambda _G$. These boundary conditions then
  lead to Ohmic hydrodynamic flow at the center of the channel \cite
  {Lucas:2017wa}.}\BibitemShut {Stop}%
\bibitem [{\citenamefont {Anninos}\ and\ \citenamefont
  {Pastras}(2009)}]{AnninosPastras}%
  \BibitemOpen
  \bibfield  {author} {\bibinfo {author} {\bibfnamefont {D.}~\bibnamefont
  {Anninos}}\ and\ \bibinfo {author} {\bibfnamefont {G.}~\bibnamefont
  {Pastras}},\ }\href {http://stacks.iop.org/1126-6708/2009/i=07/a=030}
  {\bibfield  {journal} {\bibinfo  {journal} {Journal of High Energy Physics}\
  }\textbf {\bibinfo {volume} {2009}},\ \bibinfo {pages} {030} (\bibinfo {year}
  {2009})},\ \Eprint {http://arxiv.org/abs/0807.3478} {arXiv:0807.3478
  [hep-th]} \BibitemShut {NoStop}%
\bibitem [{\citenamefont {Witczak-Krempa}\ and\ \citenamefont
  {Sachdev}(2012)}]{SachdevKrempaPRB.86.235115}%
  \BibitemOpen
  \bibfield  {author} {\bibinfo {author} {\bibfnamefont {W.}~\bibnamefont
  {Witczak-Krempa}}\ and\ \bibinfo {author} {\bibfnamefont {S.}~\bibnamefont
  {Sachdev}},\ }\href {\doibase 10.1103/PhysRevB.86.235115} {\bibfield
  {journal} {\bibinfo  {journal} {Phys. Rev. B}\ }\textbf {\bibinfo {volume}
  {86}},\ \bibinfo {pages} {235115} (\bibinfo {year} {2012})}\BibitemShut
  {NoStop}%
\bibitem [{\citenamefont {Witczak-Krempa}\ and\ \citenamefont
  {Sachdev}(2013)}]{SachdevKrempaPRB.87.155149}%
  \BibitemOpen
  \bibfield  {author} {\bibinfo {author} {\bibfnamefont {W.}~\bibnamefont
  {Witczak-Krempa}}\ and\ \bibinfo {author} {\bibfnamefont {S.}~\bibnamefont
  {Sachdev}},\ }\href {\doibase 10.1103/PhysRevB.87.155149} {\bibfield
  {journal} {\bibinfo  {journal} {Phys. Rev. B}\ }\textbf {\bibinfo {volume}
  {87}},\ \bibinfo {pages} {155149} (\bibinfo {year} {2013})}\BibitemShut
  {NoStop}%
\bibitem [{\citenamefont {Grozdanov}\ and\ \citenamefont
  {Starinets}(2017)}]{Grozdanov2017}%
  \BibitemOpen
  \bibfield  {author} {\bibinfo {author} {\bibfnamefont {S.}~\bibnamefont
  {Grozdanov}}\ and\ \bibinfo {author} {\bibfnamefont {A.~O.}\ \bibnamefont
  {Starinets}},\ }\href {\doibase 10.1007/JHEP03(2017)166} {\bibfield
  {journal} {\bibinfo  {journal} {Journal of High Energy Physics}\ }\textbf
  {\bibinfo {volume} {2017}},\ \bibinfo {pages} {166} (\bibinfo {year}
  {2017})}\BibitemShut {NoStop}%
\bibitem [{\citenamefont {Gross}\ and\ \citenamefont
  {Sloan}(1987)}]{GROSS198741}%
  \BibitemOpen
  \bibfield  {author} {\bibinfo {author} {\bibfnamefont {D.~J.}\ \bibnamefont
  {Gross}}\ and\ \bibinfo {author} {\bibfnamefont {J.~H.}\ \bibnamefont
  {Sloan}},\ }\href {\doibase https://doi.org/10.1016/0550-3213(87)90465-2}
  {\bibfield  {journal} {\bibinfo  {journal} {Nuclear Physics B}\ }\textbf
  {\bibinfo {volume} {291}},\ \bibinfo {pages} {41 } (\bibinfo {year}
  {1987})}\BibitemShut {NoStop}%
\bibitem [{\citenamefont {Jensen}\ \emph {et~al.}(2012)\citenamefont {Jensen},
  \citenamefont {Kaminski}, \citenamefont {Kovtun}, \citenamefont {Meyer},
  \citenamefont {Ritz},\ and\ \citenamefont {Yarom}}]{Jensen:2011xb}%
  \BibitemOpen
  \bibfield  {author} {\bibinfo {author} {\bibfnamefont {K.}~\bibnamefont
  {Jensen}}, \bibinfo {author} {\bibfnamefont {M.}~\bibnamefont {Kaminski}},
  \bibinfo {author} {\bibfnamefont {P.}~\bibnamefont {Kovtun}}, \bibinfo
  {author} {\bibfnamefont {R.}~\bibnamefont {Meyer}}, \bibinfo {author}
  {\bibfnamefont {A.}~\bibnamefont {Ritz}}, \ and\ \bibinfo {author}
  {\bibfnamefont {A.}~\bibnamefont {Yarom}},\ }\href {\doibase
  10.1007/JHEP05(2012)102} {\bibfield  {journal} {\bibinfo  {journal} {JHEP}\
  }\textbf {\bibinfo {volume} {05}},\ \bibinfo {pages} {102} (\bibinfo {year}
  {2012})},\ \Eprint {http://arxiv.org/abs/1112.4498} {arXiv:1112.4498
  [hep-th]} \BibitemShut {NoStop}%
\bibitem [{\citenamefont {Hoyos}(2014)}]{Hoyos:2014pba}%
  \BibitemOpen
  \bibfield  {author} {\bibinfo {author} {\bibfnamefont {C.}~\bibnamefont
  {Hoyos}},\ }\href {\doibase 10.1142/S0217979214300072} {\bibfield  {journal}
  {\bibinfo  {journal} {Int. J. Mod. Phys.}\ }\textbf {\bibinfo {volume}
  {B28}},\ \bibinfo {pages} {1430007} (\bibinfo {year} {2014})},\ \Eprint
  {http://arxiv.org/abs/1403.4739} {arXiv:1403.4739 [cond-mat.mes-hall]}
  \BibitemShut {NoStop}%
\bibitem [{\citenamefont {{Pellegrino}}\ \emph {et~al.}(2017)\citenamefont
  {{Pellegrino}}, \citenamefont {{Torre}},\ and\ \citenamefont
  {{Polini}}}]{Pellegrino:2017fd}%
  \BibitemOpen
  \bibfield  {author} {\bibinfo {author} {\bibfnamefont {F.~M.~D.}\
  \bibnamefont {{Pellegrino}}}, \bibinfo {author} {\bibfnamefont
  {I.}~\bibnamefont {{Torre}}}, \ and\ \bibinfo {author} {\bibfnamefont
  {M.}~\bibnamefont {{Polini}}},\ }\href {\doibase 10.1103/PhysRevB.96.195401}
  {\bibfield  {journal} {\bibinfo  {journal} {Physical Review B}\ }\textbf
  {\bibinfo {volume} {96}},\ \bibinfo {eid} {195401} (\bibinfo {year}
  {2017})},\ \Eprint {http://arxiv.org/abs/1706.08363} {arXiv:1706.08363
  [cond-mat.mes-hall]} \BibitemShut {NoStop}%
\bibitem [{\citenamefont {Delacretaz}\ and\ \citenamefont
  {Gromov}(2017)}]{Delacretaz:2017yia}%
  \BibitemOpen
  \bibfield  {author} {\bibinfo {author} {\bibfnamefont {L.~V.}\ \bibnamefont
  {Delacretaz}}\ and\ \bibinfo {author} {\bibfnamefont {A.}~\bibnamefont
  {Gromov}},\ }\href {\doibase 10.1103/PhysRevLett.120.079901,
  10.1103/PhysRevLett.119.226602} {\bibfield  {journal} {\bibinfo  {journal}
  {Phys. Rev. Lett.}\ }\textbf {\bibinfo {volume} {119}},\ \bibinfo {pages}
  {226602} (\bibinfo {year} {2017})},\ \bibinfo {note} {[Addendum: Phys. Rev.
  Lett.120,no.7,079901(2018)]},\ \Eprint {http://arxiv.org/abs/1706.03773}
  {arXiv:1706.03773 [cond-mat.str-el]} \BibitemShut {NoStop}%
\bibitem [{Note22()}]{Note22}%
  \BibitemOpen
  \bibinfo {note} {A black brane is a black hole with planar horizon topology,
  in this case $\protect \mathbbm {R}^2$.}\BibitemShut {Stop}%
\bibitem [{\citenamefont {de~Haro}\ \emph {et~al.}(2001)\citenamefont
  {de~Haro}, \citenamefont {Solodukhin},\ and\ \citenamefont
  {Skenderis}}]{deHaro:2000vlm}%
  \BibitemOpen
  \bibfield  {author} {\bibinfo {author} {\bibfnamefont {S.}~\bibnamefont
  {de~Haro}}, \bibinfo {author} {\bibfnamefont {S.~N.}\ \bibnamefont
  {Solodukhin}}, \ and\ \bibinfo {author} {\bibfnamefont {K.}~\bibnamefont
  {Skenderis}},\ }\href {\doibase 10.1007/s002200100381} {\bibfield  {journal}
  {\bibinfo  {journal} {Commun. Math. Phys.}\ }\textbf {\bibinfo {volume}
  {217}},\ \bibinfo {pages} {595} (\bibinfo {year} {2001})},\ \Eprint
  {http://arxiv.org/abs/hep-th/0002230} {arXiv:hep-th/0002230 [hep-th]}
  \BibitemShut {NoStop}%
\bibitem [{\citenamefont {Papadimitriou}\ and\ \citenamefont
  {Skenderis}(2004)}]{Papadimitriou:2004rz}%
  \BibitemOpen
  \bibfield  {author} {\bibinfo {author} {\bibfnamefont {I.}~\bibnamefont
  {Papadimitriou}}\ and\ \bibinfo {author} {\bibfnamefont {K.}~\bibnamefont
  {Skenderis}},\ }\href {\doibase 10.1088/1126-6708/2004/10/075} {\bibfield
  {journal} {\bibinfo  {journal} {JHEP}\ }\textbf {\bibinfo {volume} {10}},\
  \bibinfo {pages} {075} (\bibinfo {year} {2004})},\ \Eprint
  {http://arxiv.org/abs/hep-th/0407071} {arXiv:hep-th/0407071 [hep-th]}
  \BibitemShut {NoStop}%
\bibitem [{\citenamefont {Papadimitriou}\ and\ \citenamefont
  {Skenderis}(2005)}]{Papadimitriou:2005ii}%
  \BibitemOpen
  \bibfield  {author} {\bibinfo {author} {\bibfnamefont {I.}~\bibnamefont
  {Papadimitriou}}\ and\ \bibinfo {author} {\bibfnamefont {K.}~\bibnamefont
  {Skenderis}},\ }\href {\doibase 10.1088/1126-6708/2005/08/004} {\bibfield
  {journal} {\bibinfo  {journal} {JHEP}\ }\textbf {\bibinfo {volume} {08}},\
  \bibinfo {pages} {004} (\bibinfo {year} {2005})},\ \Eprint
  {http://arxiv.org/abs/hep-th/0505190} {arXiv:hep-th/0505190 [hep-th]}
  \BibitemShut {NoStop}%
\bibitem [{\citenamefont {Erdmenger}\ \emph {et~al.}(2017)\citenamefont
  {Erdmenger}, \citenamefont {Fernandez}, \citenamefont {Goulart},\ and\
  \citenamefont {Witkowski}}]{Erdmenger:2016wyp}%
  \BibitemOpen
  \bibfield  {author} {\bibinfo {author} {\bibfnamefont {J.}~\bibnamefont
  {Erdmenger}}, \bibinfo {author} {\bibfnamefont {D.}~\bibnamefont
  {Fernandez}}, \bibinfo {author} {\bibfnamefont {P.}~\bibnamefont {Goulart}},
  \ and\ \bibinfo {author} {\bibfnamefont {P.}~\bibnamefont {Witkowski}},\
  }\href {\doibase 10.1007/JHEP03(2017)147} {\bibfield  {journal} {\bibinfo
  {journal} {JHEP}\ }\textbf {\bibinfo {volume} {03}},\ \bibinfo {pages} {147}
  (\bibinfo {year} {2017})},\ \Eprint {http://arxiv.org/abs/1611.09381}
  {arXiv:1611.09381 [hep-th]} \BibitemShut {NoStop}%
\bibitem [{Note23()}]{Note23}%
  \BibitemOpen
  \bibinfo {note} {The renormalization of the Fermi velocity $v_F$ is
  negligible at most temperatures except exponentially small ones \protect
  \cite {Lucas:2017wa}.}\BibitemShut {Stop}%
\bibitem [{Note24()}]{Note24}%
  \BibitemOpen
  \bibinfo {note} {We in particular do not claim that Einstein-Maxwell
  holography \protect \eqref {eq:bulkaction} can describe a weakly interacting
  Fermi liquid regime such as the one present in e.g. graphene at $\mu /T\gg
  1$.}\BibitemShut {Stop}%
\bibitem [{\citenamefont {Balasubramanian}\ and\ \citenamefont
  {McGreevy}(2008)}]{Balasubramanian:2008dm}%
  \BibitemOpen
  \bibfield  {author} {\bibinfo {author} {\bibfnamefont {K.}~\bibnamefont
  {Balasubramanian}}\ and\ \bibinfo {author} {\bibfnamefont {J.}~\bibnamefont
  {McGreevy}},\ }\href {\doibase 10.1103/PhysRevLett.101.061601} {\bibfield
  {journal} {\bibinfo  {journal} {Phys. Rev. Lett.}\ }\textbf {\bibinfo
  {volume} {101}},\ \bibinfo {pages} {061601} (\bibinfo {year} {2008})},\
  \Eprint {http://arxiv.org/abs/0804.4053} {arXiv:0804.4053 [hep-th]}
  \BibitemShut {NoStop}%
\bibitem [{\citenamefont {Charmousis}\ \emph {et~al.}(2010)\citenamefont
  {Charmousis}, \citenamefont {Gouteraux}, \citenamefont {Kim}, \citenamefont
  {Kiritsis},\ and\ \citenamefont {Meyer}}]{Charmousis:2010zz}%
  \BibitemOpen
  \bibfield  {author} {\bibinfo {author} {\bibfnamefont {C.}~\bibnamefont
  {Charmousis}}, \bibinfo {author} {\bibfnamefont {B.}~\bibnamefont
  {Gouteraux}}, \bibinfo {author} {\bibfnamefont {B.~S.}\ \bibnamefont {Kim}},
  \bibinfo {author} {\bibfnamefont {E.}~\bibnamefont {Kiritsis}}, \ and\
  \bibinfo {author} {\bibfnamefont {R.}~\bibnamefont {Meyer}},\ }\href
  {\doibase 10.1007/JHEP11(2010)151} {\bibfield  {journal} {\bibinfo  {journal}
  {JHEP}\ }\textbf {\bibinfo {volume} {11}},\ \bibinfo {pages} {151} (\bibinfo
  {year} {2010})},\ \Eprint {http://arxiv.org/abs/1005.4690} {arXiv:1005.4690
  [hep-th]} \BibitemShut {NoStop}%
\bibitem [{\citenamefont {Aharony}\ \emph {et~al.}(2008)\citenamefont
  {Aharony}, \citenamefont {Bergman}, \citenamefont {Jafferis},\ and\
  \citenamefont {Maldacena}}]{Aharony:2008ug}%
  \BibitemOpen
  \bibfield  {author} {\bibinfo {author} {\bibfnamefont {O.}~\bibnamefont
  {Aharony}}, \bibinfo {author} {\bibfnamefont {O.}~\bibnamefont {Bergman}},
  \bibinfo {author} {\bibfnamefont {D.~L.}\ \bibnamefont {Jafferis}}, \ and\
  \bibinfo {author} {\bibfnamefont {J.}~\bibnamefont {Maldacena}},\ }\href
  {\doibase 10.1088/1126-6708/2008/10/091} {\bibfield  {journal} {\bibinfo
  {journal} {JHEP}\ }\textbf {\bibinfo {volume} {10}},\ \bibinfo {pages} {091}
  (\bibinfo {year} {2008})},\ \Eprint {http://arxiv.org/abs/0806.1218}
  {arXiv:0806.1218 [hep-th]} \BibitemShut {NoStop}%
\bibitem [{\citenamefont {Hoyos}\ \emph {et~al.}(2016)\citenamefont {Hoyos},
  \citenamefont {Rodr{\'i}guez~Fern{\'a}ndez}, \citenamefont {Jokela},\ and\
  \citenamefont {Vuorinen}}]{Hoyos:2016zke}%
  \BibitemOpen
  \bibfield  {author} {\bibinfo {author} {\bibfnamefont {C.}~\bibnamefont
  {Hoyos}}, \bibinfo {author} {\bibfnamefont {D.}~\bibnamefont
  {Rodr{\'i}guez~Fern{\'a}ndez}}, \bibinfo {author} {\bibfnamefont
  {N.}~\bibnamefont {Jokela}}, \ and\ \bibinfo {author} {\bibfnamefont
  {A.}~\bibnamefont {Vuorinen}},\ }\href {\doibase
  10.1103/PhysRevLett.117.032501} {\bibfield  {journal} {\bibinfo  {journal}
  {Phys. Rev. Lett.}\ }\textbf {\bibinfo {volume} {117}},\ \bibinfo {pages}
  {032501} (\bibinfo {year} {2016})},\ \Eprint
  {http://arxiv.org/abs/1603.02943} {arXiv:1603.02943 [hep-ph]} \BibitemShut
  {NoStop}%
\bibitem [{\citenamefont {Herzog}\ \emph {et~al.}(2007)\citenamefont {Herzog},
  \citenamefont {Kovtun}, \citenamefont {Sachdev},\ and\ \citenamefont
  {Son}}]{Herzog:2007ij}%
  \BibitemOpen
  \bibfield  {author} {\bibinfo {author} {\bibfnamefont {C.~P.}\ \bibnamefont
  {Herzog}}, \bibinfo {author} {\bibfnamefont {P.}~\bibnamefont {Kovtun}},
  \bibinfo {author} {\bibfnamefont {S.}~\bibnamefont {Sachdev}}, \ and\
  \bibinfo {author} {\bibfnamefont {D.~T.}\ \bibnamefont {Son}},\ }\href
  {\doibase 10.1103/PhysRevD.75.085020} {\bibfield  {journal} {\bibinfo
  {journal} {Phys. Rev.}\ }\textbf {\bibinfo {volume} {D75}},\ \bibinfo {pages}
  {085020} (\bibinfo {year} {2007})},\ \Eprint
  {http://arxiv.org/abs/hep-th/0701036} {arXiv:hep-th/0701036 [hep-th]}
  \BibitemShut {NoStop}%
\bibitem [{\citenamefont {Myers}\ \emph {et~al.}(2011)\citenamefont {Myers},
  \citenamefont {Sachdev},\ and\ \citenamefont {Singh}}]{Myers:2010pk}%
  \BibitemOpen
  \bibfield  {author} {\bibinfo {author} {\bibfnamefont {R.~C.}\ \bibnamefont
  {Myers}}, \bibinfo {author} {\bibfnamefont {S.}~\bibnamefont {Sachdev}}, \
  and\ \bibinfo {author} {\bibfnamefont {A.}~\bibnamefont {Singh}},\ }\href
  {\doibase 10.1103/PhysRevD.83.066017} {\bibfield  {journal} {\bibinfo
  {journal} {Phys. Rev.}\ }\textbf {\bibinfo {volume} {D83}},\ \bibinfo {pages}
  {066017} (\bibinfo {year} {2011})},\ \Eprint {http://arxiv.org/abs/1010.0443}
  {arXiv:1010.0443 [hep-th]} \BibitemShut {NoStop}%
\bibitem [{Note25()}]{Note25}%
  \BibitemOpen
  \bibinfo {note} {For example, $\protect \mathaccentV {vec}17E{v} = \protect
  \mathaccentV {vec}17E{u}/v_c$ where $\protect \mathaccentV {vec}17E{u}, v_c$
  is the fluid and characteristic velocity, respectively.}\BibitemShut {Stop}%
\bibitem [{Note26()}]{Note26}%
  \BibitemOpen
  \bibinfo {note} {Without loss of generality, we have set the external fields
  to zero and taken $\partial _\mu \epsilon = 0 =\partial _\mu u^\mu
  $.}\BibitemShut {Stop}%
\bibitem [{Note27()}]{Note27}%
  \BibitemOpen
  \bibinfo {note} {$\protect \mathaccentV {vec}17E{\protect \mathaccentV
  {tilde}07E{\Sigma }}$ after rescaling contains terms proportional to $\left
  ({v_{\protect \rm max} \over v_F}\right )^a \protect \tmspace +\thickmuskip
  {.2777em},\protect \tmspace +\thickmuskip {.2777em}a=0,2,4$.}\BibitemShut
  {Stop}%
\bibitem [{Note28()}]{Note28}%
  \BibitemOpen
  \bibinfo {note} {Recall that the fluid in particular is conformal. This is in
  contrast to common non-relativistic fluids, for which $p\ll \varepsilon $. In
  this case, $Re_{NR} = Re_{NS}$.}\BibitemShut {Stop}%
\bibitem [{\citenamefont {Lucas}\ and\ \citenamefont
  {Hartnoll}(2018)}]{Lucas:2017vlc}%
  \BibitemOpen
  \bibfield  {author} {\bibinfo {author} {\bibfnamefont {A.}~\bibnamefont
  {Lucas}}\ and\ \bibinfo {author} {\bibfnamefont {S.~A.}\ \bibnamefont
  {Hartnoll}},\ }\href {\doibase 10.1103/PhysRevB.97.045105} {\bibfield
  {journal} {\bibinfo  {journal} {Phys. Rev.}\ }\textbf {\bibinfo {volume}
  {B97}},\ \bibinfo {pages} {045105} (\bibinfo {year} {2018})},\ \Eprint
  {http://arxiv.org/abs/1706.04621} {arXiv:1706.04621 [cond-mat.str-el]}
  \BibitemShut {NoStop}%
\bibitem [{\citenamefont {Mateos}\ and\ \citenamefont
  {Trancanelli}(2011{\natexlab{a}})}]{Mateos:2011ix}%
  \BibitemOpen
  \bibfield  {author} {\bibinfo {author} {\bibfnamefont {D.}~\bibnamefont
  {Mateos}}\ and\ \bibinfo {author} {\bibfnamefont {D.}~\bibnamefont
  {Trancanelli}},\ }\href {\doibase 10.1103/PhysRevLett.107.101601} {\bibfield
  {journal} {\bibinfo  {journal} {Phys. Rev. Lett.}\ }\textbf {\bibinfo
  {volume} {107}},\ \bibinfo {pages} {101601} (\bibinfo {year}
  {2011}{\natexlab{a}})},\ \Eprint {http://arxiv.org/abs/1105.3472}
  {arXiv:1105.3472 [hep-th]} \BibitemShut {NoStop}%
\bibitem [{\citenamefont {Rebhan}\ and\ \citenamefont
  {Steineder}(2012)}]{Rebhan:2011vd}%
  \BibitemOpen
  \bibfield  {author} {\bibinfo {author} {\bibfnamefont {A.}~\bibnamefont
  {Rebhan}}\ and\ \bibinfo {author} {\bibfnamefont {D.}~\bibnamefont
  {Steineder}},\ }\href {\doibase 10.1103/PhysRevLett.108.021601} {\bibfield
  {journal} {\bibinfo  {journal} {Phys. Rev. Lett.}\ }\textbf {\bibinfo
  {volume} {108}},\ \bibinfo {pages} {021601} (\bibinfo {year} {2012})},\
  \Eprint {http://arxiv.org/abs/1110.6825} {arXiv:1110.6825 [hep-th]}
  \BibitemShut {NoStop}%
\bibitem [{\citenamefont {Mateos}\ and\ \citenamefont
  {Trancanelli}(2011{\natexlab{b}})}]{Mateos:2011tv}%
  \BibitemOpen
  \bibfield  {author} {\bibinfo {author} {\bibfnamefont {D.}~\bibnamefont
  {Mateos}}\ and\ \bibinfo {author} {\bibfnamefont {D.}~\bibnamefont
  {Trancanelli}},\ }\href {\doibase 10.1007/JHEP07(2011)054} {\bibfield
  {journal} {\bibinfo  {journal} {JHEP}\ }\textbf {\bibinfo {volume} {07}},\
  \bibinfo {pages} {054} (\bibinfo {year} {2011}{\natexlab{b}})},\ \Eprint
  {http://arxiv.org/abs/1106.1637} {arXiv:1106.1637 [hep-th]} \BibitemShut
  {NoStop}%
\bibitem [{\citenamefont {Erdmenger}\ \emph {et~al.}(2013)\citenamefont
  {Erdmenger}, \citenamefont {Fernandez},\ and\ \citenamefont
  {Zeller}}]{Erdmenger:2012zu}%
  \BibitemOpen
  \bibfield  {author} {\bibinfo {author} {\bibfnamefont {J.}~\bibnamefont
  {Erdmenger}}, \bibinfo {author} {\bibfnamefont {D.}~\bibnamefont
  {Fernandez}}, \ and\ \bibinfo {author} {\bibfnamefont {H.}~\bibnamefont
  {Zeller}},\ }\href {\doibase 10.1007/JHEP04(2013)049} {\bibfield  {journal}
  {\bibinfo  {journal} {JHEP}\ }\textbf {\bibinfo {volume} {04}},\ \bibinfo
  {pages} {049} (\bibinfo {year} {2013})},\ \Eprint
  {http://arxiv.org/abs/1212.4838} {arXiv:1212.4838 [hep-th]} \BibitemShut
  {NoStop}%
\bibitem [{\citenamefont {Brigante}\ \emph {et~al.}(2008)\citenamefont
  {Brigante}, \citenamefont {Liu}, \citenamefont {Myers}, \citenamefont
  {Shenker},\ and\ \citenamefont {Yaida}}]{Brigante:2008gz}%
  \BibitemOpen
  \bibfield  {author} {\bibinfo {author} {\bibfnamefont {M.}~\bibnamefont
  {Brigante}}, \bibinfo {author} {\bibfnamefont {H.}~\bibnamefont {Liu}},
  \bibinfo {author} {\bibfnamefont {R.~C.}\ \bibnamefont {Myers}}, \bibinfo
  {author} {\bibfnamefont {S.}~\bibnamefont {Shenker}}, \ and\ \bibinfo
  {author} {\bibfnamefont {S.}~\bibnamefont {Yaida}},\ }\href {\doibase
  10.1103/PhysRevLett.100.191601} {\bibfield  {journal} {\bibinfo  {journal}
  {Phys. Rev. Lett.}\ }\textbf {\bibinfo {volume} {100}},\ \bibinfo {pages}
  {191601} (\bibinfo {year} {2008})},\ \Eprint {http://arxiv.org/abs/0802.3318}
  {arXiv:0802.3318 [hep-th]} \BibitemShut {NoStop}%
\bibitem [{\citenamefont {Camanho}\ \emph {et~al.}(2016)\citenamefont
  {Camanho}, \citenamefont {Edelstein}, \citenamefont {Maldacena},\ and\
  \citenamefont {Zhiboedov}}]{Camanho:2014apa}%
  \BibitemOpen
  \bibfield  {author} {\bibinfo {author} {\bibfnamefont {X.~O.}\ \bibnamefont
  {Camanho}}, \bibinfo {author} {\bibfnamefont {J.~D.}\ \bibnamefont
  {Edelstein}}, \bibinfo {author} {\bibfnamefont {J.}~\bibnamefont
  {Maldacena}}, \ and\ \bibinfo {author} {\bibfnamefont {A.}~\bibnamefont
  {Zhiboedov}},\ }\href {\doibase 10.1007/JHEP02(2016)020} {\bibfield
  {journal} {\bibinfo  {journal} {JHEP}\ }\textbf {\bibinfo {volume} {02}},\
  \bibinfo {pages} {020} (\bibinfo {year} {2016})},\ \Eprint
  {http://arxiv.org/abs/1407.5597} {arXiv:1407.5597 [hep-th]} \BibitemShut
  {NoStop}%
\bibitem [{\citenamefont {Kats}\ and\ \citenamefont
  {Petrov}(2009)}]{Kats:2007mq}%
  \BibitemOpen
  \bibfield  {author} {\bibinfo {author} {\bibfnamefont {Y.}~\bibnamefont
  {Kats}}\ and\ \bibinfo {author} {\bibfnamefont {P.}~\bibnamefont {Petrov}},\
  }\href {\doibase 10.1088/1126-6708/2009/01/044} {\bibfield  {journal}
  {\bibinfo  {journal} {JHEP}\ }\textbf {\bibinfo {volume} {01}},\ \bibinfo
  {pages} {044} (\bibinfo {year} {2009})},\ \Eprint
  {http://arxiv.org/abs/0712.0743} {arXiv:0712.0743 [hep-th]} \BibitemShut
  {NoStop}%
\bibitem [{\citenamefont {Cremonini}(2011)}]{Cremonini:2011iq}%
  \BibitemOpen
  \bibfield  {author} {\bibinfo {author} {\bibfnamefont {S.}~\bibnamefont
  {Cremonini}},\ }\href {\doibase 10.1142/S0217984911027315} {\bibfield
  {journal} {\bibinfo  {journal} {Mod. Phys. Lett.}\ }\textbf {\bibinfo
  {volume} {B25}},\ \bibinfo {pages} {1867} (\bibinfo {year} {2011})},\ \Eprint
  {http://arxiv.org/abs/1108.0677} {arXiv:1108.0677 [hep-th]} \BibitemShut
  {NoStop}%
\bibitem [{\citenamefont {Grozdanov}\ \emph {et~al.}(2016)\citenamefont
  {Grozdanov}, \citenamefont {Kaplis},\ and\ \citenamefont
  {Starinets}}]{grozdanov2016strong}%
  \BibitemOpen
  \bibfield  {author} {\bibinfo {author} {\bibfnamefont {S.}~\bibnamefont
  {Grozdanov}}, \bibinfo {author} {\bibfnamefont {N.}~\bibnamefont {Kaplis}}, \
  and\ \bibinfo {author} {\bibfnamefont {A.~O.}\ \bibnamefont {Starinets}},\
  }\href@noop {} {\bibfield  {journal} {\bibinfo  {journal} {Journal of High
  Energy Physics}\ }\textbf {\bibinfo {volume} {2016}},\ \bibinfo {pages} {151}
  (\bibinfo {year} {2016})}\BibitemShut {NoStop}%
\bibitem [{\citenamefont {Strickland}(2018)}]{Strickland:2018exs}%
  \BibitemOpen
  \bibfield  {author} {\bibinfo {author} {\bibfnamefont {M.}~\bibnamefont
  {Strickland}},\ }in\ \href@noop {} {\emph {\bibinfo {booktitle} {{27th
  International Conference on Ultrarelativistic Nucleus-Nucleus Collisions
  (Quark Matter 2018) Venice, Italy, May 14-19, 2018}}}}\ (\bibinfo {year}
  {2018})\ \Eprint {http://arxiv.org/abs/1807.07191} {arXiv:1807.07191
  [nucl-th]} \BibitemShut {NoStop}%
\bibitem [{\citenamefont {{Kadanoff}}\ and\ \citenamefont
  {{Martin}}(1963)}]{1963AnPhy..24..419K}%
  \BibitemOpen
  \bibfield  {author} {\bibinfo {author} {\bibfnamefont {L.~P.}\ \bibnamefont
  {{Kadanoff}}}\ and\ \bibinfo {author} {\bibfnamefont {P.~C.}\ \bibnamefont
  {{Martin}}},\ }\href {\doibase 10.1016/0003-4916(63)90078-2} {\bibfield
  {journal} {\bibinfo  {journal} {Annals of Physics}\ }\textbf {\bibinfo
  {volume} {24}},\ \bibinfo {pages} {419} (\bibinfo {year} {1963})}\BibitemShut
  {NoStop}%
\bibitem [{\citenamefont {Kovtun}(2012)}]{Kovtun:2012rj}%
  \BibitemOpen
  \bibfield  {author} {\bibinfo {author} {\bibfnamefont {P.}~\bibnamefont
  {Kovtun}},\ }\bibfield  {booktitle} {\emph {\bibinfo {booktitle} {{INT Summer
  School on Applications of String Theory Seattle, Washington, USA, July 18-29,
  2011}}},\ }\href {\doibase 10.1088/1751-8113/45/47/473001} {\bibfield
  {journal} {\bibinfo  {journal} {J. Phys.}\ }\textbf {\bibinfo {volume}
  {A45}},\ \bibinfo {pages} {473001} (\bibinfo {year} {2012})},\ \Eprint
  {http://arxiv.org/abs/1205.5040} {arXiv:1205.5040 [hep-th]} \BibitemShut
  {NoStop}%
\end{thebibliography}%

\end{document}